\documentclass[utf8]{article}
\usepackage{natbib}
\usepackage{graphicx} 
\usepackage{tabularx}
\usepackage{amsmath, bbm, dirtytalk, makecell, dcolumn, multirow, floatrow}
\usepackage{url,hyperref,lineno,microtype,subcaption, cleveref}
\usepackage{booktabs}
\usepackage{fullpage}
\usepackage{authblk}

\title{American Views About Election Fraud in 2024}
\author{Mitchell Linegar and R. Michael Alvarez}
\affil{Ronald and Maxine Linde Center for Science, Society, and Policy, Division of Humanities and Social Sciences, 
California Institute of Technology}
\affil{Paper provisionally accepted for publication in \textit{Frontiers in Political Science}}
\date{\today}

\begin{document}




\maketitle

\begin{abstract}
What are the opinions of American registered voters about election fraud and types of election fraud as we head into the final stages of the 2024 Presidential election?  In this paper we use data from an online national survey of 2,211 U.S. registered voters interviewed between June 26 - July 3, 2024.  Respondents were asked how common they thought that ten different types of election fraud might be in the U.S.  In our analysis, we show that substantial proportions of U.S. registered voters believe that these types of election fraud are common.  Our multivariate analysis shows that partisanship correlates strongly with endorsement of types of election fraud, with Republicans consistently more likely to state that types of election fraud are common, even when we control for a wide variety of other factors.  We also find that conspiratorial thinking is strongly correlated with belief in the occurrence of types of election fraud, even when we control for partisanship.  Our results reported in this paper provide important data regarding how American registered voters perceive the prevalence of types of election fraud, just months before the 2024 Presidential election.  
\end{abstract}

\section{Introduction}

Research conducted during the past two decades has concluded that there is no evidence of significant election fraud in contemporary American national elections \citep{alvarez-katz_2008, mebane_2008, herron2019mail, eggers_fraud_2021}.  This reality has not prevented questions about the integrity of the electoral process in the United States.  Sometimes those questions arise due to misunderstandings or misperceptions about how elections are conducted in the United States.  A good example is the ``blue shift'' in reporting of election results after Election day; observers sometimes may wonder why reported election results shift from one party to another in the days or weeks after an election.  But researchers have shown that these shifts are not due to election fraud, rather they arise from distinct patterns in the timing of ballot return, processing, and tabulation \citep{foley2020, li2022election}.

But sometimes these questions about the integrity of American elections persist, and take on a life of their own, even after concerns about election outcomes have been demonstrated to have legitimate and valid explanations.  One of the most prominent examples of this occurred after the 2020 election, when then President Trump, his supporters, and some Republican officials, continued to press claims of election fraud despite evidence to the contrary.  Situations where prominent political elites continue to press false claims of election fraud can be very problematic, as millions of American voters may continue to believe that election fraud exists in American national elections, which might affect their confidence in election administration and their willingness to participate in the electoral process \citep{green_etal_2022, Berlinski_etal_2023}.  

In this paper, we analyze data collected in June and July of 2024, from a sample of 2,211 American registered voters weighted to represent the electorate, to document the extent to which concerns about various types of election fraud, or types of election fraud, persist from 2020.  We analyze in detail who among this sample believes that ten different types of types of election fraud are common or not common.  This allows us to evaluate the salience of types of election fraud as America heads into a critical presidential election as well as for whom they are salient.

\section{Past Research}\label{sec:lit}

Twenty-four years ago, many researchers and scholars were shocked by the 2000 U.S. Presidential election which exposed problems with voting technology and administrative practices.  The election was very close, with only a few hundred votes separating winner from loser in important battleground states like Florida.  The closeness of the election brought a great deal of scrutiny of election administration and voting technologies, as prior to the election there was little research on U.S. election administration and technologies \citep{Caltech_MIT_2001}.  This attention to the details of election procedures and technologies helped develop new research areas  in law, social science, and computer science \citep{alvarez2013evaluating}.

The research agenda immediately after the 2000 U.S. presidential election focused heavily on election technology.  As states and counties in the U.S. rushed to replace old lever and punchcard voting systems, some of them began acquiring and using new electronic voting machines. By 2004, concerns began to emerge about the security of electronic voting systems \citep{kohno_etal_2004}.  Others began to question whether procedural changes, like requiring government-issued photo identification in order to vote, might effect voter perceptions about election integrity, election confidence, and opinions about the extent to which types of election fraud are perpetrated in the U.S.  This was a central argument in the United States Supreme Court's decision in the 2008 \textit{Crawford v. Marion County Election Board} decision on requiring government-issued photo identification for voting \citep{ansolabehere2007vote, atkeson_etal_2014}.  All of these questions and concerns led scholars to start studying what voters themselves thought about voting system security and election fraud \citep{alvarez-hall_2008, alvarez_hall_2008}. This direction of research has become especially salient following the 2020 election and the events on January 6th, 2021.

Despite the scholarly interest in voter perceptions of election fraud, there has not been a great deal of quantitative research on the topic, nor has there been much theoretical development of a model of perceptions of election fraud.  Most of the attention of scholars has been on measuring and analyzing the cross-sectional variance in survey measures of voter confidence and trust in elections \citep{atkeson2007effect, alvarez2008americans, atkeson2015voter}.  In the literature, some studies have looked at the association between survey measures of election fraud and voter confidence, typically finding that voters more concerned about fraud have lower confidence and trust in the election process \citep{alvarez2021voting, Berlinski_etal_2023}.

That said, the limited literature on the correlates of perceptions of election fraud have provided some important guidance for our study.  First, and not surprising in a highly partisan and polarized environment, opinions about election fraud are rooted in voter partisanship and in how political parties frame discussions about election fraud \citep{beaulieu2014}. Indeed, elite messaging plays an important role in forming public opinions, and shaping polarized partisan views \citep{bowler2015election,bowler2016partisan,bowler2024confidence}.  Relatedly, it makes sense that the same ``loser effect'' seen in studies of voter confidence will be seen in perceptions of election fraud, with partisans of parties losing recent elections more likely to express concerns about election fraud \citep{anderson2005losers,sances2015partisanship, sinclair2018}, and more likely to support electoral reform \citep{hood2022getting,hood2023partisan,hood2024winners}. A particularly salient example of this effect followed the 2020 election when, after his electoral loss, Donald Trump claimed that Joe Biden's victory was fraudulent. Though untrue, a large majority of Republicans said they agree with his claim \citep{jacobson2023dimensions}.

Some have studied perceptions about election fraud over election cycles, noting that while partisanship is correlated with opinions about election fraud, the relationship seems modest \citep{ENDERS2021102366}.  Importantly these same studies have confirmed other research results, showing that conspiratorial thinking is correlated with perceptions about election fraud \citep{ENDERS2021102366, Edelson_20173}.  This underscores the general hypothesis that voters respond to their information environment, in particular what trusted elites say about election fraud \citep{cottrell2018exploration}.  These studies provide the basic framework for our subsequent analysis of our recently collected 2024 dataset, which we describe in the next section.

\section{Methodology}

We conducted an online survey of 2021 U.S. registered voters, which was fielded by YouGov June 26 to July 3, 2024.  YouGov aims to produce a dataset that would approximate a representative sample of this population.  They weighted the responses (as described in more detail in the Supplementary Material), and in the analyses reported below, we use the weights provided by YouGov.  The margin of error for the survey is approximately 2.4\%.  

In our survey, we asked subjects to tell us how commonly they believe that ten examples of election fraud occur in the United States.  Subjects could indicate that they believed for each of these types of election fraud that ``it is very common'', ``it occurs occassionally'', ``it occurs infrequently'', ``it almost never occurs'', or ``I'm not sure.''  We provide in the Supplementary Material the complete data for all response options to each of the ten election fraud questions.  However, for analytical purposes and ease of exposition, in the paper we focus on the responses aggregated to common (it is very common or it occurs occasionally) or not common (it occurs infrequently or it almost never occurs).  We provide additional methodological details about the survey in the Supplementary Materials Sections 1 and 2.  

These are similar to, and in some cases identical to, questions that have been included in past surveys, in particular the \textit{The Survey of the Performance of American Elections} \citep{stewart_spae_2022}.  This allows us to compare our responses from the summer of 2024 to those from previous surveys, specifically those from 2012, 2014, 2016, 2020, and 2022 \citep{stewart2012spae, stewart2014spae, stewart2016spae, stewart2020spae, stewart2022spae}.  The ten election fraud questions (and the labels we use for them) are:
\begin{itemize}
\item People voting more than once in an election (\textit{Multiple Voting})
\item People stealing or tampering with ballots that have been counted (\textit{Ballot Tampering})
\item People pretending to be someone else when going to vote (\textit{Impersonation})
\item People voting who are not U.S. citizens (\textit{Non-Citizen Voting})
\item People taking advantage of absentee or mail balloting to engage in vote fraud (\textit{Mail Ballot Fraud})
\item Officials changing the reported vote count in a way that is not a true reflection of the ballots that were actually counted (\textit{Official Tampering})
\item Vote counting software manipulated in a way to not count ballots as intended (\textit{Software Hacking})
\item Paying voters to cast a ballot for a particular candidate (\textit{Paying Voters})
\item Voting under fraudulent voter registrations that use a fake name and a fake address (\textit{Registration Fraud})
\item People submitting too many ballots in drop boxes on behalf of others (\textit{Dropbox Fraud})
\end{itemize}

Our analysis of these data proceeds as follows.  We first show graphically the responses we received in our survey, and we also compare ours to those from the SPAE.  We then discuss the simple bivariate associations we have in our data, between each of the election fraud questions, and our demographic and attitudinal features (the detailed tables are provided in the Supplementary Materials).  Next, we use multivariate regression models to examine the covariates from our survey which most strongly associate with belief in election fraud.  In the main text below, we focus on a summary scale of the number of types of election fraud that each subject in our sample endorses; we use this summary regression for ease of presentation and interpretation.  In the main text we show marginal effect plots to clarify the associations between each covariate and the summary scale of the types of election fraud.  We provide regression results for each of the election fraud beliefs in the Supplementary Materials (Section 4). Our survey contains quite a few covariates of interest:  age, gender, race and ethnicity, education, region of residence, urban or rural residence, partisanship, ideology, political interest.  

The survey also contains batteries of questions designed to produce measures of beliefs in conspiratorial thinking and populism, the wordings of these questions are provided in the Supplementary Materials (Section 1). To form conspiracy and populism scores for each participant, we ask six conspiracy and populism questions, rated on a five-point scale, which we then sum to form each of these measures. These measures thus range from a minimum of 6 to a maximum of 30. Finally, we bin these scores in 5-point intervals to present the results. 

\section{Results}

\subsection{Graphical Results}

\begin{figure}[ht]
    \centering
    \includegraphics[width=.85\linewidth]{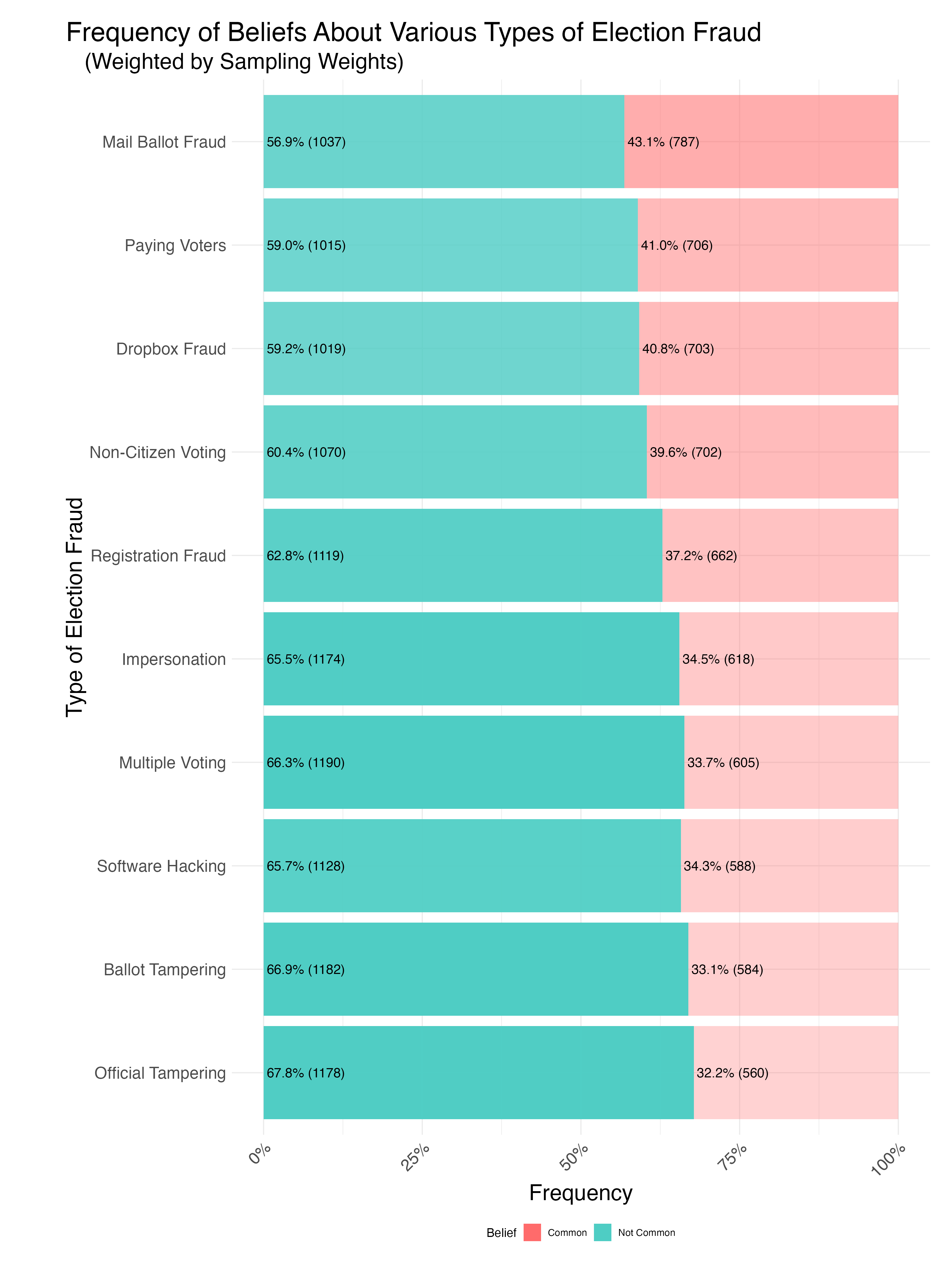}
    \caption{Concerns about types of election fraud, June-July 2024 survey.  Weighted frequencies of U.S. registered voters saying that the particular type of election fraud is very common or that it occurs occasionally.}
    \label{crime_freqs}
\end{figure}

We begin by discussing the weighted frequencies for each of the ten election fraud questions included in our survey.  The frequencies we provide in Figure~\ref{crime_freqs} are the weighted proportion of registered voters in our sample saying that the particular type of fraud `` is very common'' or ``occurs occasionally''.  We aggregate these survey responses in this way to make presentation, analysis, and interpretation easier; this is also consistent with how past studies have presented survey responses about election fraud \citep{stewart_spae_2022}.
Readers interested in seeing the full range of responses to each election fraud question can see the Supplementary Material (Table S2).  

We array the types of election fraud from highest frequency of concern to the lowest.  The first important conclusion to take from Figure~\ref{crime_freqs} is that each of these types of election fraud is considered to be common or occasional for a third or more of American registered voters.  That result alone is cause for concern, as based on previous research discussed earlier, it is also highly likely that registered voters who believe that any of these types of election fraud are common or occasional are less likely to participate in elections and are less likely to have confidence in the electoral process.

Second, we see in Figure~\ref{crime_freqs} a cluster of four types of election fraud that are viewed as common or occasional by at least four in ten U.S. registered voters:  mail ballot fraud (0.43), paying voters (0.41), dropbox fraud (0.41) and non-citizen voting (0.40).  Of the remaining types of election fraud, there are three with only a third of U.S. registered voters saying they are common or occasional:  multiple voting (0.34), ballot tampering (0.33), and tampering by election officials (0.32).  This leads to a second conclusion, some of these types of election fraud are seen by U.S. registered voters to be more prevalent, while others are seen as less likely to be prevalent.  

\subsection{Over-time Comparison}

\begin{figure}[ht]
    \centering
    \includegraphics[width=.9\linewidth]{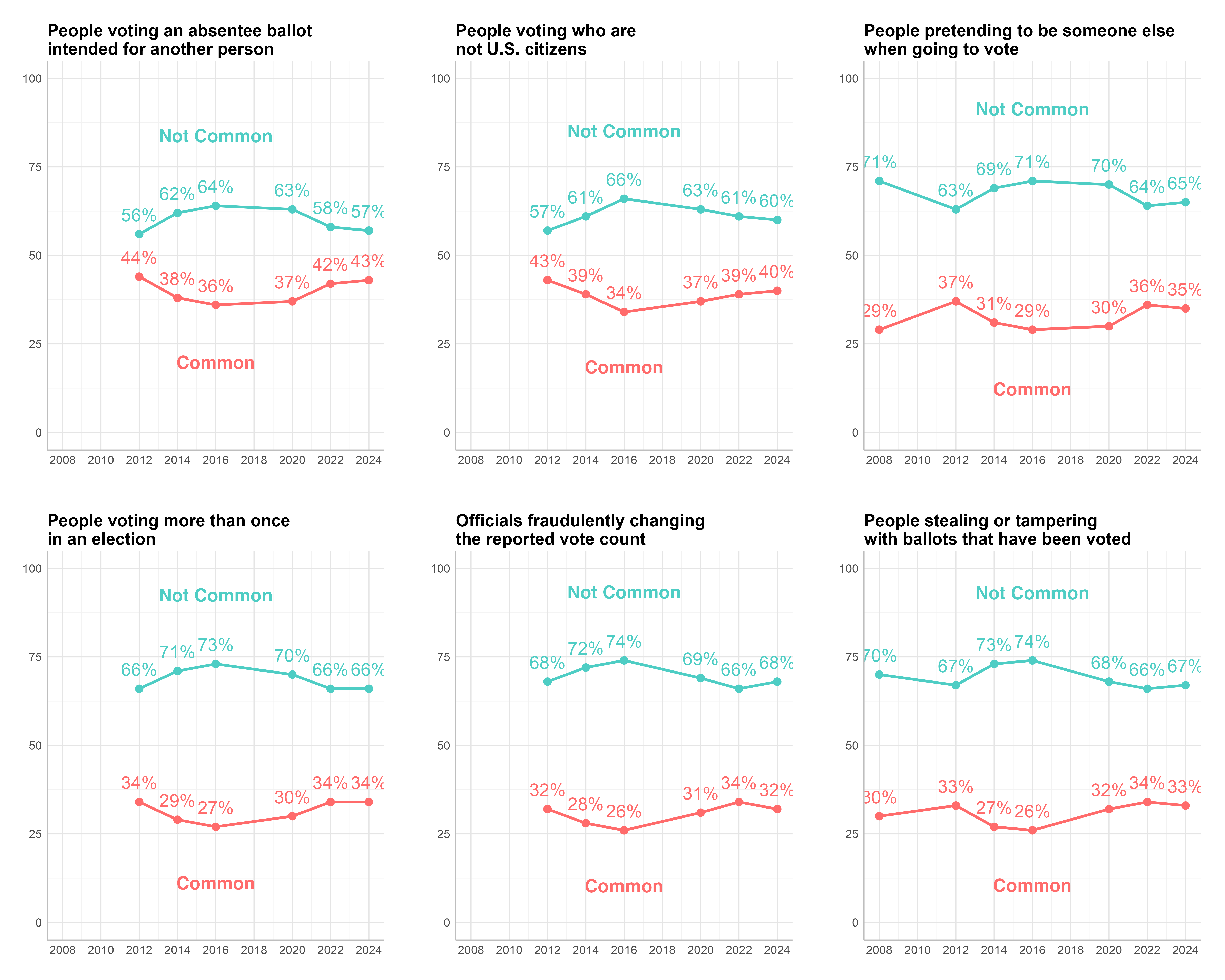}
    \caption{Concerns about types of election fraud, data from 2008-2022 are from the Survey of the Performance of American Elections.  Data from 2024 are from our survey discussed in the text.}
    \label{crime_combo}
\end{figure}

As we noted above, the questions that we asked in our survey are generally similar to questions that have been asked in recent versions of the Survey of the Performance of American Elections \cite{stewart_spae_2022}.  Importantly, this gives us the ability to add our 2024 data to those from past versions of SPAE and to examine how American opinions about these types of election fraud may have changed since 2022.\footnote{We offer the comparison of our 2024 types of election fraud survey responses to those from the SPAE while noting that there are methodological differences between the SPAE and our survey.  Importantly, the SPAE surveys, like our survey, are conducted online and are weighted to produce a sample of U.S. registered voters. The main difference is that the underlying design of the SPAE is to produce samples of registered voters in each state, which can then be weighted nationally.  Ours is a national sample.}  We present our the time series for six of the types of election fraud questions in Figure~\ref{crime_combo}, the six that have been included in SPAE surveys since at least 2012.

We have six panels in Figure~\ref{crime_combo}.  Starting in the upper left, with people voting an absentee ballot intended for another person, we see relative stability since 2022, with a 1\% increase in the proportion saying this is a common occurrence.  Note, though, that this is a 6\% increase since 2020. The upper middle panel of Figure~\ref{crime_combo} gives the results for endorsement of the notion that people are voting who are not U.S. citizens.  Note that again our 2024 estimates are similar to what the SPAE estimated in 2022 (40\% in 2024 believing this is a common occurrence, relative to 39\% in 2022).  But it is also important to point out that the proportion of the population believing that non-citizen voting is common has steadily risen since 2016 from 34\% to 40\%, a 6\% increase.  The upper right-hand panel in the Figures shows the results over time for the claim that people are pretending to be someone else when they go to vote.  Here our 2024 estimate is slightly lower than the 2022 SPAE estimate (36\% to 35\%).  Both 2022 and 2024 are higher than during the 2014-2020, when the percentage thinking this concern was common hovered between 29\% and 31\%.  

In the lower-left panel of Figure~\ref{crime_combo} we provide the time-series for the claim that people vote more than once in an election.  While we see stability in our 2024 estimate of the percentage endorsing this claim relative to 2022 (both 34\%), we also want to point out that there has been a slight increase in the percentage seeing this as common in recent years (30\% in 2020).  

The middle lower panel in this Figure provides the results overtime for the claim that officials fraudulently change the reported vote count.  Our estimate for 2024 for the percentage endorsing this claim is 2 points lower than 2022 (32\% relative to 34\%), and our 2024 estimate for those believing this is common is close to the percentage the SPAE estimated for 2020.  A similar conclusion can be taken from the lower-right panel of Figure~\ref{crime_combo}:  our 2024 estimate for the percentage endorsing this claim (32\%) is quite similar to the SPAE estimates for 2020 (32\%) and 2022 (34\%).

\subsection{Bivariate Results}\label{sec:biv}

As we show in the Supplementary Materials (Sections 3.1 and 3.2) in detail, two variables have a striking correlation with belief in types of election fraud: party and level of belief in non-election conspiracy theories.
\footnote{In the Supplementary Materials, we provide tables showing the bivariate relationships between each of the election fraud covariates and a number of demographic, attitudinal, and behavioral covariates: gender, education, region of residence, urban or rural residence, political interest, conspiratorial beliefs, and partisanship.}
The former may be related to the ``loser effect" discussed in Section ~\ref{sec:lit}, while the latter makes some intuitive sense: given that actual frequencies of the types of election fraud studied in this paper appear to be quite low, belief in them being common is likely to be correlated with believing that other infrequent events are also common\footnote{There is also substantial correlation between party identification and belief in non-election conspiracies. We leave the question of whether conspiratorial partisans would be more likely to believe in the integrity of the election if their preferred party won to future work.}.

\begin{figure}[ht]
    \centering
    \includegraphics[width=.85\linewidth]{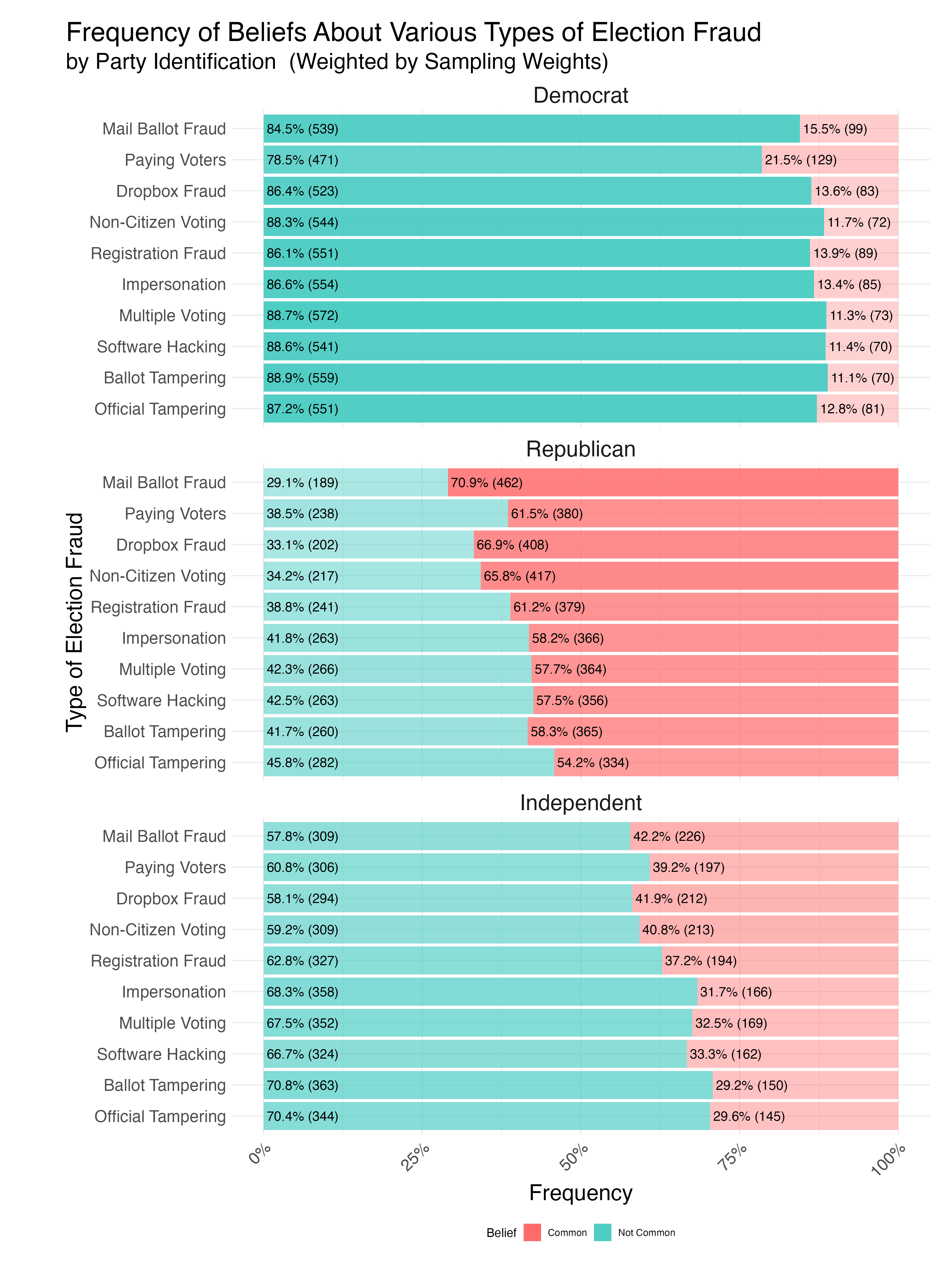}
    \caption{Concerns about types of election fraud by party.}
    \label{fig:crimes_by_party}
\end{figure}

\begin{figure}[ht]
    \centering
    \includegraphics[width=.9\linewidth]{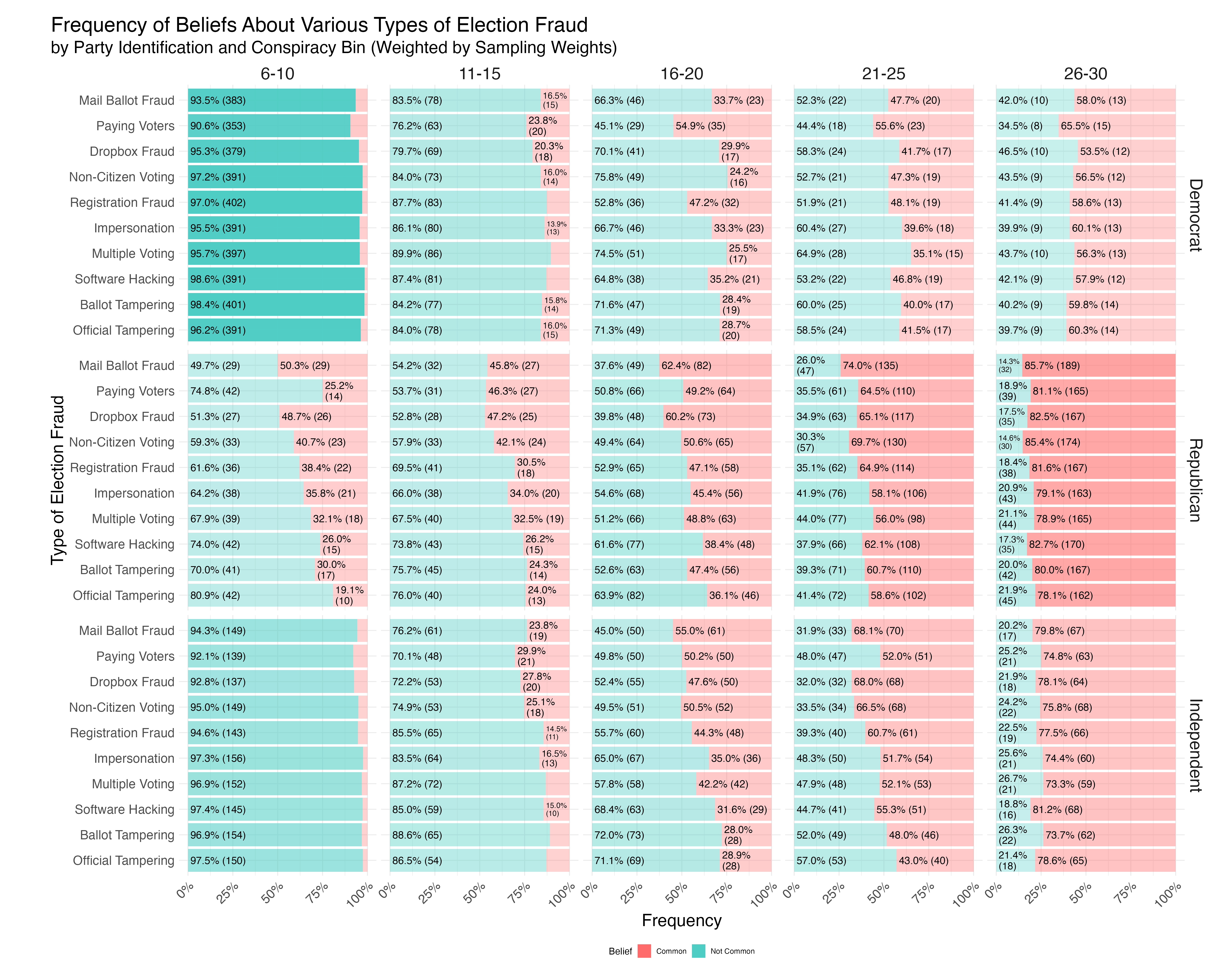}
    \caption{Concerns about types of election fraud by level of belief in non-election conspiracy theories. Respondents are grouped by party and their total level of agreement (on a 1-5 scale) with six conspiracy questions. Higher numbers represent a higher average agreement with the conspiracy theory questions. We include questions pertaining to COVID-19, the United Nations, secret societies, scientists misleading the public, and the deliberate spread of certain diseases. Please refer to the Supplementary Materials for specific wording.}
    \label{fig:crimes_by_party_conspiracy}
\end{figure}

Agreement with populist statements is positively correlated with belief that types of election fraud are common: among the strongest endorsers of populist statements, at least half believe any given election rumor. As we show in Section ~\ref{sec:miv}, this is robust even when controlling for a number of factors, including susceptibility to misinformation and endorsement of conspiracy theories. As we show in the Supplementary Material (Section 4.1), however, this is due in part to Republicans being more likely to strongly agree with populist statements, while Democrats tend to agree less strongly (i.e., a quadratic specification maybe appropriate for populism). This indicates that there is a need for additional research on how populism interacts with other identities like partisanship to associate with opinions about types of election fraud.

Otherwise, as can be seen in Section ~\ref{sec:miv} and the Supplementary Material, few covariates have strong associations with belief in election rumors once political party, populism, and level of non-election conspiratorial thinking are controlled for. After controlling for them, there appear to be no significant differences in belief in types of election fraud across age, gender, race, level of education, or geographic location\footnote{If we omit conspiracy and populism from our specification, our results are similar with the rest of the literature, in particular, education has a protective effect against belief in election fraud. The difference is due to correlation between party, education, populism, and non-election conspiratorial thinking.}. In some cases this is is due to significant correlation between . While some of these variables (like education) display significant heterogeneity, these differences are primarily driven by political party and non-election conspiratorial beliefs. For example, as we show in the Supplementary Material, Republicans with post-graduate education have approximately the same level of belief that types of election fraud are common as do their Republican peers with a high school education or less.


These bivariate correlations can only do so much to illuminate what drives belief in types of election fraud. For this, we turn to the multivariate analysis of the next section.

\subsection{Regression Results}\label{sec:miv}

In this section we present individual-level results on the total number and proportion of types of election fraud endorsed by respondents. We focus on the OLS results in the first column of Table ~\ref{tab:total_law_reg}, which has the raw number of types of election fraud respondents believe are common or occasional. As we noted before, we aggregate the responses in this way to ease analysis and interpretation. These results are similar to the binomial regression results in the second column, which has the proportion of types of election fraud respondents believe are common or occasional. For more analysis of belief in individual types of election fraud, please refer to the Supplementary Materials (Section 4).

\begin{table}[h] \centering 
  \caption{Regression Results for Aggregate Answers to Election Crime Questions} 
  \label{tab:total_law_reg} 
\footnotesize 
\begin{tabular}{@{\extracolsep{5pt}}lcc} 
\\[-1.8ex]\hline 
\hline \\[-1.8ex] 
 & \multicolumn{2}{c}{\textit{Dependent variable:}} \\ 
\cline{2-3} 
\\[-1.8ex] & Number of Election Crimes Believed & Proportion of Election Crimes Believed \\ 
\\[-1.8ex] & \textit{survey-weighted} & \textit{survey-weighted} \\ 
 & \textit{normal} & \textit{logistic} \\ 
\\[-1.8ex] & (1) & (2)\\ 
\hline \\[-1.8ex] 
 30-44 & $-$0.006 & $-$0.007 \\ 
  & (0.278) & (0.190) \\ 
  45-64 & $-$0.208 & $-$0.143 \\ 
  & (0.272) & (0.184) \\ 
  65+ & $-$0.372 & $-$0.241 \\ 
  & (0.271) & (0.187) \\ 
  Female & 0.143 & 0.118 \\ 
  & (0.167) & (0.107) \\ 
  Black & 0.147 & 0.257 \\ 
  & (0.278) & (0.183) \\ 
  Hispanic & 0.054 & 0.086 \\ 
  & (0.296) & (0.199) \\ 
  Other & 0.603 & 0.398 \\ 
  & (0.361) & (0.228) \\ 
  Some college & 0.100 & 0.028 \\ 
  & (0.237) & (0.142) \\ 
  College grad & 0.157 & 0.057 \\ 
  & (0.249) & (0.152) \\ 
  Postgrad & 0.434 & 0.241 \\ 
  & (0.261) & (0.175) \\ 
  Republican & 1.423$^{***}$ & 0.913$^{***}$ \\ 
  & (0.345) & (0.201) \\ 
  Independent & 0.380 & 0.413$^{**}$ \\ 
  & (0.200) & (0.154) \\ 
  Moderate & 0.051 & 0.230 \\ 
  & (0.193) & (0.169) \\ 
  Conservative & 0.125 & 0.256 \\ 
  & (0.328) & (0.206) \\ 
  Midwest & 0.091 & 0.033 \\ 
  & (0.245) & (0.162) \\ 
  South & 0.018 & 0.006 \\ 
  & (0.226) & (0.147) \\ 
  West & 0.126 & 0.066 \\ 
  & (0.239) & (0.157) \\ 
  Suburb & $-$0.012 & $-$0.027 \\ 
  & (0.204) & (0.140) \\ 
  Town & 0.117 & 0.022 \\ 
  & (0.262) & (0.168) \\ 
  Rural area & $-$0.579$^{*}$ & $-$0.377$^{*}$ \\ 
  & (0.244) & (0.156) \\ 
  Follows politics some of the time & $-$0.558$^{**}$ & $-$0.244 \\ 
  & (0.199) & (0.125) \\ 
  Follows politics only now and then & $-$0.766$^{**}$ & $-$0.321 \\ 
  & (0.283) & (0.180) \\ 
  Follows politics hardly at all & $-$0.919$^{*}$ & $-$0.409 \\ 
  & (0.401) & (0.234) \\ 
  Populism Score & 0.045$^{*}$ & 0.024 \\ 
  & (0.023) & (0.015) \\ 
  Conspiracy Score & 0.248$^{***}$ & 0.137$^{***}$ \\ 
  & (0.016) & (0.010) \\ 
  Constant & $-$2.413$^{***}$ & $-$4.429$^{***}$ \\ 
  & (0.585) & (0.419) \\ 
 \hline \\[-1.8ex] 
Observations & 2,075 & 2,075 \\ 
Log Likelihood & $-$5,359.648 &  \\ 
Akaike Inf. Crit. & 10,771.300 &  \\ 
\hline 
\hline \\[-1.8ex] 
\textit{Note:}  & \multicolumn{2}{r}{$^{*}$p$<$0.05; $^{**}$p$<$0.01; $^{***}$p$<$0.001} \\ 
\end{tabular} 
\end{table}

A few results are immediately clear: the predictors with the largest effects are belonging to the Republican party, living in a rural area, infrequently following politics, and belief in non-election conspiracy theories. Republican respondents say they believe around 1.4 more types of election fraud are common than do similar Democratic respondents. This result appears in each of our specifications, and, as shown in detail in the Supplementary Materials, identifying with the Republican party is frequently the largest single predictor of belief in election conspiracies. 

One of the few covariates with as strong a correlation with belief in election conspiracies is belief in non-election conspiracies. In Table ~\ref{tab:total_law_reg}, we see that each additional point in the``conspiracy score" is associated with belief in 0.248 additional types of election fraud (again, ``conspiracy score" is the sum of conspiracy confidence questions; these six questions are each rated on scale from 1 to 5). Put another way, an increase of one standard deviation in conspiracy beliefs is associated with belief in two additional types of election fraud. As shown in Figure ~\ref{fig:crimes_by_party_conspiracy}, Republicans with low conspiracy scores display lower rates of belief in election conspiracies than do Democrats with high conspiracy scores (it should be noted that these are both small sub-populations). 

Perhaps surprisingly, we find that relative to urban respondents, those living in rural areas believe 0.5 \textit{fewer} types of election fraud. This runs counter to recent literature arguing that rural people are less trusting of government, e.g., \citet{kirk2024}. In the supplementary materials, we show that this appears to be driven by rural conservatives, who believe an average of one fewer election conspiracies than urban conservatives. 

Relative to respondents who follow politics ``most of the time", all other respondents are \textit{less} likely to believe rumors. This effect gets stronger the less interested in politics respondents are: those following ``some of the time" believe in 0.5 fewer types of election fraud, while those that follow ``hardly at all" believe in 0.9 fewer. As we show in the Supplementary Material, this result is driven by politically disengaged Democrats believing more rumors than their engaged peers, and due to a strong majority of Republicans who report following politics ``most of the time", and who believe in types of election fraud at significantly higher rates than the rest of the survey population. 

\begin{figure}[ht]
    \centering
    \includegraphics[width=.85\linewidth]{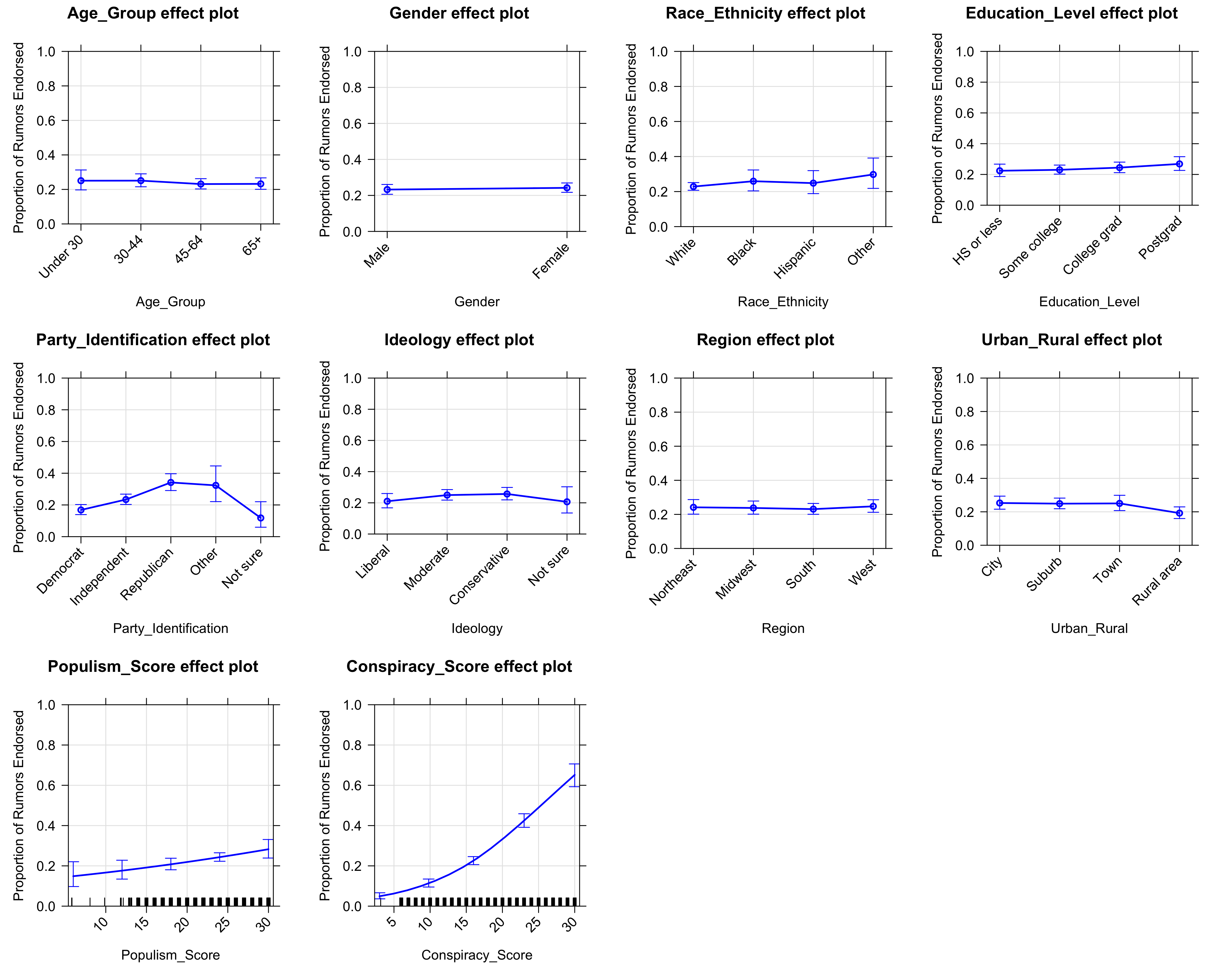}
    \caption{Marginal effects plot for multivariate regression on the proportion of types of election fraud believed.}
    \label{fig:law_me}
\end{figure}

These results are echoed in the marginal effects plot in Figure ~\ref{fig:law_me}, where we see that a greater level of agreement with conspiracy questions is easily the strongest mover of belief in types of election fraud on the margin. Similar to the discussion of regression results, one striking takeaway is how few variables seem to matter on the margin. All of age, race, gender, region, and even ideology seem to have only minor marginal contributions to election beliefs. Education appears to contribute to a net \textit{increase} on the margin. Of course, it is important to keep in mind for all of these marginal effects that other important covariates (like conspiracy score) are implicitly set to their average value, thus ignoring the correlations between them.

\section{Discussion and Conclusion}

Our analysis of perceptions of types of election fraud among American registered voters in 2024 has several findings with implications for democratic governance and electoral integrity in the United States. First and foremost, our results demonstrate that beliefs in various forms of election fraud remain widespread, despite a lack of evidence for their occurrence at any meaningful scale. This persistence of election fraud beliefs, even years after the contentious 2020 election, suggests that these perceptions may have become deeply entrenched in the American political psyche.

One of the most striking findings is the extent to which party affiliation correlates with beliefs in types of election fraud. Republicans consistently express higher levels of belief in various forms of election fraud compared to Democrats, even when controlling for other factors such as demographics, political engagement, and general conspiracy thinking. This partisan divide in perceptions of electoral integrity is concerning, as it may contribute to further polarization and undermine the legitimacy of election outcomes for a significant portion of the electorate.  Whether this is evidence of a ``loser effect'' or might be a more lasting reflection of the partisan polarization in the United States today will require additional research after the 2024 election. 

The strong association between belief in non-election conspiracies and acceptance of election fraud narratives is also noteworthy. This relationship suggests that susceptibility to election fraud beliefs may be part of a broader pattern of conspiratorial thinking. As such, efforts to address election fraud myths may need to consider broader strategies for improving critical thinking and media literacy among the general public.

Interestingly, our findings challenge some common assumptions about the rural-urban divide in political trust. Contrary to expectations, we found that rural respondents, particularly rural conservatives, were less likely to believe in types of election fraud compared to their urban counterparts. This unexpected result warrants further investigation and may have implications for how we understand the geographic distribution of political attitudes and beliefs.

Another counterintuitive finding is the relationship between political engagement and belief in types of election fraud. Our results indicate that those who follow politics less frequently are actually less likely to believe in election fraud. This suggests that increased exposure to political information, particularly among Republicans, may paradoxically increase susceptibility to election fraud narratives. This finding raises important questions about the quality and sources of political information consumed by engaged citizens and the potential echo chamber effects in partisan media ecosystems.

Our findings underscore the need for new approaches to address entrenched beliefs in election fraud. Experimental studies should explore multifaceted interventions that combine information provision with techniques drawn from research on countering conspiracy theories. For instance, researchers should continue to test the effectiveness of ``inoculation" methods, exposing individuals to weakened forms of misinformation to build resistance against future encounters (see, for example \cite{vanderlinden2017,vanderlinden2021,vanderlinden2022,LEWANDOWSKY2023101711, careyetal2024}). Additionally, longitudinal studies tracking changes in election fraud beliefs over time, particularly before and after major elections, could provide valuable insights into the durability of these perceptions and the factors that influence their evolution.

This study has provided a comprehensive snapshot of American voters' perceptions of types of election fraud in 2024, revealing the persistent and partisan nature of these beliefs. By examining a wide range of election fraud narratives and their correlates, we have illuminated the interplay between partisanship, conspiracy thinking, political engagement, and geographic factors in shaping these perceptions. Our analysis reveals that despite the passage of time since the contentious 2020 election, beliefs in various forms of election fraud remain widespread, challenging assumptions about rural-urban divides and the effects of political engagement. These findings highlight the challenges facing American democracy as it contends with pervasive skepticism about electoral integrity, even in the absence of evidence for significant fraud.

As we approach future elections, the task of rebuilding trust in the American electoral system is both urgent and formidable. Yet, throughout its history, American democracy has faced and overcome significant challenges. By deepening our understanding of the roots of election fraud beliefs and developing evidence-based strategies to address them, we can work towards a future where faith in the democratic process transcends partisan divides.

\clearpage

\clearpage

\section*{Acknowledgements} 
 The authors thank Shreya Nag for her assistance.  

\section*{Author Contributions}
The authors contributed equally to all aspects of the production of this paper.  

\section*{Ethical Consideration}
The collection and analysis of the survey data reported in this paper was reviewed by the Institutional Review Board at the California Institute of Technology, IR22-1220.

\section*{Funding}
Funding for this project was provided by the The John Randolph Haynes and Dora Haynes Foundation.  

\section*{Conflict of Interest}
The authors declare no conflicts of interest.

\section*{Data Availability}
The data and code needed to replicate the results reported in this paper will be available in a public repository upon publication.

\bibliographystyle{Frontiers-Harvard}

\clearpage

\end{document}


\onecolumn

\title{Supplementary Material}

\maketitle


\section{Survey Details}

\subsection{Survey Methodology}
Researchers at Caltech designed a national survey which contained a series of questions about election fraud.  The survey was conducted by YouGov, using a sample from their online panel that was designed to be  representative of U.S. registered voters. The survey contains responses from 2,211 U.S. registered voters and was fielded between June 26 and July 3, 2024.  YouGov provided weights taking into consideration gender, age, race, and education (using information from the American Community Survey); they also used information from the 2020 presidential election vote, party identification, and 2022 Congressional vote.  The sample weights range from 0.1 to 5.0, with a mean of 1.0 and standard deviation of 0.6. According to information provided by YouGov, the margin of error for this survey is approximately 2.4\%.

\begin{table}[h]
  \centering
        \caption{Survey marginals, Unweighted and Weighted.}
    \label{tab:election_crimes_marginals}

    \begin{tabular}{r|cc|}
\multicolumn{1}{c}{} & \multicolumn{1}{c}{Unweighted} & 
\multicolumn{1}{c}{Weighted} \\ \hline
\multicolumn{1}{c}{Age} & \multicolumn{2}{c}{} \\ \hline\hline
Under 30 & .13 & .15 \\
30 to 44 & .20 & .23 \\
45 to 64 & .37 & .35 \\
65 and over & .30 & .37 \\ \hline
\multicolumn{1}{c}{Gender} & \multicolumn{2}{c}{} \\ \hline\hline
Man & .45 & .46 \\
Woman & .55 & .54 \\ \hline
\multicolumn{1}{c}{Race/Ethnicity} & \multicolumn{2}{c}{} \\ \hline\hline
White & .72 & .71 \\
Black & .11 & .12 \\
Hispanic & .12 & .10 \\
Other & .06 & .06 \\ \hline
\multicolumn{1}{c}{Education} & \multicolumn{2}{c}{} \\ \hline\hline
HS or less & .27 & .30 \\
Some College & .32 & .30 \\
College Grad & .26 & .25 \\
Post-Grad & .16 & .15 \\ \hline
\multicolumn{1}{c}{Partisanship} & \multicolumn{2}{c}{} \\ \hline\hline
Democrat & .37 & .34 \\
Republican & .32 & .35 \\
Independent & .28 & .27 \\
Other & .03 & .03 \\
Not Sure & .01 & .01 \\ \hline
\multicolumn{1}{c}{Ideology} & \multicolumn{2}{c}{} \\ \hline\hline
Liberal & .30 & .28 \\
Moderate & .32 & .31 \\
Conservative & .34 & .37 \\
Not Sure & .04 & .04 \\ \hline
\multicolumn{1}{c}{Region} & \multicolumn{2}{c}{} \\ \hline\hline
Northease & .18 & .18 \\
Midwest & .23 & .22 \\
South & .37 & .37 \\
West & .22 & .23 \\ \hline
\multicolumn{1}{c}{Urban/rural} & \multicolumn{2}{c}{} \\ \hline\hline
City & .26 & .26 \\
Suburb & .41 & .41 \\
Town & .13 & .14 \\
Rural Area & .20 & .19 \\ \hline

   \end{tabular}
\end{table}

We provide in Table~\ref{tab:election_crimes_marginals} the unweighted and weighted survey marginals for the important demographic and attitudinal features that we included in our analyses:  age, gender, race/ethnicity, educational attainment, partisanship, ideology, region of residence and urban/rural status.  It is clear from Table~\ref{tab:election_crimes_marginals} that the unweighted sample provided by YouGov is quite similar to the weighted sample.  Despite that similarity, we use the sample weights provided by YouGov in the analyses reported in the paper.  

\subsection{Election Fraud}

The election fraud questions are listed next.  In italics are the labels that we will use to refer to specific election crime topics.  These are similar to, and in some cases identical to, questions that have been included in other surveys, in particular the \textit{The Survey of the Performance of American Elections}.\footnote{See \url{https://electionlab.mit.edu/research/projects/survey-performance-american-elections}.}
\begin{itemize}
\item People voting more than once in an election (\textit{Multiple Voting})
\item People stealing or tampering with ballots that have been counted (\textit{Ballot Tampering})
\item People pretending to be someone else when going to vote (\textit{Impersonation})
\item People voting who are not U.S. citizens (\textit{Non-Citizen Voting})
\item People taking advantage of absentee or mail balloting to engage in vote fraud (\textit{Mail Ballot Fraud})
\item Officials changing the reported vote count in a way that is not a true reflection of the ballots that were actually counted (\textit{Official Tampering})
\item Vote counting software manipulated in a way to not count ballots as intended (\textit{Software Hacking})
\item Paying voters to cast a ballot for a particular candidate (\textit{Paying Voters})
\item Voting under fraudulent voter registrations that use a fake name and a fake address (\textit{Registration Fraud})
\item People submitting too many ballots in drop boxes on behalf of others (\textit{Dropbox Fraud})
\end{itemize}
Subjects could indicate that they believed for each of these election fraud that ``it is very common'', ``it occurs occassionally'', ``it occurs infrequently'', ``it almost never occurs'', or ``I'm not sure.''

\clearpage







Table~\ref{tab:election_crimes} provides the weighted frequencies for responses to these questions.  Table~\ref{tab:election_crimes2} presents similar results, only here we have aggregated the responses ``it is very common'' and ``it occurs occassionally'' to ``Concerned", and ``it occurs infrequently'' and ``it almost never occurs'' to ``Not Concerned.''

\begin{table}[ht]
    \centering
    \caption{Election Fraud}
    \label{tab:election_crimes}

    \begin{tabular}{r|ccccc|}
\multicolumn{1}{c}{Problem} & \multicolumn{1}{c}{Common} & \multicolumn{1}{c}{Occasionally} & \multicolumn{1}{c}{Infrequently} & \multicolumn{1}{c}{Never} & \multicolumn{1}{c}{Not Sure}\\ \hline
Multiple Voting & 0.11 & 0.18 & 0.17 & 0.40 & 0.14 \\ 
Ballot Tampering & 0.12 & 0.16 & 0.15 & 0.42 & 0.16  \\
Impersonation & 0.11 & 0.18 & 0.17 & 0.39 & 0.15  \\
Non-Citizen Voting & 0.18 & 0.15 & 0.14 & 0.37 & 0.16  \\
Mail Ballot Fraud & 0.19 & 0.18 & 0.15 & 0.35 & 0.13  \\
Official Tampering & 0.12 & 0.14 & 0.14 & 0.43 & 0.17  \\
Software Hacking & 0.13 & 0.15 & 0.13 & 0.41 & 0.18  \\
Paying Voters & 0.14 & 0.19 & 0.15 & 0.34 & 0.18  \\
Registration Fraud & 0.14 & 0.17 & 0.16 & 0.37 & 0.15  \\
Dropbox Fraud & 0.16 & 0.17 & 0.14 & 0.35 & 0.18  \\ \hline
    \end{tabular}
\end{table}



\clearpage

\subsection{Conspiracy Theory and Populism Beliefs}

This survey contained a six-question battery regarding different conspiracy theories. Survey respondents could say they strongly agree (1), somewhat agree (2), neither agree nor disagree (3), somewhat disagree (4), or strongly disagree  (5)regarding the following six statements:
\begin{enumerate}
    \item The rapid spread of certain viruses and/or diseases is the result of the deliberate, concealed efforts of some organization.
    \item Groups of scientists deliberately attempt to create panic about future risks because it is in their interests to do so.
    \item Many well-known celebrities, politicians, and wealthy people are members of a secret society, which has control over our lives.
    \item The United Nations is part of the new world order, planning to undermine American sovereignty and aiming to introduce a global currency to end the U.S. dollar.
    \item Governments use COVID-19 testing to collect genetic information about their citizens.
    \item The government is manipulating official laboratory tests and techniques, so that it looks like a lot of people have fallen victim to COVID-19.
\end{enumerate}

To form conspiracy and populism scores for each participant, we ask six conspiracy and populism questions, rated on a five-point scale, which we then sum to form each of these measures. These measures thus range from a minimum of 6 to a maximum of 30. Finally, we bin these scores in 5-point bins to present the results. 

In Table~\ref{tab:conspiracy_counts} we provide the raw counts for each of the responses, for each conspiracy theory question. 

\begin{table}
\begin{center}
\begin{tabular}{lrrrrrr}
\toprule
Conspiracy & 1 & 2 & 3 & 4 & 5 & Skip \\ \hline
1 & 374 & 414 & 439 & 254 & 730 & NA \\
2 & 414 & 460 & 351 & 269 & 717 & NA \\
3 & 258 & 402 & 442 & 290 & 817 & 2 \\
4 & 369 & 337 & 426 & 243 & 834 & 2 \\
5 & 271 & 346 & 414 & 289 & 889 & 2 \\
6 & 492 & 330 & 338 & 203 & 847 & 1 \\ \hline
\end{tabular}
\caption{Response counts from the survey for each of the conspiracy theory questions.  The conspiracy theory row numbers correspond to the numbers in the enumerated list above.  The numeric response numbers correspond to those in the text (strongly agree, 1), (somewhat agree, 2), (neither agree nor disagree, 3), (somewhat disagree, 4), (strongly disagree, 5), and skipped.}  
\label{tab:conspiracy_counts}
\end{center}
\end{table}

This survey also contained a six-question battery with different statements that might indicate populist belief. Survey respondents could say they strongly agree (1), somewhat agree (2), neither agree nor disagree (3), somewhat disagree (4), or strongly disagree  (5), regarding the following six statements:
\begin{enumerate}
\item The politicians in the U.S. Congress need to follow the will of the people. 
\item  The people, and not politicians, should make our most important policy decisions. 
\item The political differences between the elite and the people are larger than the differences among the people. 
\item I would rather be represented by a citizen than by a specialized politician. \item Elected officials talk too much and take too little action. 
\item What people call ``compromise'' in politics is really just selling out on one’s principles
\end{enumerate}

In Table~\ref{tab:populism_counts} we provide the raw counts for each of the responses, for each conspiracy theory question.
\begin{table}
\begin{center}
\begin{tabular}{lrrrrrr}
\toprule
Variable & 1 & 2 & 3 & 4 & 5 & Skip \\ \hline
1 & 1084 & 798 & 258 & 47 & 23 & 1 \\
2 & 625 & 701 & 581 & 237 & 67 & NA \\
3 & 728 & 802 & 534 & 107 & 40 & NA \\
4 & 633 & 662 & 651 & 218 & 46 & 1 \\
5 & 1114 & 782 & 243 & 54 & 18 & NA \\
6 & 336 & 573 & 640 & 457 & 205 & NA \\ \hline
\end{tabular}
\caption{Response counts from the survey for each of the populism questions.  The conspiracy theory row numbers correspond to the numbers in the enumerated list above.  The numeric response numbers correspond to those in the text (strongly agree, 1), (somewhat agree, 2), (neither agree nor disagree, 3), (somewhat disagree, 4), (strongly disagree, 5), and skipped.}  
\label{tab:populism_counts}
\end{center}
\end{table}

\newpage
\clearpage

\section{Summary Statistics}

{ \let\oldcentering\centering \renewcommand\centering{\tiny\oldcentering}
\begin{table}[!h]
\centering
\caption{\label{tab:tab:summary_statistics_law.tex}Summary Statistics - Election Crimes}
\centering
\begin{tabular}[t]{>{}l>{}llrr}
\toprule
\textbf{Variable} & \textbf{N} & \textbf{NA} & \textbf{Not Common} & \textbf{Common}\\
\midrule
\textbf{Multiple Voting} & \textbf{2,075} & 0.116 & 0.610 & 0.275\\
\textbf{Ballot Tampering} & \textbf{2,075} & 0.126 & 0.607 & 0.267\\
\textbf{Impersonation} & \textbf{2,075} & 0.117 & 0.600 & 0.283\\
\textbf{Non Citizen Voting} & \textbf{2,075} & 0.129 & 0.553 & 0.318\\
\textbf{Mail Ballot Fraud} & \textbf{2,075} & 0.099 & 0.543 & 0.358\\
\addlinespace
\textbf{Official Tampering} & \textbf{2,075} & 0.143 & 0.605 & 0.252\\
\textbf{Software Hacking} & \textbf{2,075} & 0.147 & 0.589 & 0.264\\
\textbf{Paying Voters} & \textbf{2,075} & 0.150 & 0.530 & 0.320\\
\textbf{Registration Fraud} & \textbf{2,075} & 0.121 & 0.581 & 0.298\\
\textbf{Dropbox Fraud} & \textbf{2,075} & 0.147 & 0.536 & 0.318\\
\bottomrule
\end{tabular}
\end{table}

}

{ \let\oldcentering\centering \renewcommand\centering{\tiny\oldcentering}
\begin{table}[!h]
\centering
\caption{\label{tab:tab:summary_statistics_numeric_weighted.tex}Summary Statistics - Weighted Numeric Variables}
\centering
\begin{threeparttable}
\begin{tabular}[t]{>{}l>{}lllllllll}
\toprule
\textbf{Variable} & \textbf{N} & \textbf{Mean} & \textbf{Min} & \textbf{Q10} & \textbf{Q25} & \textbf{Median} & \textbf{Q75} & \textbf{Q90} & \textbf{Max}\\
\midrule
\textbf{Populism Score} & \textbf{2,075} & 23.222 & 6.000 & 18.000 & 20.000 & 23.000 & 26.000 & 29.000 & 30.000\\
\textbf{Conspiracy Score} & \textbf{2,075} & 16.513 & 3.000 & 6.000 & 8.000 & 17.000 & 23.000 & 28.000 & 30.000\\
\textbf{Total Election Crimes} & \textbf{2,075} & 3.164 & 0.000 & 0.000 & 0.000 & 1.000 & 7.000 & 10.000 & 10.000\\
\bottomrule
\end{tabular}
\begin{tablenotes}[para]
\item \textit{Note:} 
\item 
\end{tablenotes}
\end{threeparttable}
\end{table}

}

{ \let\oldcentering\centering \renewcommand\centering{\tiny\oldcentering}
\begin{table}[!h]
\centering
\caption{\label{tab:tab:summary_statistics_numeric_unweighted.tex}Summary Statistics - Unweighted Numeric Variables}
\centering
\begin{tabular}[t]{>{}l>{}llllllllll}
\toprule
\textbf{Variable} & \textbf{N} & \textbf{Mean} & \textbf{Var} & \textbf{Min} & \textbf{Q10} & \textbf{Q25} & \textbf{Median} & \textbf{Q75} & \textbf{Q90} & \textbf{Max}\\
\midrule
\textbf{Populism Score} & \textbf{2,075} & 23.220 & 16.038 & 6.000 & 18.000 & 20.000 & 23.000 & 26.000 & 29.000 & 30.000\\
\textbf{Conspiracy Score} & \textbf{2,075} & 15.966 & 65.178 & 3.000 & 6.000 & 7.000 & 16.000 & 23.000 & 28.000 & 30.000\\
\textbf{Total Election Crimes} & \textbf{2,075} & 2.952 & 14.345 & 0.000 & 0.000 & 0.000 & 0.000 & 6.000 & 10.000 & 10.000\\
\bottomrule
\end{tabular}
\end{table}

}

{ \let\oldcentering\centering \renewcommand\centering{\tiny\oldcentering}
\begin{table}[!h]
\centering
\caption{\label{tab:tab:summary_statistics_age_group.tex}Summary Statistics - Age Group}
\centering
\begin{tabular}[t]{>{}l>{}lrrrr}
\toprule
\textbf{Variable} & \textbf{N} & \textbf{Under 30} & \textbf{30-44} & \textbf{45-64} & \textbf{65+}\\
\midrule
\textbf{Age Group} & \textbf{2,075} & 0.131 & 0.194 & 0.368 & 0.307\\
\bottomrule
\end{tabular}
\end{table}

}

{ \let\oldcentering\centering \renewcommand\centering{\tiny\oldcentering}
\begin{table}[!h]
\centering
\caption{\label{tab:tab:summary_statistics_education_level.tex}Summary Statistics - Education Level}
\centering
\begin{tabular}[t]{>{}l>{}lrrrr}
\toprule
\textbf{Variable} & \textbf{N} & \textbf{HS or less} & \textbf{Some college} & \textbf{College grad} & \textbf{Postgrad}\\
\midrule
\textbf{Education Level} & \textbf{2,075} & 0.258 & 0.318 & 0.265 & 0.159\\
\bottomrule
\end{tabular}
\end{table}

}

{ \let\oldcentering\centering \renewcommand\centering{\tiny\oldcentering}
\begin{table}[!h]
\centering
\caption{\label{tab:tab:summary_statistics_gender.tex}Summary Statistics - Gender}
\centering
\begin{tabular}[t]{>{}l>{}lrr}
\toprule
\textbf{Variable} & \textbf{N} & \textbf{Male} & \textbf{Female}\\
\midrule
\textbf{Gender} & \textbf{2,075} & 0.462 & 0.538\\
\bottomrule
\end{tabular}
\end{table}

}

{ \let\oldcentering\centering \renewcommand\centering{\tiny\oldcentering}
\begin{table}[!h]
\centering
\caption{\label{tab:tab:summary_statistics_ideology.tex}Summary Statistics - Ideology}
\centering
\begin{tabular}[t]{>{}l>{}lrrr}
\toprule
\textbf{Variable} & \textbf{N} & \textbf{Liberal} & \textbf{Moderate} & \textbf{Conservative}\\
\midrule
\textbf{Ideology} & \textbf{2,075} & 0.313 & 0.330 & 0.357\\
\bottomrule
\end{tabular}
\end{table}

}


{ \let\oldcentering\centering \renewcommand\centering{\tiny\oldcentering}
\begin{table}[!h]
\centering
\caption{\label{tab:tab:summary_statistics_party_identification.tex}Summary Statistics - Party Identification}
\centering
\begin{tabular}[t]{>{}l>{}lrrr}
\toprule
\textbf{Variable} & \textbf{N} & \textbf{Democrat} & \textbf{Republican} & \textbf{Independent}\\
\midrule
\textbf{Party Identification} & \textbf{2,075} & 0.371 & 0.321 & 0.308\\
\bottomrule
\end{tabular}
\end{table}

}

{ \let\oldcentering\centering \renewcommand\centering{\tiny\oldcentering}
\begin{table}[!h]
\centering
\caption{\label{tab:tab:summary_statistics_political_interest.tex}Summary Statistics - Political Interest}
\centering
\begin{tabular}[t]{>{}l>{}lrrrr}
\toprule
\multicolumn{6}{c}{Follows politics:} \\
\textbf{Variable} & \textbf{N} & \textbf{Most of the time} & \textbf{Some of the time} & \textbf{Now and then} & \textbf{Hardly at all}\\
\midrule
\textbf{Political Interest} & \textbf{2,075} & 0.569 & 0.281 & 0.097 & 0.053\\
\bottomrule
\end{tabular}
\end{table}

}

{ \let\oldcentering\centering \renewcommand\centering{\tiny\oldcentering}
\begin{table}[!h]
\centering
\caption{\label{tab:tab:summary_statistics_race_ethnicity.tex}Summary Statistics - Race Ethnicity}
\centering
\begin{tabular}[t]{>{}l>{}lrrrr}
\toprule
\textbf{Variable} & \textbf{N} & \textbf{White} & \textbf{Black} & \textbf{Hispanic} & \textbf{Other}\\
\midrule
\textbf{Race Ethnicity} & \textbf{2,075} & 0.728 & 0.096 & 0.115 & 0.061\\
\bottomrule
\end{tabular}
\end{table}

}

{ \let\oldcentering\centering \renewcommand\centering{\tiny\oldcentering}
\begin{table}[!h]
\centering
\caption{\label{tab:tab:summary_statistics_region.tex}Summary Statistics - Region}
\centering
\begin{tabular}[t]{>{}l>{}lrrrr}
\toprule
\textbf{Variable} & \textbf{N} & \textbf{Northeast} & \textbf{Midwest} & \textbf{South} & \textbf{West}\\
\midrule
\textbf{Region} & \textbf{2,075} & 0.175 & 0.230 & 0.373 & 0.221\\
\bottomrule
\end{tabular}
\end{table}

}

{ \let\oldcentering\centering \renewcommand\centering{\tiny\oldcentering}
\begin{table}[!h]
\centering
\caption{\label{tab:tab:summary_statistics_urban_rural.tex}Summary Statistics - Urban Rural}
\centering
\begin{tabular}[t]{>{}l>{}lrrrr}
\toprule
\textbf{Variable} & \textbf{N} & \textbf{City} & \textbf{Suburb} & \textbf{Town} & \textbf{Rural area}\\
\midrule
\textbf{Urban Rural} & \textbf{2,075} & 0.256 & 0.412 & 0.133 & 0.200\\
\bottomrule
\end{tabular}
\end{table}

}

\newpage
\clearpage

\section{Graphical Results}

\subsection{Bivariate Plots}

\begin{figure}[ht]
    \centering
    \includegraphics[width=0.8\linewidth]{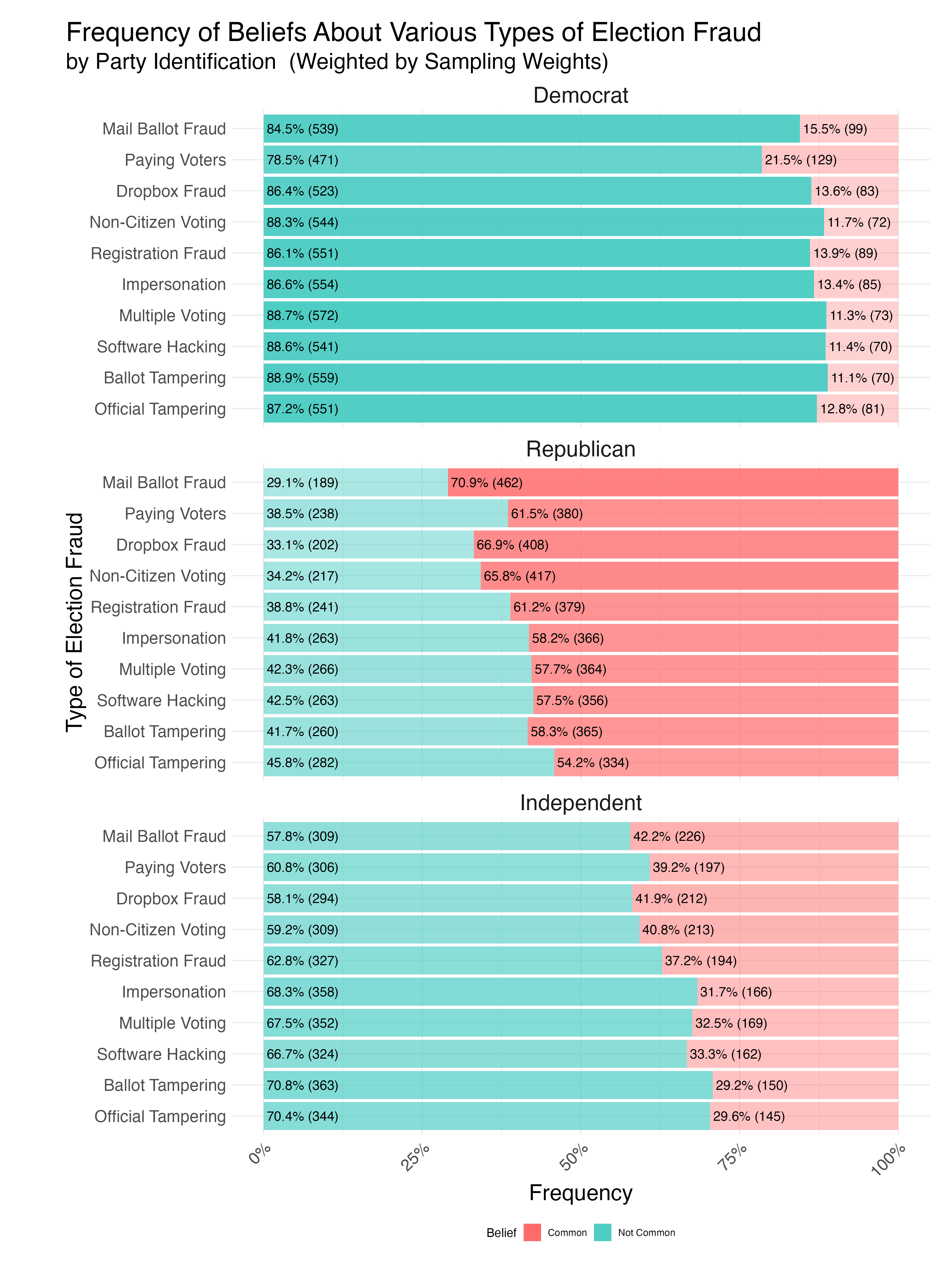}
    \caption{Election Fraud, weighted row frequencies.}
    \label{crime_freqs_party}
\end{figure}

\clearpage

\begin{figure}[ht]
    \centering
    \includegraphics[width=0.8\linewidth]{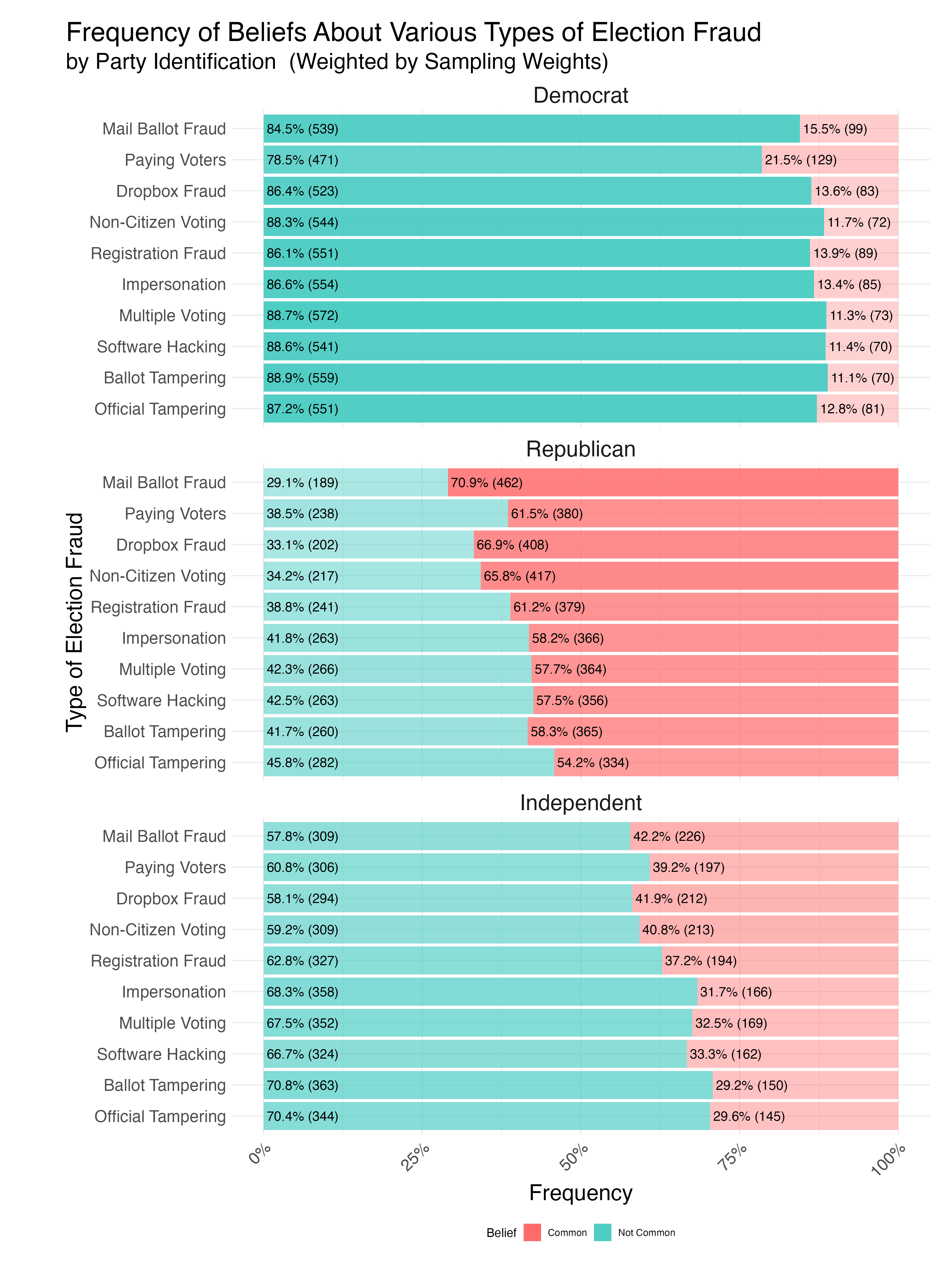}
    \caption{Election Fraud, weighted row frequencies.}
    \label{crime_freqs_ideology}
\end{figure}

\clearpage

\begin{figure}[ht]
    \centering
    \includegraphics[width=0.8\linewidth]{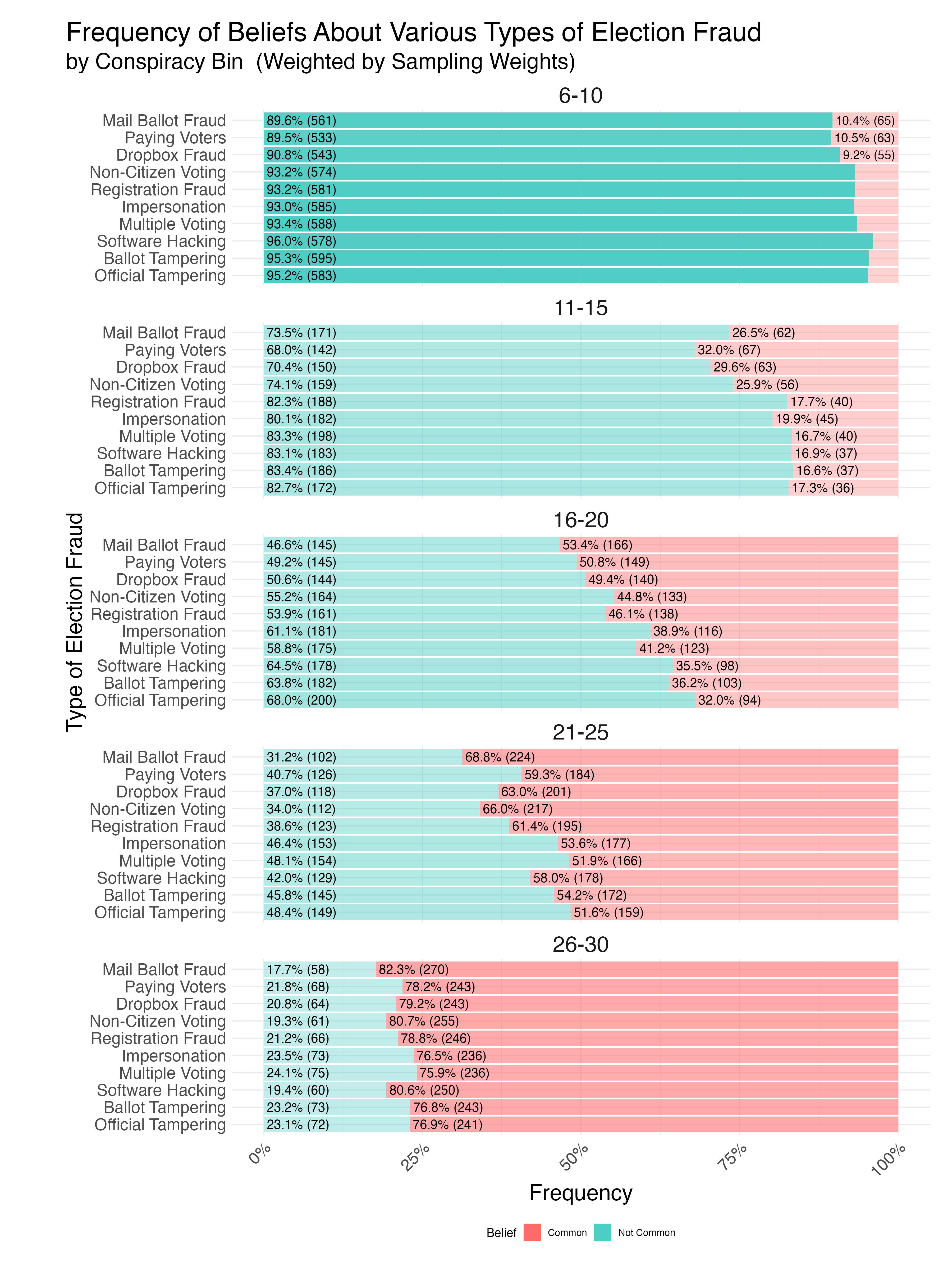}
    \caption{Concerns about election fraud by level of belief in non-election conspiracy theories. Respondents are grouped by party and their total level of agreement (on a 1-5 scale) with six conspiracy questions. Higher numbers represent a higher average agreement with the conspiracy theory questions.}
    \label{crime_freqs_conspiracy}
\end{figure}

\clearpage

\begin{figure}[ht]
    \centering
    \includegraphics[width=0.8\linewidth]{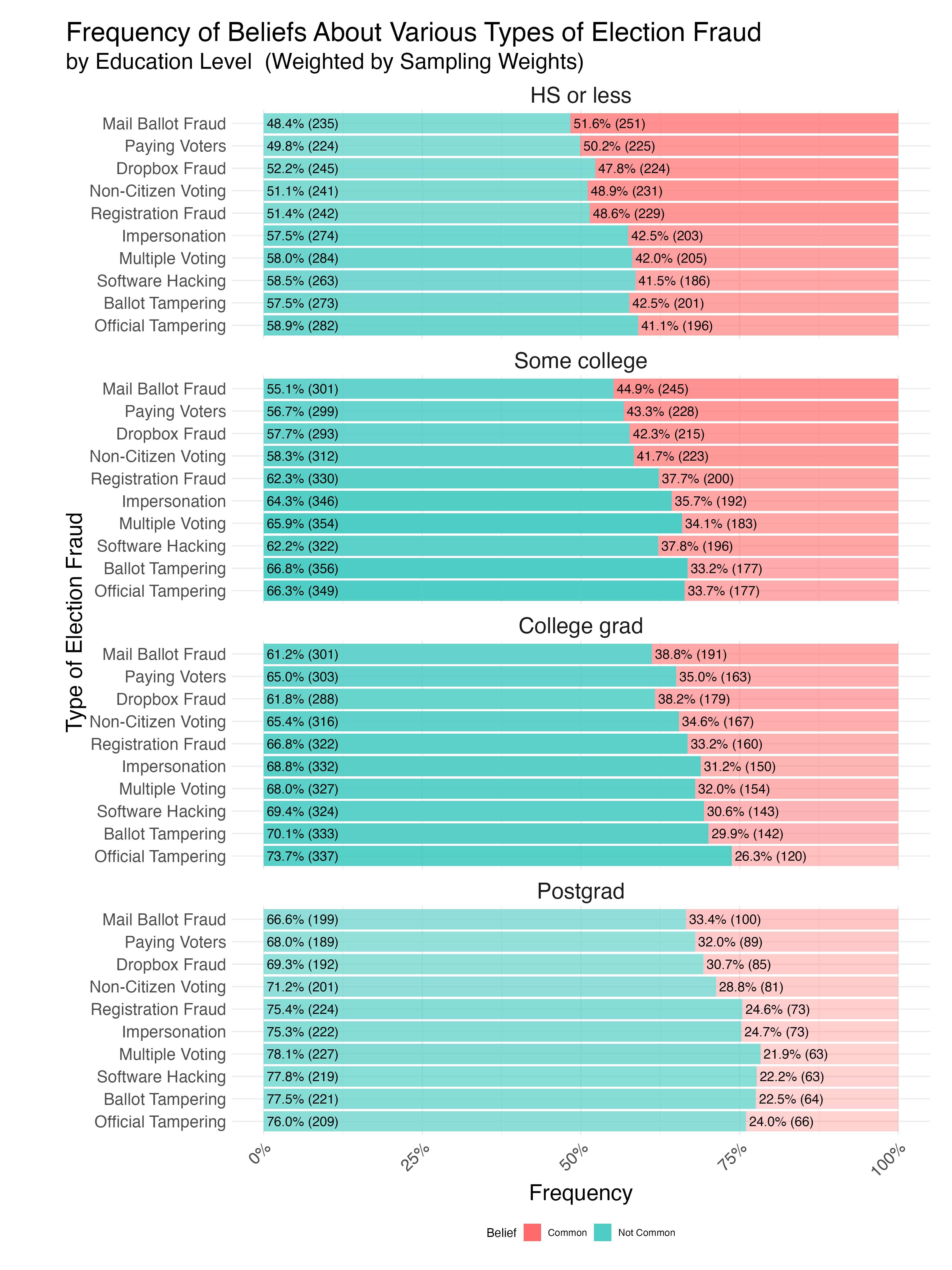}
    \caption{Election Fraud, weighted row frequencies.}
    \label{crime_freqs_education}
\end{figure}

\clearpage

\begin{figure}[ht]
    \centering
    \includegraphics[width=0.8\linewidth]{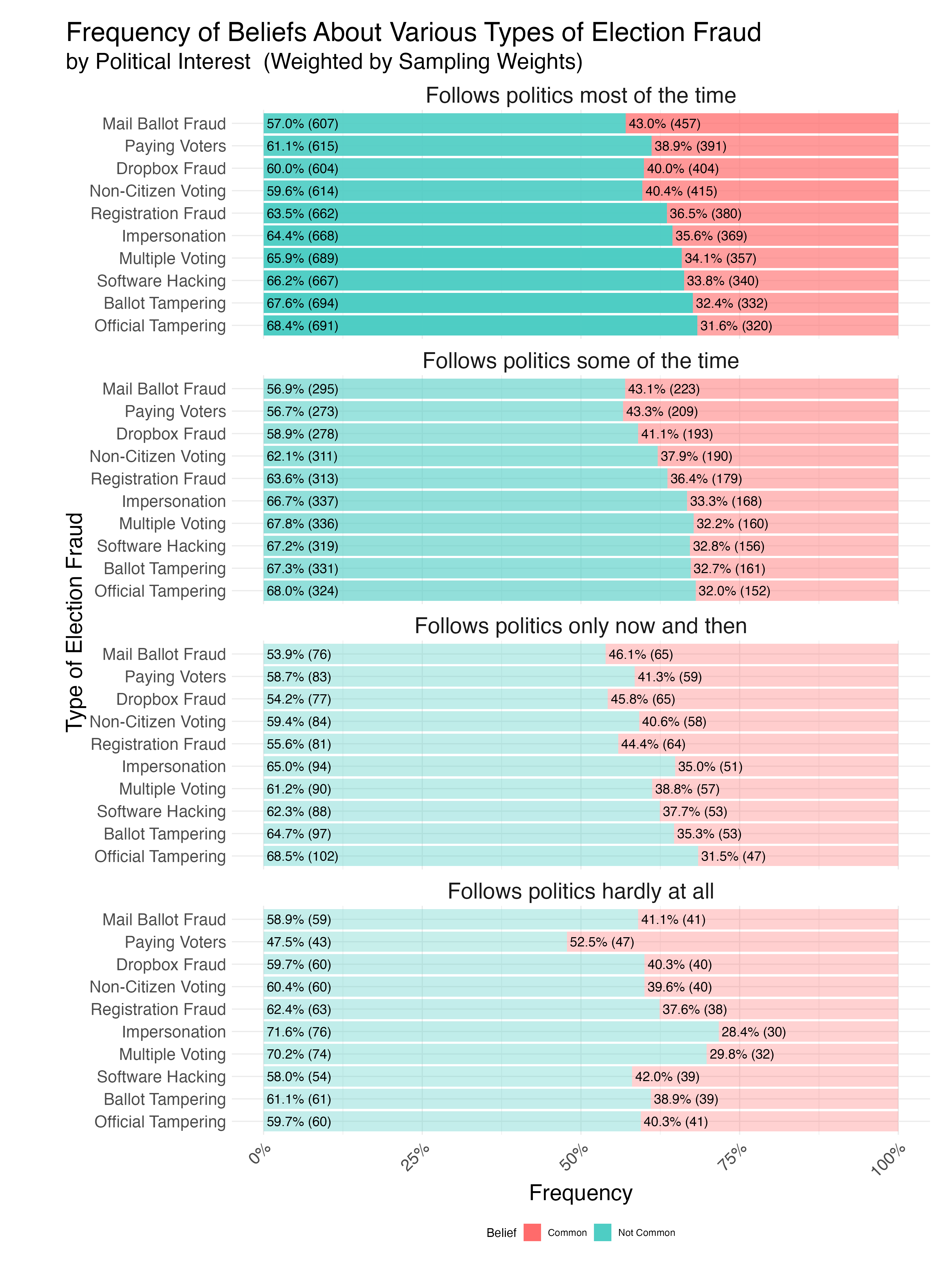}
    \caption{Election Fraud, weighted row frequencies.}
    \label{crime_freqs_interest}
\end{figure}

\clearpage

\begin{figure}[ht]
    \centering
    \includegraphics[width=0.8\linewidth]{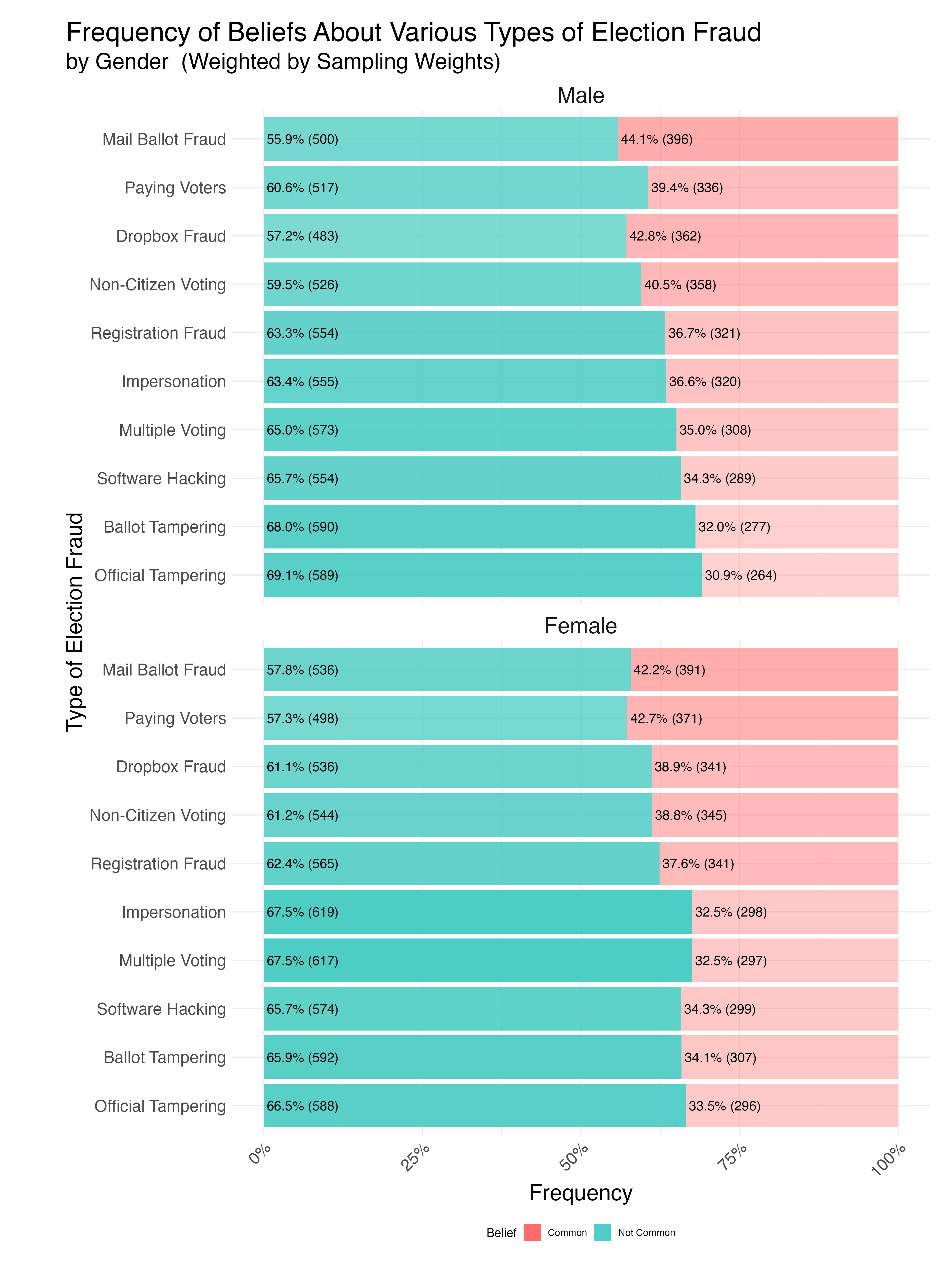}
    \caption{Election Fraud, weighted row frequencies.}
    \label{crime_freqs_gender}
\end{figure}

\clearpage

\begin{figure}[ht]
    \centering
    \includegraphics[width=0.8\linewidth]{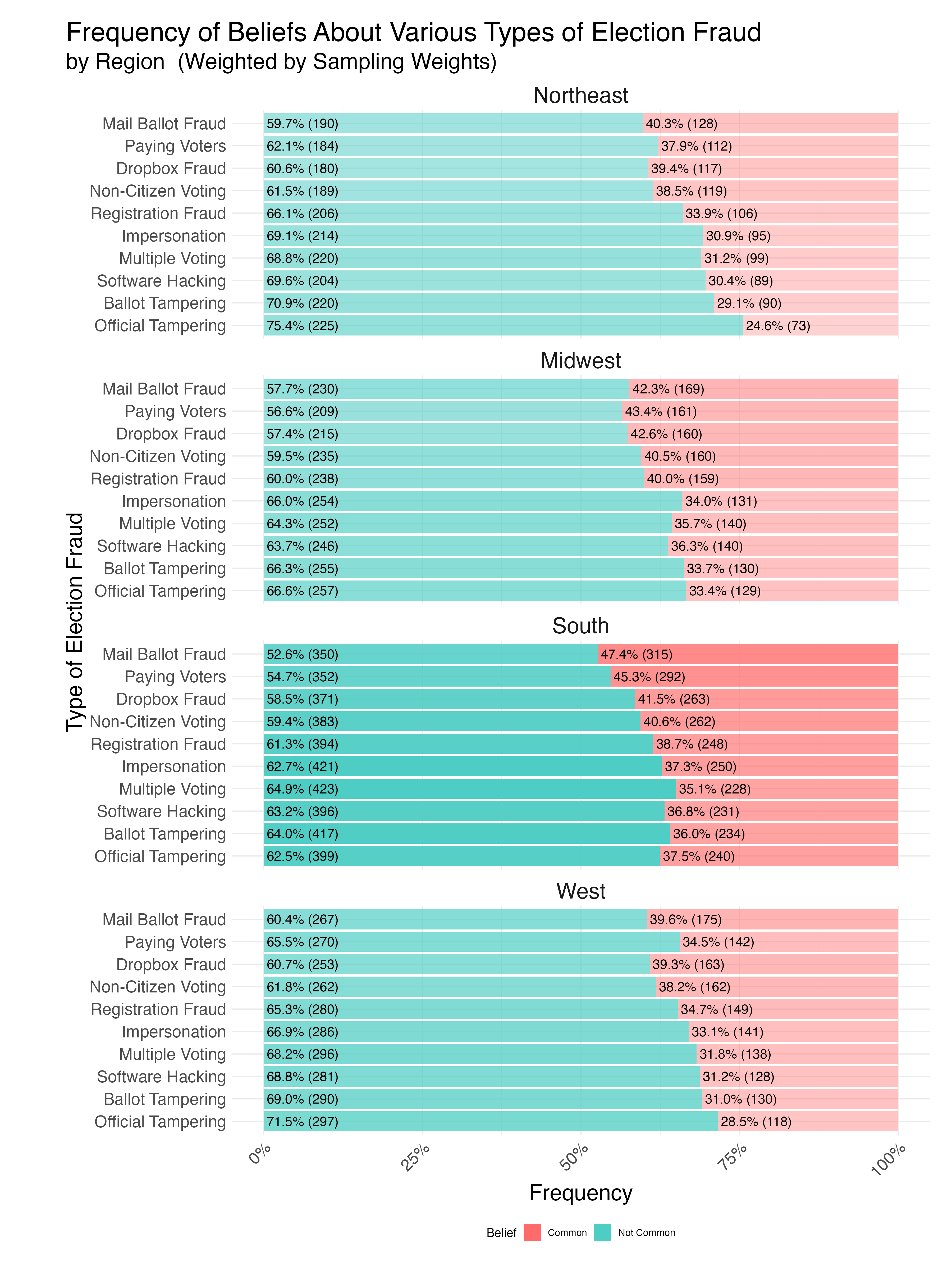}
    \caption{Election Fraud, weighted row frequencies.}
    \label{crime_freqs_region}
\end{figure}

\clearpage

\begin{figure}[ht]
    \centering
    \includegraphics[width=0.8\linewidth]{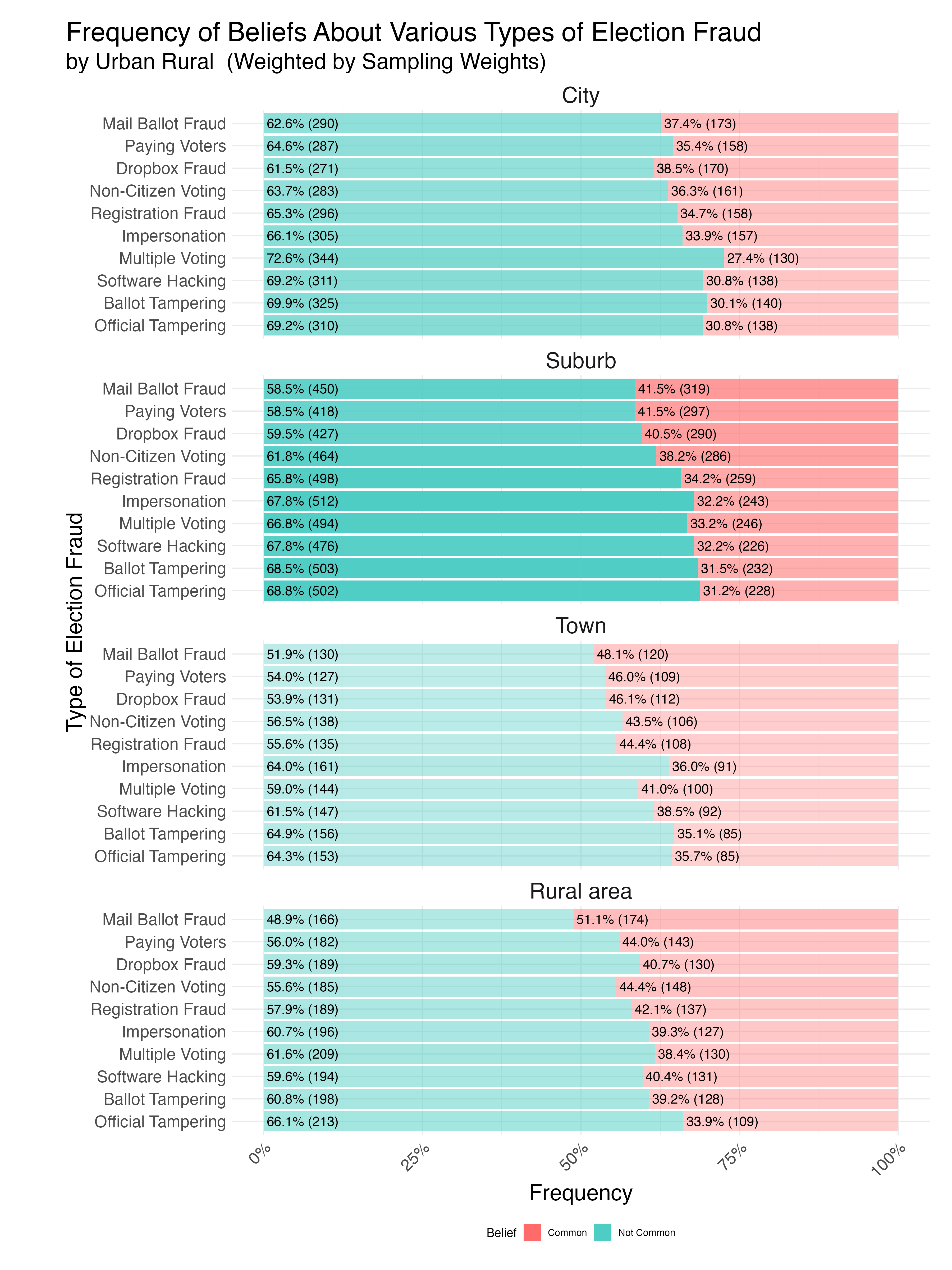}
    \caption{Election Fraud, weighted row frequencies.}
    \label{crime_freqs_urban}
\end{figure}

\clearpage

\begin{table}[]
    \centering
    \begin{tabular}{r|cc|cccc|}
\multicolumn{1}{c}{} & \multicolumn{2}{c}{Gender} & 
\multicolumn{4}{c}{Age} \\ \hline
\multicolumn{1}{c}{Problem} & \multicolumn{1}{c}{Female} & \multicolumn{1}{c}{Male} & \multicolumn{1}{c}{Under 30} & \multicolumn{1}{c}{30 to 44} & \multicolumn{1}{c}{45 to 64} & 
\multicolumn{1}{c}{65 and Over} \\ \hline\hline
Multiple Voting & .33 & .35 & .25 & .30 & .38 & .36 \\ 
Ballot Tampering & .34 & .32 & .29 & .36 & .35 & .31  \\
Impersonation & .33 & .37 & .27 & .34 & .38 & .35  \\
Non-Citizen Voting & .39 & .40 & .32 & .36 & .45 & .41  \\
Mail Ballot Fraud & .43 & .44 & .35 & .45 & .46 & .42  \\
Official Tampering & .34 & .31 & .26 & .32 & .36 & .32 \\
Software Hacking & .35 & .34 & .29 & .37 & .36 & .34 \\
Paying Voters & .43 & .39 & .38 & .44 & .41 & .41  \\
Registration Fraud & .38 & .37 & .32 & .35 & .39 & .39  \\
Dropbox Fraud & .39 & .43 & .37 & .43 & .42 & .39  \\ \hline\hline
    \end{tabular}
    \caption{Election Fraud:  Gender and Age, weighted row frequencies.}
    \label{tab:election_crimes1}
\end{table}

\clearpage

\begin{table}[]
    \centering
    \begin{tabular}{r|cccc|cccc|}
\multicolumn{1}{c}{} & \multicolumn{4}{c}{Education} & 
\multicolumn{4}{c}{Race/Ethnicity} \\ \hline
\multicolumn{1}{c}{Problem} & \multicolumn{1}{c}{HS or less} & \multicolumn{1}{c}{Some College} & \multicolumn{1}{c}{College} & \multicolumn{1}{c}{Post-grad} & \multicolumn{1}{c}{White} & 
\multicolumn{1}{c}{Black} & \multicolumn{1}{c}{Hispanic} & 
\multicolumn{1}{c}{Other} \\ \hline\hline
Multiple Voting &  .42 & .34 & .32 & .22 & .35 & .30 & .32 & .35\\ 
Ballot Tampering &  .43 & .33 & .30 & .22 & .33 & .34 & .37 & .34 \\
Impersonation & .43 & .35 & .32 & .25 & .35 & .31 & .31 & .38  \\
Non-Citizen Voting &  .49 & .41 & .35 & .29 & .41 & .30 & .33 & .45 \\
Mail Ballot Fraud & .52 & .45 & .39 & .33 & .44 & .37 & .44 & .48  \\
Official Tampering & .41 & .34 & .26 & .24 & .33 & .28 & .33 & .32 \\
Software Hacking & .43 & .38 & .31 & .22 & .34 & .35 & .37 & .40 \\
Paying Voters & .51 & .43 & .35 & .31 & .40 & .47 & .40 & .42  \\
Registration Fraud &  .49 & .37 & .34 & .24 & .37 & .39 & .34 & .40 \\
Dropbox Fraud & .48 & .42 & .38 & .31 & .42 & .39 & .36 & .43  \\ \hline\hline
    \end{tabular}
    \caption{Election Fraud:  education and race/ethnicity, weighted row frequencies.}
    \label{tab:election_crimes2}
\end{table}

\begin{table}[]
    \centering
    \begin{tabular}{r|cccc|cccc|}
\multicolumn{1}{c}{} & \multicolumn{4}{c}{Region} & 
\multicolumn{4}{c}{Urban/Rural} \\ \hline
\multicolumn{1}{c}{Problem} & \multicolumn{1}{c}{NE} & \multicolumn{1}{c}{MW} & \multicolumn{1}{c}{S} & \multicolumn{1}{c}{West} & \multicolumn{1}{c}{City} & 
\multicolumn{1}{c}{Suburb} & \multicolumn{1}{c}{Town} & 
\multicolumn{1}{c}{Rural} \\ \hline\hline
Multiple Voting & .31 & .36 & .36 & .32 & .27 & .33 & .40 & .39 \\  
Ballot Tampering & .31 & .33 & .36 & .31 & .30 & .32 & .36 & .39 \\
Impersonation & .31 & .34 & .37 & .33 & .34 & .32 & .35 & .40 \\
Non-Citizen Voting &  .38 & .41 & .41 & .38 & .37 & .38 & .43 & .45 \\
Mail Ballot Fraud & .41 & .43 & .37 & .40 & .38 & .41 & .49 & .51 \\
Official Tampering & .24 & .33 & .37 & .29 & .31 & .31 & .36 & .34 \\
Software Hacking & .32 & .36 & .37 & .31 & .31 & .32 & .40 & .41 \\
Paying Voters & .39 & .43 & .45 & .34 & .35 & .42 & .46 & .44 \\
Registration Fraud &  .34 & .40 & .39 & .35 & .35 & .34 & .44 & .42 \\
Dropbox Fraud & .41 & .42 & .41 & .39 & .39 & .41 & .46 & .41  \\ \hline\hline
    \end{tabular}
    \caption{Election Fraud:  region and urban/rural residency, weighted row frequencies.}
    \label{tab:election_crimes3}
\end{table}

\begin{table}[]
    \centering
    \begin{tabular}{r|ccccc|}
\multicolumn{1}{c}{} & \multicolumn{5}{c}{Partisanship} \\ \hline
\multicolumn{1}{c}{Problem} & \multicolumn{1}{c}{Dem} & \multicolumn{1}{c}{Rep} & \multicolumn{1}{c}{Ind} & \multicolumn{1}{c}{Other} & \multicolumn{1}{c}{NS} \\ \hline\hline
Multiple Voting & .12 & .58 & .32 & .40 & .25  \\  
Ballot Tampering & .11 & .58 & .29 & .38 & .30  \\
Impersonation &  .13 & .58 & .32 & .40 & .09 \\
Non-Citizen Voting & .12 & .66  & .41 & .47 & .23   \\
Mail Ballot Fraud & .16 & .70 & .43 & .47 & .44 \\
Official Tampering & .13 & .54 & .28 & .43 & .29  \\
Software Hacking & .12 & .57 & .33 & .42 & .50 \\
Paying Voters & .22 & .62 & .39 & .44 & .18 \\
Registration Fraud & .14 & .61 & .37 & .45 & .18   \\
Dropbox Fraud & .14 & .67 & .42 & .47 & .17  \\ \hline\hline
    \end{tabular}
    \caption{Election Fraud:  Partisanship, weighted row frequencies.}
    \label{tab:election_crimes4}
\end{table}

\begin{table}[]
    \centering
    \begin{tabular}{r|cccc|}
\multicolumn{1}{c}{} & \multicolumn{4}{c}{Ideology} \\ \hline
\multicolumn{1}{c}{Problem} & \multicolumn{1}{c}{Lib} & \multicolumn{1}{c}{Mod} & \multicolumn{1}{c}{Cons} & \multicolumn{1}{c}{NS} \\ \hline\hline
Multiple Voting & .10 & .29 & .57 & .38  \\  
Ballot Tampering & .10 & .27 & .56 & .43  \\
Impersonation & .10 & .32 & .56 & .40  \\
Non-Citizen Voting & .11 & .35 & .66 & .44   \\
Mail Ballot Fraud & .13 & .38 & .70 & .51 \\
Official Tampering & .11 & .27 & .54 & .32  \\
Software Hacking & .11 & .26 & .59 & .50 \\
Paying Voters & .17 & .38 & .61 & .44 \\
Registration Fraud & .12 & .33 & .61 & .43   \\
Dropbox Fraud & .14 & .35 & .67 & .45  \\ \hline\hline
    \end{tabular}
    \caption{Election Fraud:  Ideology, weighted row frequencies.}
    \label{tab:election_crimes5}
\end{table}

\clearpage

\subsection{Multivariate Plots}

\begin{figure}[ht]
    \centering
    \includegraphics[width=0.8\linewidth]{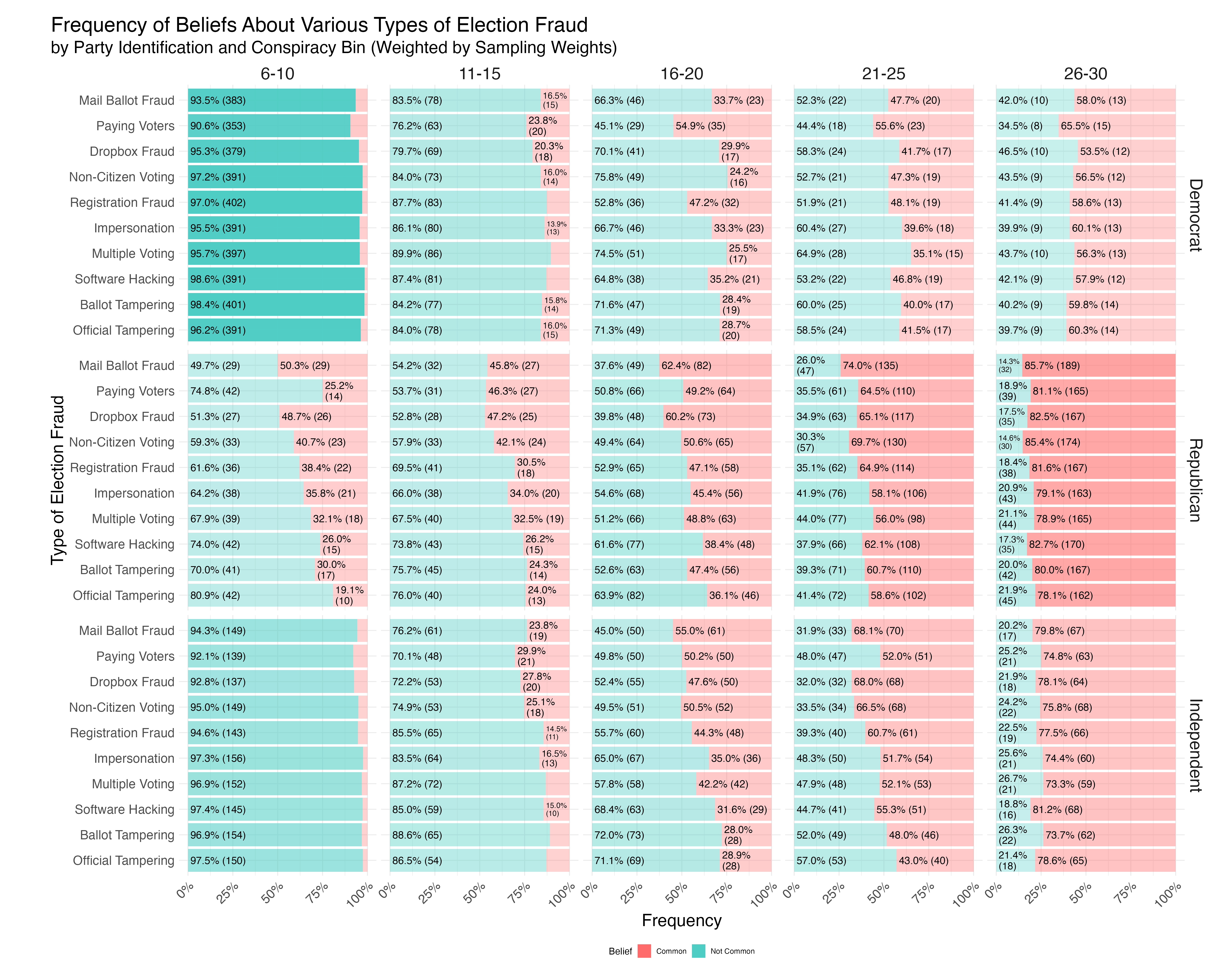}
    \caption{Concerns about election fraud by party and level of belief in non-election conspiracy theories. Respondents are grouped by party and their total level of agreement (on a 1-5 scale) with six conspiracy questions. Higher numbers represent a higher average agreement with the conspiracy theory questions.}
    \label{crime_freqs_party_conspiracy}
\end{figure}

\clearpage

\newpage

\begin{figure}[ht]
    \centering
    \includegraphics[width=0.8\linewidth]{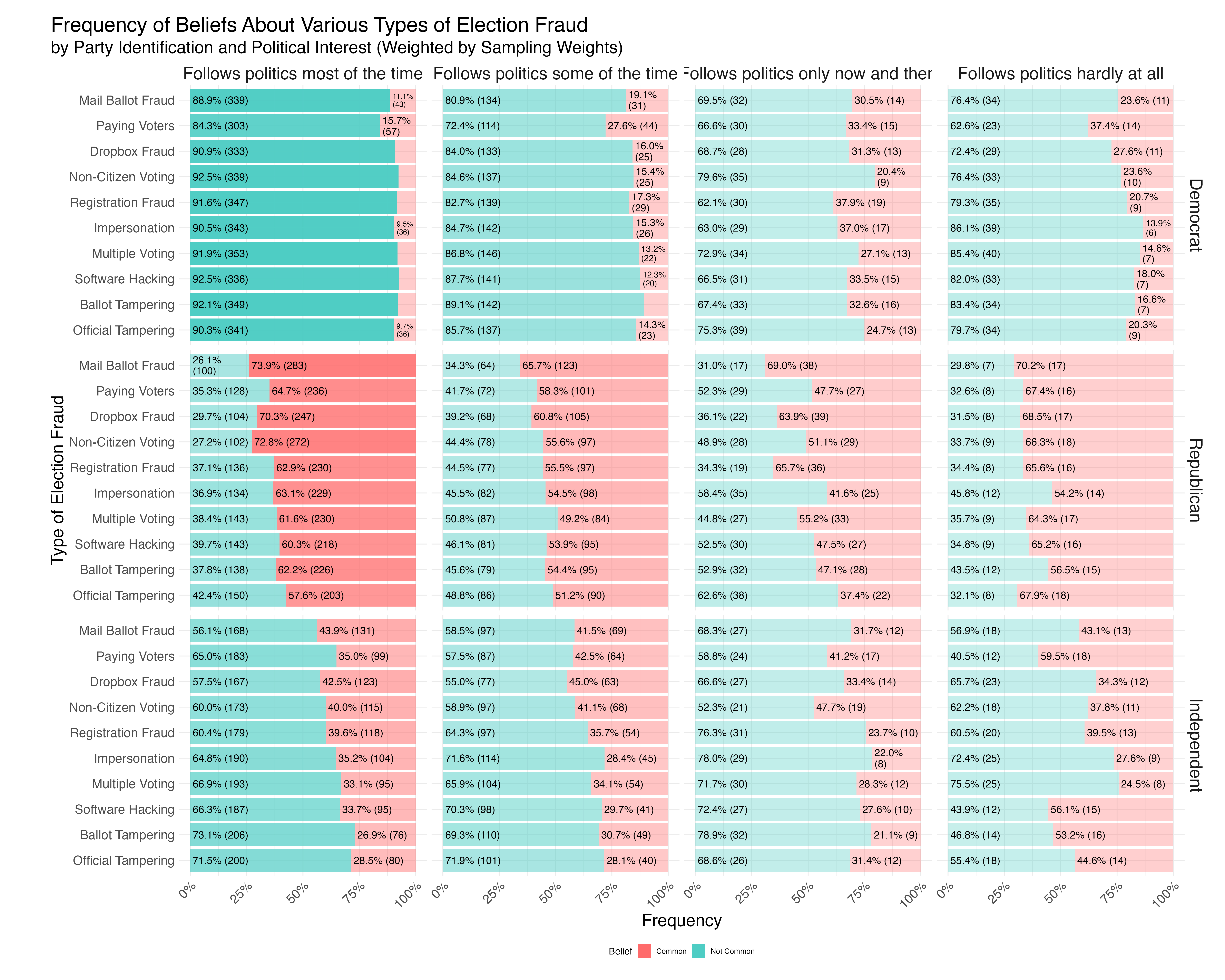}
    \caption{Concerns about election fraud by party and stated level of political interest.}
    \label{crime_freqs_party_interest}
\end{figure}

\newpage

\clearpage

\begin{figure}[ht]
    \centering
    \includegraphics[width=0.8\linewidth]{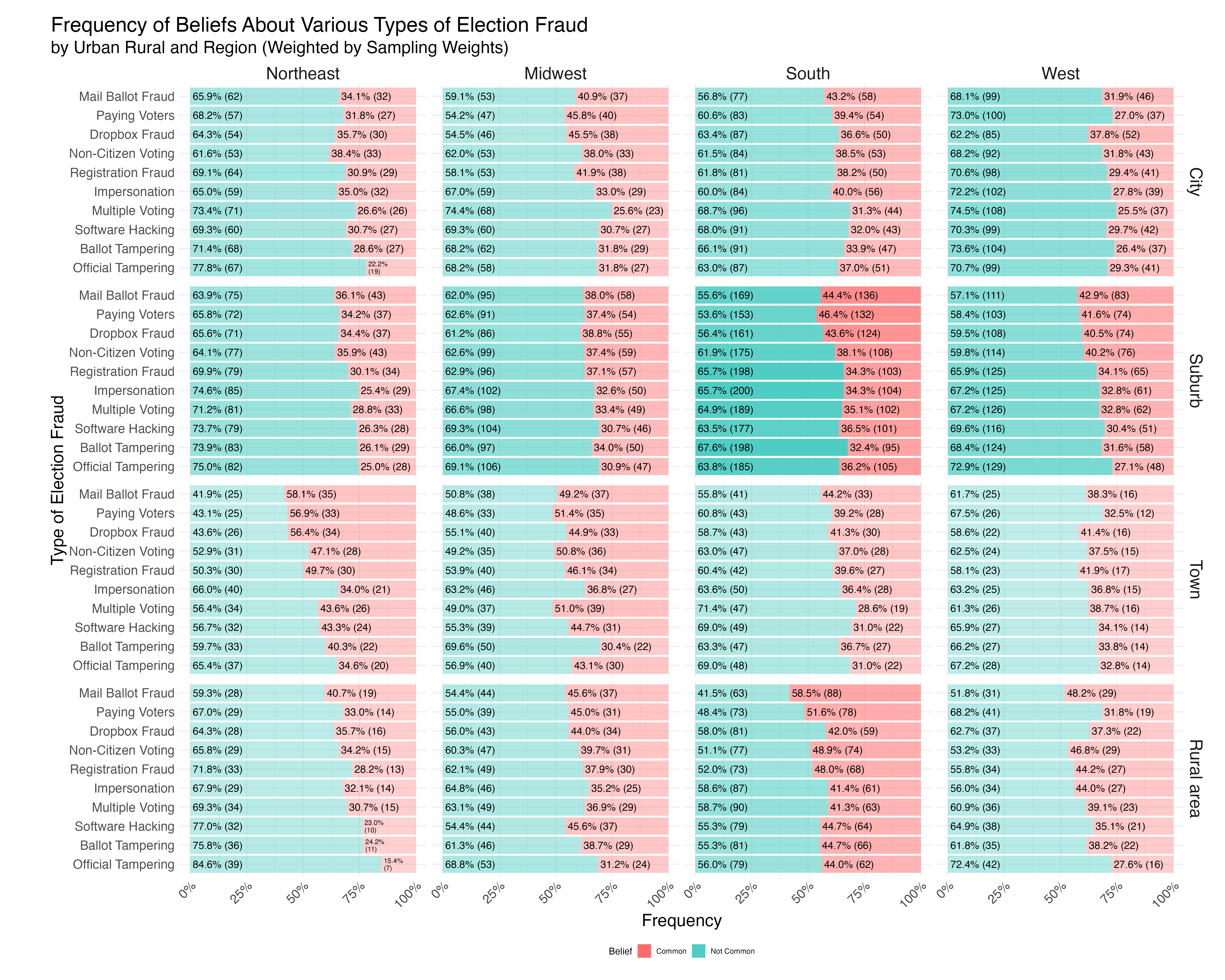}
    \caption{Concerns about election fraud by urban/rural status.}
    \label{crime_freqs_urban_region}
\end{figure}

\newpage

\clearpage

\begin{figure}[ht]
    \centering
    \includegraphics[width=0.8\linewidth]{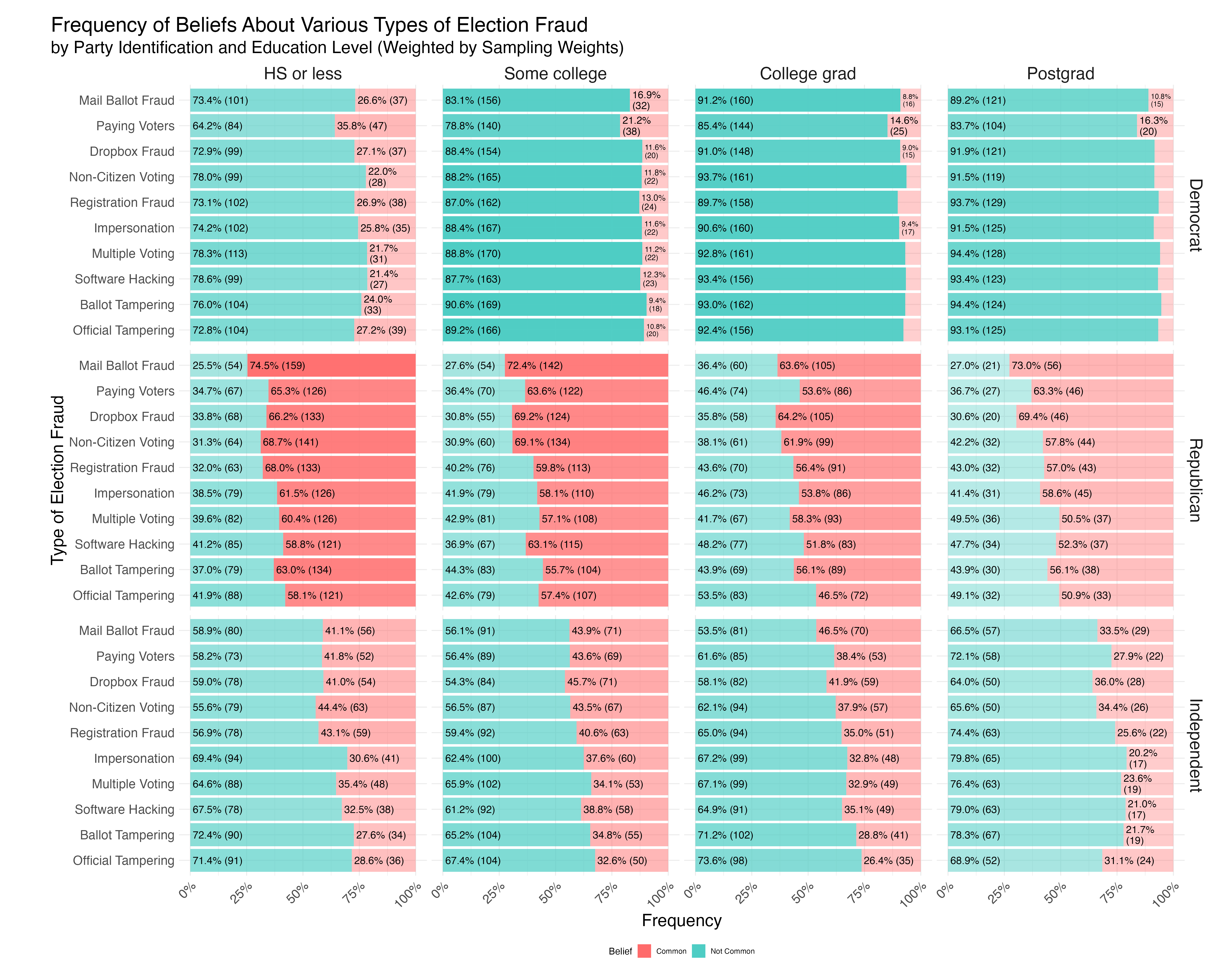}
    \caption{Concerns about election fraud by party and level of education.}
    \label{crime_freqs_party_education}
\end{figure}

\newpage 

\clearpage

\begin{figure}[ht]
    \centering
    \includegraphics[width=0.8\linewidth]{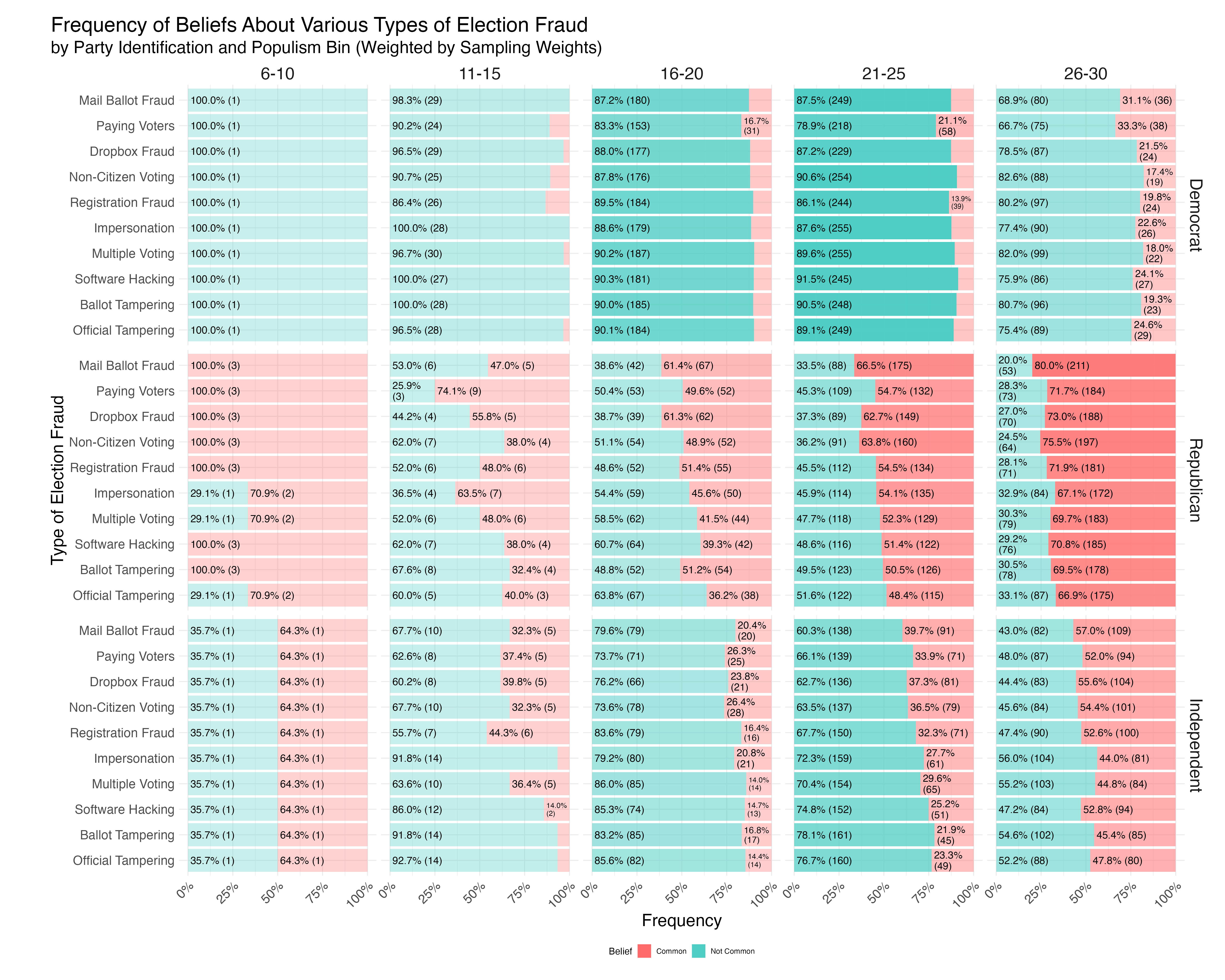}
    \caption{Concerns about election fraud by party and level of populism. Respondents are grouped by party and their total level of agreement (on a 1-5 scale) with six populism questions. Higher numbers represent a higher average agreement with populist sentiment.}
    \label{crime_freqs_party_populism}
\end{figure}

\newpage

\clearpage

\section{Regression Tables}

{ \let\oldcentering\centering \renewcommand\centering{\tiny\oldcentering}

\begin{table}[h] \centering 
  \caption{Logistic Regression Results for Election Crime Questions} 
  \label{tab:law_regs_table1} 
\footnotesize 
\begin{tabular}{@{\extracolsep{5pt}}lcccc} 
\\[-1.8ex]\hline 
\hline \\[-1.8ex] 
 & \multicolumn{4}{c}{\textit{Dependent variable:}} \\ 
\cline{2-5} 
\\[-1.8ex] & Multiple Voting & Ballot Tampering & Impersonation & Non-Citizen Voting \\ 
\\[-1.8ex] & (1) & (2) & (3) & (4)\\ 
\hline \\[-1.8ex] 
 30-44 & $-$0.137 & 0.141 & 0.063 & $-$0.213 \\ 
  & (0.305) & (0.279) & (0.286) & (0.286) \\ 
  45-64 & 0.255 & $-$0.072 & 0.191 & 0.171 \\ 
  & (0.292) & (0.264) & (0.268) & (0.262) \\ 
  65+ & 0.258 & $-$0.184 & 0.112 & $-$0.028 \\ 
  & (0.294) & (0.273) & (0.275) & (0.275) \\ 
  Female & 0.127 & 0.413$^{*}$ & 0.028 & 0.209 \\ 
  & (0.155) & (0.163) & (0.150) & (0.157) \\ 
  Black & 0.476 & 0.561 & 0.311 & 0.117 \\ 
  & (0.271) & (0.311) & (0.260) & (0.289) \\ 
  Hispanic & 0.276 & 0.555$^{*}$ & 0.039 & $-$0.179 \\ 
  & (0.307) & (0.261) & (0.284) & (0.271) \\ 
  Other & 0.283 & 0.251 & 0.355 & 0.401 \\ 
  & (0.317) & (0.343) & (0.302) & (0.341) \\ 
  Some college & $-$0.293 & $-$0.242 & $-$0.241 & $-$0.111 \\ 
  & (0.207) & (0.207) & (0.202) & (0.203) \\ 
  College grad & 0.004 & $-$0.076 & $-$0.122 & $-$0.256 \\ 
  & (0.224) & (0.220) & (0.217) & (0.217) \\ 
  Postgrad & $-$0.173 & 0.100 & $-$0.071 & 0.014 \\ 
  & (0.249) & (0.256) & (0.247) & (0.257) \\ 
  Republican & 1.165$^{***}$ & 1.350$^{***}$ & 1.011$^{***}$ & 1.411$^{***}$ \\ 
  & (0.287) & (0.313) & (0.279) & (0.286) \\ 
  Independent & 0.563$^{*}$ & 0.548$^{*}$ & 0.263 & 0.975$^{***}$ \\ 
  & (0.223) & (0.249) & (0.219) & (0.224) \\ 
  Moderate & 0.239 & $-$0.075 & 0.555$^{*}$ & 0.268 \\ 
  & (0.246) & (0.269) & (0.245) & (0.245) \\ 
  Conservative & 0.191 & $-$0.022 & 0.300 & 0.411 \\ 
  & (0.311) & (0.334) & (0.305) & (0.304) \\ 
  Midwest & $-$0.046 & 0.036 & 0.001 & $-$0.170 \\ 
  & (0.242) & (0.261) & (0.242) & (0.257) \\ 
  South & $-$0.087 & 0.027 & 0.078 & $-$0.325 \\ 
  & (0.234) & (0.232) & (0.226) & (0.222) \\ 
  West & $-$0.002 & 0.036 & 0.112 & $-$0.132 \\ 
  & (0.242) & (0.249) & (0.236) & (0.234) \\ 
  Suburb & 0.303 & 0.009 & $-$0.233 & $-$0.188 \\ 
  & (0.216) & (0.215) & (0.200) & (0.214) \\ 
  Town & 0.364 & $-$0.164 & $-$0.361 & $-$0.385 \\ 
  & (0.253) & (0.272) & (0.240) & (0.259) \\ 
  Rural area & $-$0.214 & $-$0.288 & $-$0.567$^{*}$ & $-$0.618$^{*}$ \\ 
  & (0.243) & (0.246) & (0.234) & (0.243) \\ 
  Follows politics some of the time & $-$0.014 & $-$0.089 & $-$0.187 & $-$0.130 \\ 
  & (0.187) & (0.185) & (0.180) & (0.180) \\ 
  Follows politics only now and then & 0.222 & $-$0.150 & $-$0.276 & $-$0.065 \\ 
  & (0.283) & (0.316) & (0.283) & (0.290) \\ 
  Follows politics hardly at all & $-$0.324 & 0.033 & $-$0.639$^{*}$ & $-$0.244 \\ 
  & (0.316) & (0.333) & (0.314) & (0.368) \\ 
  Populism Score & 0.036 & 0.020 & 0.015 & 0.010 \\ 
  & (0.022) & (0.020) & (0.022) & (0.020) \\ 
  Conspiracy Score & 0.150$^{***}$ & 0.174$^{***}$ & 0.149$^{***}$ & 0.156$^{***}$ \\ 
  & (0.015) & (0.015) & (0.014) & (0.014) \\ 
  Constant & $-$5.312$^{***}$ & $-$5.206$^{***}$ & $-$4.231$^{***}$ & $-$4.072$^{***}$ \\ 
  & (0.619) & (0.618) & (0.550) & (0.584) \\ 
 \hline \\[-1.8ex] 
Observations & 1,835 & 1,814 & 1,832 & 1,807 \\ 
\hline 
\hline \\[-1.8ex] 
\textit{Note:}  & \multicolumn{4}{r}{$^{*}$p$<$0.05; $^{**}$p$<$0.01; $^{***}$p$<$0.001} \\ 
\end{tabular} 
\end{table} 

}

\newpage
\clearpage

{ \let\oldcentering\centering \renewcommand\centering{\tiny\oldcentering}

\begin{table}[h] \centering 
  \caption{Logistic Regression Results for Election Crime Questions} 
  \label{tab:law_regs_table2} 
\footnotesize 
\begin{tabular}{@{\extracolsep{5pt}}lccc} 
\\[-1.8ex]\hline 
\hline \\[-1.8ex] 
 & \multicolumn{3}{c}{\textit{Dependent variable:}} \\ 
\cline{2-4} 
\\[-1.8ex] & Mail Ballot Fraud & Official Tampering & Software Hacking \\ 
\\[-1.8ex] & (1) & (2) & (3)\\ 
\hline \\[-1.8ex] 
 30-44 & 0.119 & $-$0.119 & 0.107 \\ 
  & (0.274) & (0.299) & (0.300) \\ 
  45-64 & $-$0.042 & 0.114 & $-$0.111 \\ 
  & (0.262) & (0.289) & (0.284) \\ 
  65+ & $-$0.186 & 0.012 & $-$0.116 \\ 
  & (0.270) & (0.298) & (0.291) \\ 
  Female & 0.196 & 0.330$^{*}$ & 0.295 \\ 
  & (0.151) & (0.160) & (0.174) \\ 
  Black & 0.380 & $-$0.153 & 0.504 \\ 
  & (0.259) & (0.280) & (0.300) \\ 
  Hispanic & 0.267 & 0.224 & 0.416 \\ 
  & (0.263) & (0.302) & (0.310) \\ 
  Other & 0.555 & 0.096 & 0.676$^{*}$ \\ 
  & (0.296) & (0.343) & (0.315) \\ 
  Some college & $-$0.094 & $-$0.309 & 0.100 \\ 
  & (0.204) & (0.215) & (0.224) \\ 
  College grad & $-$0.112 & $-$0.425 & $-$0.0005 \\ 
  & (0.215) & (0.228) & (0.235) \\ 
  Postgrad & 0.216 & 0.152 & 0.127 \\ 
  & (0.234) & (0.279) & (0.260) \\ 
  Republican & 1.348$^{***}$ & 0.657$^{*}$ & 1.043$^{**}$ \\ 
  & (0.266) & (0.314) & (0.331) \\ 
  Independent & 0.657$^{**}$ & 0.232 & 0.685$^{**}$ \\ 
  & (0.200) & (0.249) & (0.253) \\ 
  Moderate & 0.304 & $-$0.155 & $-$0.368 \\ 
  & (0.219) & (0.272) & (0.271) \\ 
  Conservative & 0.467 & $-$0.034 & 0.067 \\ 
  & (0.286) & (0.336) & (0.342) \\ 
  Midwest & $-$0.181 & 0.370 & 0.063 \\ 
  & (0.230) & (0.242) & (0.266) \\ 
  South & $-$0.025 & 0.555$^{*}$ & $-$0.095 \\ 
  & (0.203) & (0.216) & (0.248) \\ 
  West & $-$0.107 & 0.200 & $-$0.002 \\ 
  & (0.218) & (0.241) & (0.268) \\ 
  Suburb & 0.108 & $-$0.144 & $-$0.006 \\ 
  & (0.192) & (0.209) & (0.228) \\ 
  Town & 0.112 & $-$0.305 & $-$0.101 \\ 
  & (0.230) & (0.259) & (0.270) \\ 
  Rural area & $-$0.106 & $-$0.799$^{**}$ & $-$0.327 \\ 
  & (0.232) & (0.249) & (0.252) \\ 
  Follows politics some of the time & $-$0.080 & 0.084 & 0.003 \\ 
  & (0.174) & (0.194) & (0.197) \\ 
  Follows politics only now and then & 0.013 & $-$0.184 & 0.120 \\ 
  & (0.271) & (0.293) & (0.307) \\ 
  Follows politics hardly at all & $-$0.311 & 0.300 & 0.196 \\ 
  & (0.352) & (0.332) & (0.362) \\ 
  Populism Score & 0.050$^{*}$ & 0.064$^{**}$ & 0.053$^{**}$ \\ 
  & (0.020) & (0.022) & (0.020) \\ 
  Conspiracy Score & 0.136$^{***}$ & 0.178$^{***}$ & 0.186$^{***}$ \\ 
  & (0.012) & (0.015) & (0.017) \\ 
  Constant & $-$4.856$^{***}$ & $-$5.948$^{***}$ & $-$6.086$^{***}$ \\ 
  & (0.555) & (0.626) & (0.678) \\ 
 \hline \\[-1.8ex] 
Observations & 1,869 & 1,779 & 1,769 \\ 
\hline 
\hline \\[-1.8ex] 
\textit{Note:}  & \multicolumn{3}{r}{$^{*}$p$<$0.05; $^{**}$p$<$0.01; $^{***}$p$<$0.001} \\ 
\end{tabular} 
\end{table} 

}

\newpage
\clearpage

{ \let\oldcentering\centering \renewcommand\centering{\tiny\oldcentering}

\begin{table}[h] \centering 
  \caption{Logistic Regression Results for Election Crime Questions} 
  \label{tab:law_regs_table3} 
\footnotesize 
\begin{tabular}{@{\extracolsep{5pt}}lccc} 
\\[-1.8ex]\hline 
\hline \\[-1.8ex] 
 & \multicolumn{3}{c}{\textit{Dependent variable:}} \\ 
\cline{2-4} 
\\[-1.8ex] & Paying Voters & Registration Fraud & Dropbox Fraud \\ 
\\[-1.8ex] & (1) & (2) & (3)\\ 
\hline \\[-1.8ex] 
 30-44 & $-$0.013 & $-$0.236 & $-$0.035 \\ 
  & (0.261) & (0.281) & (0.290) \\ 
  45-64 & $-$0.227 & $-$0.105 & $-$0.195 \\ 
  & (0.243) & (0.262) & (0.279) \\ 
  65+ & $-$0.209 & 0.100 & $-$0.404 \\ 
  & (0.256) & (0.268) & (0.286) \\ 
  Female & 0.369$^{**}$ & 0.328$^{*}$ & $-$0.004 \\ 
  & (0.142) & (0.157) & (0.152) \\ 
  Black & 0.631$^{*}$ & 0.657$^{*}$ & 0.405 \\ 
  & (0.259) & (0.270) & (0.268) \\ 
  Hispanic & 0.316 & 0.218 & $-$0.116 \\ 
  & (0.245) & (0.270) & (0.304) \\ 
  Other & 0.236 & 0.263 & 0.271 \\ 
  & (0.287) & (0.305) & (0.303) \\ 
  Some college & $-$0.103 & $-$0.442$^{*}$ & $-$0.056 \\ 
  & (0.194) & (0.200) & (0.209) \\ 
  College grad & $-$0.293 & $-$0.302 & $-$0.087 \\ 
  & (0.202) & (0.216) & (0.217) \\ 
  Postgrad & 0.089 & $-$0.333 & 0.209 \\ 
  & (0.247) & (0.236) & (0.247) \\ 
  Republican & 0.615$^{*}$ & 1.094$^{***}$ & 1.466$^{***}$ \\ 
  & (0.264) & (0.295) & (0.272) \\ 
  Independent & 0.112 & 0.627$^{**}$ & 1.004$^{***}$ \\ 
  & (0.203) & (0.225) & (0.209) \\ 
  Moderate & 0.270 & 0.257 & $-$0.016 \\ 
  & (0.212) & (0.242) & (0.214) \\ 
  Conservative & 0.274 & 0.311 & 0.329 \\ 
  & (0.281) & (0.308) & (0.280) \\ 
  Midwest & 0.108 & 0.088 & $-$0.186 \\ 
  & (0.213) & (0.246) & (0.233) \\ 
  South & 0.034 & $-$0.006 & $-$0.312 \\ 
  & (0.190) & (0.224) & (0.218) \\ 
  West & $-$0.252 & $-$0.006 & $-$0.142 \\ 
  & (0.213) & (0.241) & (0.219) \\ 
  Suburb & 0.371$^{*}$ & $-$0.191 & $-$0.069 \\ 
  & (0.184) & (0.199) & (0.200) \\ 
  Town & 0.254 & $-$0.002 & $-$0.217 \\ 
  & (0.225) & (0.236) & (0.237) \\ 
  Rural area & $-$0.165 & $-$0.519$^{*}$ & $-$0.730$^{**}$ \\ 
  & (0.217) & (0.249) & (0.239) \\ 
  Follows politics some of the time & 0.113 & $-$0.053 & $-$0.019 \\ 
  & (0.170) & (0.180) & (0.183) \\ 
  Follows politics only now and then & $-$0.165 & 0.217 & 0.114 \\ 
  & (0.277) & (0.280) & (0.259) \\ 
  Follows politics hardly at all & 0.248 & $-$0.263 & $-$0.174 \\ 
  & (0.334) & (0.322) & (0.348) \\ 
  Populism Score & 0.023 & 0.017 & 0.021 \\ 
  & (0.020) & (0.019) & (0.022) \\ 
  Conspiracy Score & 0.139$^{***}$ & 0.161$^{***}$ & 0.135$^{***}$ \\ 
  & (0.012) & (0.014) & (0.013) \\ 
  Constant & $-$4.009$^{***}$ & $-$4.421$^{***}$ & $-$3.691$^{***}$ \\ 
  & (0.527) & (0.565) & (0.540) \\ 
 \hline \\[-1.8ex] 
Observations & 1,763 & 1,824 & 1,771 \\ 
\hline 
\hline \\[-1.8ex] 
\textit{Note:}  & \multicolumn{3}{r}{$^{*}$p$<$0.05; $^{**}$p$<$0.01; $^{***}$p$<$0.001} \\ 
\end{tabular} 
\end{table} 

}

\newpage
\clearpage

{ \let\oldcentering\centering \renewcommand\centering{\tiny\oldcentering}

\begin{table}[h] \centering 
  \caption{Regression Results for Aggregate Answers to Election Crime Questions} 
  \label{tab:total_law_reg} 
\footnotesize 
\begin{tabular}{@{\extracolsep{5pt}}lcc} 
\\[-1.8ex]\hline 
\hline \\[-1.8ex] 
 & \multicolumn{2}{c}{\textit{Dependent variable:}} \\ 
\cline{2-3} 
\\[-1.8ex] & Number of Election Crimes Believed & Proportion of Election Crimes Believed \\ 
\\[-1.8ex] & \textit{survey-weighted} & \textit{survey-weighted} \\ 
 & \textit{normal} & \textit{logistic} \\ 
\\[-1.8ex] & (1) & (2)\\ 
\hline \\[-1.8ex] 
 30-44 & $-$0.006 & $-$0.007 \\ 
  & (0.278) & (0.190) \\ 
  45-64 & $-$0.208 & $-$0.143 \\ 
  & (0.272) & (0.184) \\ 
  65+ & $-$0.372 & $-$0.241 \\ 
  & (0.271) & (0.187) \\ 
  Female & 0.143 & 0.118 \\ 
  & (0.167) & (0.107) \\ 
  Black & 0.147 & 0.257 \\ 
  & (0.278) & (0.183) \\ 
  Hispanic & 0.054 & 0.086 \\ 
  & (0.296) & (0.199) \\ 
  Other & 0.603 & 0.398 \\ 
  & (0.361) & (0.228) \\ 
  Some college & 0.100 & 0.028 \\ 
  & (0.237) & (0.142) \\ 
  College grad & 0.157 & 0.057 \\ 
  & (0.249) & (0.152) \\ 
  Postgrad & 0.434 & 0.241 \\ 
  & (0.261) & (0.175) \\ 
  Republican & 1.423$^{***}$ & 0.913$^{***}$ \\ 
  & (0.345) & (0.201) \\ 
  Independent & 0.380 & 0.413$^{**}$ \\ 
  & (0.200) & (0.154) \\ 
  Moderate & 0.051 & 0.230 \\ 
  & (0.193) & (0.169) \\ 
  Conservative & 0.125 & 0.256 \\ 
  & (0.328) & (0.206) \\ 
  Midwest & 0.091 & 0.033 \\ 
  & (0.245) & (0.162) \\ 
  South & 0.018 & 0.006 \\ 
  & (0.226) & (0.147) \\ 
  West & 0.126 & 0.066 \\ 
  & (0.239) & (0.157) \\ 
  Suburb & $-$0.012 & $-$0.027 \\ 
  & (0.204) & (0.140) \\ 
  Town & 0.117 & 0.022 \\ 
  & (0.262) & (0.168) \\ 
  Rural area & $-$0.579$^{*}$ & $-$0.377$^{*}$ \\ 
  & (0.244) & (0.156) \\ 
  Follows politics some of the time & $-$0.558$^{**}$ & $-$0.244 \\ 
  & (0.199) & (0.125) \\ 
  Follows politics only now and then & $-$0.766$^{**}$ & $-$0.321 \\ 
  & (0.283) & (0.180) \\ 
  Follows politics hardly at all & $-$0.919$^{*}$ & $-$0.409 \\ 
  & (0.401) & (0.234) \\ 
  Populism Score & 0.045$^{*}$ & 0.024 \\ 
  & (0.023) & (0.015) \\ 
  Conspiracy Score & 0.248$^{***}$ & 0.137$^{***}$ \\ 
  & (0.016) & (0.010) \\ 
  Constant & $-$2.413$^{***}$ & $-$4.429$^{***}$ \\ 
  & (0.585) & (0.419) \\ 
 \hline \\[-1.8ex] 
Observations & 2,075 & 2,075 \\ 
Log Likelihood & $-$5,359.648 &  \\ 
Akaike Inf. Crit. & 10,771.300 &  \\ 
\hline 
\hline \\[-1.8ex] 
\textit{Note:}  & \multicolumn{2}{r}{$^{*}$p$<$0.05; $^{**}$p$<$0.01; $^{***}$p$<$0.001} \\ 
\end{tabular} 
\end{table} 

}

{ \let\oldcentering\centering \renewcommand\centering{\tiny\oldcentering}

\begin{table}[h] \centering 
  \caption{Regression Results for Aggregate Answers to Election Crime Questions} 
  \label{tab:total_law_reg_mini} 
\footnotesize 
\begin{tabular}{@{\extracolsep{5pt}}lcc} 
\\[-1.8ex]\hline 
\hline \\[-1.8ex] 
 & \multicolumn{2}{c}{\textit{Dependent variable:}} \\ 
\cline{2-3} 
\\[-1.8ex] & Number of Election Crimes Believed & Proportion of Election Crimes Believed \\ 
\\[-1.8ex] & \textit{survey-weighted} & \textit{survey-weighted} \\ 
 & \textit{normal} & \textit{logistic} \\ 
\\[-1.8ex] & (1) & (2)\\ 
\hline \\[-1.8ex] 
 30-44 & 0.185 & 0.103 \\ 
  & (0.307) & (0.173) \\ 
  45-64 & 0.186 & 0.102 \\ 
  & (0.283) & (0.161) \\ 
  65+ & 0.039 & 0.025 \\ 
  & (0.291) & (0.166) \\ 
  Female & $-$0.083 & $-$0.045 \\ 
  & (0.177) & (0.094) \\ 
  Black & 0.952$^{**}$ & 0.600$^{***}$ \\ 
  & (0.300) & (0.177) \\ 
  Hispanic & 0.186 & 0.102 \\ 
  & (0.288) & (0.161) \\ 
  Other & 1.118$^{*}$ & 0.608$^{**}$ \\ 
  & (0.444) & (0.228) \\ 
  Some college & 0.039 & 0.023 \\ 
  & (0.248) & (0.128) \\ 
  College grad & $-$0.261 & $-$0.135 \\ 
  & (0.256) & (0.134) \\ 
  Postgrad & $-$0.389 & $-$0.219 \\ 
  & (0.278) & (0.156) \\ 
  Republican & 4.148$^{***}$ & 2.206$^{***}$ \\ 
  & (0.204) & (0.125) \\ 
  Independent & 1.847$^{***}$ & 1.199$^{***}$ \\ 
  & (0.203) & (0.131) \\ 
  Constant & 0.970$^{**}$ & $-$2.170$^{***}$ \\ 
  & (0.339) & (0.216) \\ 
 \hline \\[-1.8ex] 
Observations & 2,075 & 2,075 \\ 
Log Likelihood & $-$5,648.683 &  \\ 
Akaike Inf. Crit. & 11,323.370 &  \\ 
\hline 
\hline \\[-1.8ex] 
\textit{Note:}  & \multicolumn{2}{r}{$^{*}$p$<$0.05; $^{**}$p$<$0.01; $^{***}$p$<$0.001} \\ 
\end{tabular} 
\end{table} 

}

\newpage
\clearpage

\newpage
\clearpage

\subsection{Regression Tables with Interactions (Party ID and Political Interest)}

{ \let\oldcentering\centering \renewcommand\centering{\tiny\oldcentering}

\begin{table}[h] \centering 
  \caption{Logistic Regression Results for Election Crime Questions} 
  \label{tab:law_regs_table1_int} 
\footnotesize 
\begin{tabular}{@{\extracolsep{5pt}}lcccc} 
\\[-1.8ex]\hline 
\hline \\[-1.8ex] 
 & \multicolumn{4}{c}{\textit{Dependent variable:}} \\ 
\cline{2-5} 
\\[-1.8ex] & Multiple Voting & Ballot Tampering & Impersonation & Non-Citizen Voting \\ 
\\[-1.8ex] & (1) & (2) & (3) & (4)\\ 
\hline \\[-1.8ex] 
 30-44 & $-$0.138 & 0.163 & 0.030 & $-$0.203 \\ 
  & (0.304) & (0.279) & (0.285) & (0.270) \\ 
  45-64 & 0.255 & $-$0.069 & 0.150 & 0.161 \\ 
  & (0.291) & (0.266) & (0.268) & (0.251) \\ 
  65+ & 0.260 & $-$0.187 & 0.080 & $-$0.065 \\ 
  & (0.294) & (0.275) & (0.275) & (0.268) \\ 
  Female & 0.130 & 0.410$^{*}$ & 0.026 & 0.213 \\ 
  & (0.155) & (0.163) & (0.150) & (0.158) \\ 
  Black & 0.416 & 0.526 & 0.238 & 0.014 \\ 
  & (0.269) & (0.316) & (0.258) & (0.288) \\ 
  Hispanic & 0.264 & 0.595$^{*}$ & 0.069 & $-$0.186 \\ 
  & (0.311) & (0.272) & (0.282) & (0.271) \\ 
  Other & 0.226 & 0.245 & 0.295 & 0.322 \\ 
  & (0.322) & (0.359) & (0.302) & (0.338) \\ 
  Some college & $-$0.313 & $-$0.240 & $-$0.259 & $-$0.115 \\ 
  & (0.207) & (0.205) & (0.201) & (0.198) \\ 
  College grad & $-$0.001 & $-$0.083 & $-$0.145 & $-$0.279 \\ 
  & (0.224) & (0.222) & (0.216) & (0.215) \\ 
  Postgrad & $-$0.203 & 0.105 & $-$0.077 & $-$0.022 \\ 
  & (0.249) & (0.258) & (0.249) & (0.257) \\ 
  Republican & 1.348$^{***}$ & 1.558$^{***}$ & 1.355$^{***}$ & 2.037$^{***}$ \\ 
  & (0.368) & (0.372) & (0.349) & (0.344) \\ 
  Independent & 0.665$^{*}$ & 0.478 & 0.620$^{*}$ & 1.194$^{***}$ \\ 
  & (0.312) & (0.321) & (0.297) & (0.292) \\ 
  Moderate & 0.242 & $-$0.031 & 0.539$^{*}$ & 0.243 \\ 
  & (0.249) & (0.276) & (0.247) & (0.239) \\ 
  Conservative & 0.189 & 0.011 & 0.265 & 0.379 \\ 
  & (0.314) & (0.344) & (0.305) & (0.295) \\ 
  Midwest & $-$0.024 & 0.091 & 0.010 & $-$0.163 \\ 
  & (0.238) & (0.262) & (0.239) & (0.253) \\ 
  South & $-$0.047 & 0.043 & 0.089 & $-$0.312 \\ 
  & (0.234) & (0.236) & (0.225) & (0.221) \\ 
  West & 0.014 & 0.073 & 0.115 & $-$0.151 \\ 
  & (0.240) & (0.255) & (0.233) & (0.231) \\ 
  Suburb & 0.297 & 0.015 & $-$0.241 & $-$0.196 \\ 
  & (0.218) & (0.216) & (0.200) & (0.209) \\ 
  Town & 0.368 & $-$0.091 & $-$0.340 & $-$0.317 \\ 
  & (0.251) & (0.268) & (0.239) & (0.253) \\ 
  Rural area & $-$0.198 & $-$0.269 & $-$0.546$^{*}$ & $-$0.608$^{*}$ \\ 
  & (0.244) & (0.248) & (0.235) & (0.243) \\ 
  Follows politics some of the time & 0.157 & $-$0.155 & 0.135 & 0.576 \\ 
  & (0.395) & (0.406) & (0.375) & (0.347) \\ 
  Follows politics only now and then & 0.657 & 0.812 & 0.996$^{*}$ & 0.297 \\ 
  & (0.442) & (0.468) & (0.459) & (0.478) \\ 
  Follows politics hardly at all & 0.021 & $-$0.127 & $-$0.387 & 0.858 \\ 
  & (0.597) & (0.702) & (0.660) & (0.610) \\ 
  Populism Score & 0.036 & 0.020 & 0.017 & 0.009 \\ 
  & (0.022) & (0.020) & (0.023) & (0.020) \\ 
  Conspiracy Score & 0.149$^{***}$ & 0.172$^{***}$ & 0.147$^{***}$ & 0.158$^{***}$ \\ 
  & (0.015) & (0.015) & (0.014) & (0.014) \\ 
  Republican: Follows politics some of the time & $-$0.422 & $-$0.120 & $-$0.371 & $-$1.300$^{**}$ \\ 
  & (0.478) & (0.491) & (0.455) & (0.446) \\ 
  Independent: Follows politics some of the time & 0.082 & 0.377 & $-$0.452 & $-$0.367 \\ 
  & (0.471) & (0.506) & (0.450) & (0.438) \\ 
  Republican: Follows politics only now and then & $-$0.574 & $-$1.560$^{*}$ & $-$1.871$^{**}$ & $-$1.144 \\ 
  & (0.608) & (0.650) & (0.573) & (0.639) \\ 
  Independent: Follows politics only now and then & $-$0.595 & $-$0.969 & $-$1.570$^{*}$ & 0.446 \\ 
  & (0.591) & (0.685) & (0.641) & (0.631) \\ 
  Republican: Follows politics hardly at all & 0.074 & $-$0.199 & $-$0.029 & $-$1.436 \\ 
  & (0.773) & (0.870) & (0.817) & (0.813) \\ 
  Independent: Follows politics hardly at all & $-$0.941 & 0.734 & $-$0.541 & $-$1.577$^{*}$ \\ 
  & (0.751) & (0.851) & (0.802) & (0.789) \\ 
  Constant & $-$5.425$^{***}$ & $-$5.326$^{***}$ & $-$4.444$^{***}$ & $-$4.407$^{***}$ \\ 
  & (0.615) & (0.603) & (0.554) & (0.563) \\ 
 \hline \\[-1.8ex] 
Observations & 1,835 & 1,814 & 1,832 & 1,807 \\ 
\hline 
\hline \\[-1.8ex] 
\textit{Note:}  & \multicolumn{4}{r}{$^{*}$p$<$0.05; $^{**}$p$<$0.01; $^{***}$p$<$0.001} \\ 
\end{tabular} 
\end{table} 

}

\clearpage

{ \let\oldcentering\centering \renewcommand\centering{\tiny\oldcentering}

\begin{table}[h] \centering 
  \caption{Logistic Regression Results for Election Crime Questions} 
  \label{tab:law_regs_table2_int} 
\footnotesize 
\begin{tabular}{@{\extracolsep{5pt}}lccc} 
\\[-1.8ex]\hline 
\hline \\[-1.8ex] 
 & \multicolumn{3}{c}{\textit{Dependent variable:}} \\ 
\cline{2-4} 
\\[-1.8ex] & Mail Ballot Fraud & Official Tampering & Software Hacking \\ 
\\[-1.8ex] & (1) & (2) & (3)\\ 
\hline \\[-1.8ex] 
 30-44 & 0.091 & $-$0.106 & 0.092 \\ 
  & (0.272) & (0.292) & (0.297) \\ 
  45-64 & $-$0.057 & 0.101 & $-$0.136 \\ 
  & (0.260) & (0.283) & (0.281) \\ 
  65+ & $-$0.204 & $-$0.005 & $-$0.135 \\ 
  & (0.268) & (0.292) & (0.289) \\ 
  Female & 0.206 & 0.317$^{*}$ & 0.283 \\ 
  & (0.150) & (0.161) & (0.173) \\ 
  Black & 0.304 & $-$0.155 & 0.464 \\ 
  & (0.264) & (0.286) & (0.308) \\ 
  Hispanic & 0.265 & 0.231 & 0.447 \\ 
  & (0.265) & (0.302) & (0.317) \\ 
  Other & 0.507 & 0.091 & 0.664$^{*}$ \\ 
  & (0.292) & (0.341) & (0.317) \\ 
  Some college & $-$0.104 & $-$0.322 & 0.104 \\ 
  & (0.204) & (0.213) & (0.223) \\ 
  College grad & $-$0.115 & $-$0.449$^{*}$ & $-$0.018 \\ 
  & (0.214) & (0.228) & (0.235) \\ 
  Postgrad & 0.197 & 0.136 & 0.133 \\ 
  & (0.233) & (0.277) & (0.262) \\ 
  Republican & 1.599$^{***}$ & 0.825$^{*}$ & 1.242$^{**}$ \\ 
  & (0.338) & (0.385) & (0.399) \\ 
  Independent & 0.875$^{**}$ & 0.219 & 0.838$^{*}$ \\ 
  & (0.268) & (0.344) & (0.331) \\ 
  Moderate & 0.308 & $-$0.172 & $-$0.367 \\ 
  & (0.220) & (0.272) & (0.277) \\ 
  Conservative & 0.458 & $-$0.057 & 0.043 \\ 
  & (0.285) & (0.337) & (0.347) \\ 
  Midwest & $-$0.175 & 0.375 & 0.056 \\ 
  & (0.229) & (0.240) & (0.264) \\ 
  South & $-$0.012 & 0.569$^{**}$ & $-$0.108 \\ 
  & (0.204) & (0.216) & (0.250) \\ 
  West & $-$0.102 & 0.206 & $-$0.005 \\ 
  & (0.218) & (0.241) & (0.269) \\ 
  Suburb & 0.103 & $-$0.134 & $-$0.001 \\ 
  & (0.193) & (0.211) & (0.230) \\ 
  Town & 0.113 & $-$0.238 & $-$0.069 \\ 
  & (0.229) & (0.257) & (0.268) \\ 
  Rural area & $-$0.104 & $-$0.773$^{**}$ & $-$0.309 \\ 
  & (0.232) & (0.252) & (0.252) \\ 
  Follows politics some of the time & 0.216 & 0.169 & 0.114 \\ 
  & (0.336) & (0.394) & (0.365) \\ 
  Follows politics only now and then & 0.485 & 0.241 & 0.900 \\ 
  & (0.452) & (0.456) & (0.488) \\ 
  Follows politics hardly at all & 0.201 & 0.294 & 0.203 \\ 
  & (0.647) & (0.664) & (0.770) \\ 
  Populism Score & 0.050$^{*}$ & 0.064$^{**}$ & 0.053$^{**}$ \\ 
  & (0.020) & (0.023) & (0.020) \\ 
  Conspiracy Score & 0.135$^{***}$ & 0.179$^{***}$ & 0.185$^{***}$ \\ 
  & (0.012) & (0.015) & (0.017) \\ 
  Republican: Follows politics some of the time & $-$0.485 & $-$0.211 & $-$0.082 \\ 
  & (0.425) & (0.478) & (0.475) \\ 
  Independent: Follows politics some of the time & $-$0.268 & 0.054 & $-$0.225 \\ 
  & (0.416) & (0.486) & (0.472) \\ 
  Republican: Follows politics only now and then & $-$0.535 & $-$1.105 & $-$1.256 \\ 
  & (0.618) & (0.577) & (0.647) \\ 
  Independent: Follows politics only now and then & $-$0.803 & 0.317 & $-$0.738 \\ 
  & (0.600) & (0.635) & (0.648) \\ 
  Republican: Follows politics hardly at all & $-$0.596 & 0.253 & $-$0.151 \\ 
  & (0.836) & (0.832) & (0.915) \\ 
  Independent: Follows politics hardly at all & $-$0.838 & $-$0.084 & 0.238 \\ 
  & (0.790) & (0.809) & (0.940) \\ 
  Constant & $-$5.010$^{***}$ & $-$6.028$^{***}$ & $-$6.186$^{***}$ \\ 
  & (0.550) & (0.622) & (0.659) \\ 
 \hline \\[-1.8ex] 
Observations & 1,869 & 1,779 & 1,769 \\ 
\hline 
\hline \\[-1.8ex] 
\textit{Note:}  & \multicolumn{3}{r}{$^{*}$p$<$0.05; $^{**}$p$<$0.01; $^{***}$p$<$0.001} \\ 
\end{tabular} 
\end{table} 

}

\clearpage

{ \let\oldcentering\centering \renewcommand\centering{\tiny\oldcentering}

\begin{table}[h] \centering 
  \caption{Logistic Regression Results for Election Crime Questions} 
  \label{tab:law_regs_table3_int} 
\footnotesize 
\begin{tabular}{@{\extracolsep{5pt}}lccc} 
\\[-1.8ex]\hline 
\hline \\[-1.8ex] 
 & \multicolumn{3}{c}{\textit{Dependent variable:}} \\ 
\cline{2-4} 
\\[-1.8ex] & Paying Voters & Registration Fraud & Dropbox Fraud \\ 
\\[-1.8ex] & (1) & (2) & (3)\\ 
\hline \\[-1.8ex] 
 30-44 & 0.014 & $-$0.290 & $-$0.076 \\ 
  & (0.253) & (0.284) & (0.286) \\ 
  45-64 & $-$0.234 & $-$0.126 & $-$0.226 \\ 
  & (0.237) & (0.263) & (0.276) \\ 
  65+ & $-$0.221 & 0.075 & $-$0.426 \\ 
  & (0.252) & (0.273) & (0.285) \\ 
  Female & 0.361$^{*}$ & 0.343$^{*}$ & $-$0.001 \\ 
  & (0.142) & (0.158) & (0.153) \\ 
  Black & 0.595$^{*}$ & 0.569$^{*}$ & 0.272 \\ 
  & (0.266) & (0.269) & (0.274) \\ 
  Hispanic & 0.317 & 0.234 & $-$0.121 \\ 
  & (0.253) & (0.274) & (0.307) \\ 
  Other & 0.241 & 0.189 & 0.173 \\ 
  & (0.289) & (0.298) & (0.301) \\ 
  Some college & $-$0.108 & $-$0.464$^{*}$ & $-$0.079 \\ 
  & (0.193) & (0.200) & (0.209) \\ 
  College grad & $-$0.322 & $-$0.312 & $-$0.094 \\ 
  & (0.201) & (0.217) & (0.217) \\ 
  Postgrad & 0.064 & $-$0.347 & 0.175 \\ 
  & (0.248) & (0.236) & (0.243) \\ 
  Republican & 0.855$^{**}$ & 1.470$^{***}$ & 1.848$^{***}$ \\ 
  & (0.319) & (0.376) & (0.353) \\ 
  Independent & 0.032 & 1.155$^{***}$ & 1.295$^{***}$ \\ 
  & (0.270) & (0.306) & (0.300) \\ 
  Moderate & 0.264 & 0.259 & $-$0.015 \\ 
  & (0.216) & (0.244) & (0.216) \\ 
  Conservative & 0.271 & 0.295 & 0.306 \\ 
  & (0.288) & (0.315) & (0.278) \\ 
  Midwest & 0.117 & 0.090 & $-$0.180 \\ 
  & (0.214) & (0.245) & (0.232) \\ 
  South & 0.033 & 0.013 & $-$0.291 \\ 
  & (0.194) & (0.224) & (0.218) \\ 
  West & $-$0.256 & $-$0.0003 & $-$0.131 \\ 
  & (0.216) & (0.241) & (0.220) \\ 
  Suburb & 0.373$^{*}$ & $-$0.198 & $-$0.069 \\ 
  & (0.184) & (0.201) & (0.201) \\ 
  Town & 0.320 & $-$0.039 & $-$0.199 \\ 
  & (0.226) & (0.234) & (0.237) \\ 
  Rural area & $-$0.161 & $-$0.516$^{*}$ & $-$0.719$^{**}$ \\ 
  & (0.218) & (0.248) & (0.239) \\ 
  Follows politics some of the time & 0.259 & 0.495 & 0.368 \\ 
  & (0.297) & (0.349) & (0.372) \\ 
  Follows politics only now and then & $-$0.019 & 1.175$^{**}$ & 0.827 \\ 
  & (0.417) & (0.442) & (0.434) \\ 
  Follows politics hardly at all & 0.532 & 0.342 & 0.834 \\ 
  & (0.606) & (0.592) & (0.632) \\ 
  Populism Score & 0.022 & 0.018 & 0.022 \\ 
  & (0.020) & (0.019) & (0.022) \\ 
  Conspiracy Score & 0.139$^{***}$ & 0.160$^{***}$ & 0.135$^{***}$ \\ 
  & (0.012) & (0.014) & (0.013) \\ 
  Republican: Follows politics some of the time & $-$0.443 & $-$0.665 & $-$0.685 \\ 
  & (0.385) & (0.441) & (0.456) \\ 
  Independent: Follows politics some of the time & 0.119 & $-$0.723 & $-$0.234 \\ 
  & (0.413) & (0.447) & (0.448) \\ 
  Republican: Follows politics only now and then & $-$0.779 & $-$0.861 & $-$0.931 \\ 
  & (0.571) & (0.627) & (0.568) \\ 
  Independent: Follows politics only now and then & 0.618 & $-$1.953$^{**}$ & $-$0.928 \\ 
  & (0.619) & (0.633) & (0.591) \\ 
  Republican: Follows politics hardly at all & $-$0.722 & $-$0.396 & $-$1.106 \\ 
  & (0.777) & (0.775) & (0.807) \\ 
  Independent: Follows politics hardly at all & $-$0.133 & $-$1.144 & $-$1.608$^{*}$ \\ 
  & (0.770) & (0.768) & (0.779) \\ 
  Constant & $-$4.052$^{***}$ & $-$4.719$^{***}$ & $-$3.940$^{***}$ \\ 
  & (0.519) & (0.570) & (0.533) \\ 
 \hline \\[-1.8ex] 
Observations & 1,763 & 1,824 & 1,771 \\ 
\hline 
\hline \\[-1.8ex] 
\textit{Note:}  & \multicolumn{3}{r}{$^{*}$p$<$0.05; $^{**}$p$<$0.01; $^{***}$p$<$0.001} \\ 
\end{tabular} 
\end{table} 

}

\clearpage

{ \let\oldcentering\centering \renewcommand\centering{\tiny\oldcentering}

\begin{table}[h] \centering 
  \caption{Regression Results for Aggregate Answers to Election Crime Questions} 
  \label{tab:total_law_reg_int} 
\footnotesize 
\begin{tabular}{@{\extracolsep{5pt}}lcc} 
\\[-1.8ex]\hline 
\hline \\[-1.8ex] 
 & \multicolumn{2}{c}{\textit{Dependent variable:}} \\ 
\cline{2-3} 
\\[-1.8ex] & Number of Election Crimes Believed & Proportion of Election Crimes Believed \\ 
\\[-1.8ex] & \textit{survey-weighted} & \textit{survey-weighted} \\ 
 & \textit{normal} & \textit{logistic} \\ 
\\[-1.8ex] & (1) & (2)\\ 
\hline \\[-1.8ex] 
 30-44 & $-$0.006 & $-$0.019 \\ 
  & (0.276) & (0.187) \\ 
  45-64 & $-$0.198 & $-$0.154 \\ 
  & (0.268) & (0.181) \\ 
  65+ & $-$0.367 & $-$0.262 \\ 
  & (0.269) & (0.186) \\ 
  Female & 0.141 & 0.125 \\ 
  & (0.166) & (0.107) \\ 
  Black & 0.054 & 0.187 \\ 
  & (0.279) & (0.184) \\ 
  Hispanic & 0.042 & 0.085 \\ 
  & (0.301) & (0.202) \\ 
  Other & 0.554 & 0.357 \\ 
  & (0.357) & (0.226) \\ 
  Some college & 0.067 & 0.010 \\ 
  & (0.237) & (0.142) \\ 
  College grad & 0.131 & 0.041 \\ 
  & (0.248) & (0.152) \\ 
  Postgrad & 0.393 & 0.216 \\ 
  & (0.261) & (0.175) \\ 
  Republican & 1.820$^{***}$ & 1.257$^{***}$ \\ 
  & (0.401) & (0.249) \\ 
  Independent & 0.458 & 0.626$^{**}$ \\ 
  & (0.256) & (0.210) \\ 
  Moderate & 0.047 & 0.225 \\ 
  & (0.196) & (0.172) \\ 
  Conservative & 0.092 & 0.239 \\ 
  & (0.332) & (0.209) \\ 
  Midwest & 0.113 & 0.042 \\ 
  & (0.243) & (0.161) \\ 
  South & 0.041 & 0.013 \\ 
  & (0.225) & (0.147) \\ 
  West & 0.145 & 0.072 \\ 
  & (0.238) & (0.157) \\ 
  Suburb & $-$0.012 & $-$0.031 \\ 
  & (0.204) & (0.140) \\ 
  Town & 0.157 & 0.048 \\ 
  & (0.262) & (0.169) \\ 
  Rural area & $-$0.555$^{*}$ & $-$0.367$^{*}$ \\ 
  & (0.243) & (0.157) \\ 
  Follows politics some of the time & $-$0.309 & 0.107 \\ 
  & (0.267) & (0.269) \\ 
  Follows politics only now and then & $-$0.207 & 0.291 \\ 
  & (0.366) & (0.281) \\ 
  Follows politics hardly at all & $-$0.384 & 0.131 \\ 
  & (0.566) & (0.451) \\ 
  Populism Score & 0.045 & 0.025 \\ 
  & (0.023) & (0.015) \\ 
  Conspiracy Score & 0.245$^{***}$ & 0.136$^{***}$ \\ 
  & (0.016) & (0.010) \\ 
  Republican: Follows politics some of the time & $-$0.727 & $-$0.566 \\ 
  & (0.459) & (0.327) \\ 
  Independent: Follows politics some of the time & 0.049 & $-$0.234 \\ 
  & (0.408) & (0.326) \\ 
  Republican: Follows politics only now and then & $-$1.228$^{*}$ & $-$0.972$^{**}$ \\ 
  & (0.613) & (0.370) \\ 
  Independent: Follows politics only now and then & $-$0.333 & $-$0.532 \\ 
  & (0.563) & (0.384) \\ 
  Republican: Follows politics hardly at all & $-$0.641 & $-$0.616 \\ 
  & (1.056) & (0.595) \\ 
  Independent: Follows politics hardly at all & $-$0.894 & $-$0.741 \\ 
  & (0.811) & (0.551) \\ 
  Constant & $-$2.510$^{***}$ & $-$4.626$^{***}$ \\ 
  & (0.571) & (0.407) \\ 
 \hline \\[-1.8ex] 
Observations & 2,075 & 2,075 \\ 
Log Likelihood & $-$5,354.161 &  \\ 
Akaike Inf. Crit. & 10,772.320 &  \\ 
\hline 
\hline \\[-1.8ex] 
\textit{Note:}  & \multicolumn{2}{r}{$^{*}$p$<$0.05; $^{**}$p$<$0.01; $^{***}$p$<$0.001} \\ 
\end{tabular} 
\end{table} 

}

\newpage
\clearpage

\subsection{Regression Tables with Interactions (Urban/Rural and Ideology)}

{ \let\oldcentering\centering \renewcommand\centering{\tiny\oldcentering}

\begin{table}[h] \centering 
  \caption{Logistic Regression Results for Election Crime Questions} 
  \label{tab:law_regs_table1_int2} 
\footnotesize 
\begin{tabular}{@{\extracolsep{5pt}}lcccc} 
\\[-1.8ex]\hline 
\hline \\[-1.8ex] 
 & \multicolumn{4}{c}{\textit{Dependent variable:}} \\ 
\cline{2-5} 
\\[-1.8ex] & Multiple Voting & Ballot Tampering & Impersonation & Non-Citizen Voting \\ 
\\[-1.8ex] & (1) & (2) & (3) & (4)\\ 
\hline \\[-1.8ex] 
 30-44 & $-$0.152 & 0.127 & 0.096 & $-$0.190 \\ 
  & (0.306) & (0.285) & (0.281) & (0.286) \\ 
  45-64 & 0.252 & $-$0.094 & 0.194 & 0.165 \\ 
  & (0.294) & (0.273) & (0.265) & (0.264) \\ 
  65+ & 0.258 & $-$0.209 & 0.134 & $-$0.016 \\ 
  & (0.293) & (0.277) & (0.271) & (0.277) \\ 
  Female & 0.124 & 0.405$^{*}$ & 0.024 & 0.214 \\ 
  & (0.157) & (0.163) & (0.149) & (0.157) \\ 
  Black & 0.492 & 0.596 & 0.316 & 0.111 \\ 
  & (0.273) & (0.313) & (0.263) & (0.288) \\ 
  Hispanic & 0.277 & 0.571$^{*}$ & 0.064 & $-$0.150 \\ 
  & (0.314) & (0.274) & (0.284) & (0.263) \\ 
  Other & 0.290 & 0.273 & 0.369 & 0.419 \\ 
  & (0.316) & (0.343) & (0.299) & (0.346) \\ 
  Some college & $-$0.283 & $-$0.235 & $-$0.250 & $-$0.118 \\ 
  & (0.207) & (0.207) & (0.201) & (0.202) \\ 
  College grad & $-$0.010 & $-$0.098 & $-$0.137 & $-$0.256 \\ 
  & (0.224) & (0.221) & (0.216) & (0.216) \\ 
  Postgrad & $-$0.156 & 0.117 & $-$0.068 & 0.013 \\ 
  & (0.250) & (0.256) & (0.248) & (0.259) \\ 
  Republican & 1.149$^{***}$ & 1.292$^{***}$ & 1.027$^{***}$ & 1.429$^{***}$ \\ 
  & (0.284) & (0.313) & (0.276) & (0.281) \\ 
  Independent & 0.547$^{*}$ & 0.505$^{*}$ & 0.266 & 0.974$^{***}$ \\ 
  & (0.219) & (0.249) & (0.216) & (0.223) \\ 
  Moderate & 0.015 & $-$0.252 & 0.429 & 0.094 \\ 
  & (0.418) & (0.425) & (0.378) & (0.392) \\ 
  Conservative & 0.408 & 0.330 & 0.642 & 0.498 \\ 
  & (0.509) & (0.525) & (0.474) & (0.492) \\ 
  Midwest & $-$0.034 & 0.060 & $-$0.005 & $-$0.174 \\ 
  & (0.239) & (0.255) & (0.237) & (0.254) \\ 
  South & $-$0.065 & 0.046 & 0.072 & $-$0.317 \\ 
  & (0.234) & (0.230) & (0.222) & (0.220) \\ 
  West & 0.001 & 0.021 & 0.096 & $-$0.137 \\ 
  & (0.240) & (0.244) & (0.231) & (0.229) \\ 
  Suburb & 0.254 & $-$0.039 & 0.026 & $-$0.267 \\ 
  & (0.482) & (0.466) & (0.453) & (0.422) \\ 
  Town & 0.346 & 0.177 & $-$1.334 & $-$1.320 \\ 
  & (0.545) & (0.539) & (0.721) & (0.868) \\ 
  Rural area & 0.011 & $-$0.008 & $-$0.019 & 0.215 \\ 
  & (0.640) & (0.628) & (0.562) & (0.493) \\ 
  Follows politics some of the time & $-$0.024 & $-$0.111 & $-$0.194 & $-$0.126 \\ 
  & (0.190) & (0.187) & (0.181) & (0.182) \\ 
  Follows politics only now and then & 0.197 & $-$0.176 & $-$0.248 & $-$0.025 \\ 
  & (0.279) & (0.317) & (0.281) & (0.288) \\ 
  Follows politics hardly at all & $-$0.338 & 0.025 & $-$0.653$^{*}$ & $-$0.278 \\ 
  & (0.313) & (0.335) & (0.313) & (0.361) \\ 
  Populism Score & 0.035 & 0.017 & 0.012 & 0.006 \\ 
  & (0.022) & (0.020) & (0.023) & (0.020) \\ 
  Conspiracy Score & 0.151$^{***}$ & 0.177$^{***}$ & 0.151$^{***}$ & 0.157$^{***}$ \\ 
  & (0.015) & (0.015) & (0.014) & (0.013) \\ 
  Suburb: Moderate & 0.263 & 0.168 & $-$0.151 & 0.270 \\ 
  & (0.570) & (0.564) & (0.533) & (0.514) \\ 
  Town: Moderate & $-$0.144 & $-$0.109 & $-$0.541 & $-$0.082 \\ 
  & (0.587) & (0.579) & (0.557) & (0.562) \\ 
  Rural area: Moderate & $-$0.082 & $-$0.066 & 1.333 & 1.134 \\ 
  & (0.662) & (0.696) & (0.808) & (0.948) \\ 
  Suburb: Conservative & $-$0.020 & $-$0.687 & 0.765 & 0.955 \\ 
  & (0.677) & (0.670) & (0.815) & (0.977) \\ 
  Town: Conservative & 0.505 & 0.549 & 0.030 & $-$0.664 \\ 
  & (0.716) & (0.713) & (0.641) & (0.603) \\ 
  Rural area: Conservative & $-$0.719 & $-$0.838 & $-$1.075 & $-$1.106 \\ 
  & (0.734) & (0.729) & (0.658) & (0.622) \\ 
  Constant & $-$5.294$^{***}$ & $-$5.213$^{***}$ & $-$4.258$^{***}$ & $-$4.005$^{***}$ \\ 
  & (0.690) & (0.683) & (0.610) & (0.613) \\ 
 \hline \\[-1.8ex] 
Observations & 1,835 & 1,814 & 1,832 & 1,807 \\ 
\hline 
\hline \\[-1.8ex] 
\textit{Note:}  & \multicolumn{4}{r}{$^{*}$p$<$0.05; $^{**}$p$<$0.01; $^{***}$p$<$0.001} \\ 
\end{tabular} 
\end{table} 

}

\clearpage

{ \let\oldcentering\centering \renewcommand\centering{\tiny\oldcentering}

\begin{table}[h] \centering 
  \caption{Logistic Regression Results for Election Crime Questions} 
  \label{tab:law_regs_table2_int2} 
\footnotesize 
\begin{tabular}{@{\extracolsep{5pt}}lccc} 
\\[-1.8ex]\hline 
\hline \\[-1.8ex] 
 & \multicolumn{3}{c}{\textit{Dependent variable:}} \\ 
\cline{2-4} 
\\[-1.8ex] & Mail Ballot Fraud & Official Tampering & Software Hacking \\ 
\\[-1.8ex] & (1) & (2) & (3)\\ 
\hline \\[-1.8ex] 
 30-44 & 0.125 & $-$0.103 & 0.115 \\ 
  & (0.275) & (0.298) & (0.304) \\ 
  45-64 & $-$0.050 & 0.118 & $-$0.124 \\ 
  & (0.263) & (0.290) & (0.290) \\ 
  65+ & $-$0.182 & 0.021 & $-$0.117 \\ 
  & (0.270) & (0.299) & (0.296) \\ 
  Female & 0.196 & 0.324$^{*}$ & 0.297 \\ 
  & (0.150) & (0.161) & (0.175) \\ 
  Black & 0.384 & $-$0.148 & 0.507 \\ 
  & (0.260) & (0.278) & (0.303) \\ 
  Hispanic & 0.273 & 0.246 & 0.431 \\ 
  & (0.268) & (0.304) & (0.315) \\ 
  Other & 0.576 & 0.107 & 0.719$^{*}$ \\ 
  & (0.299) & (0.341) & (0.316) \\ 
  Some college & $-$0.096 & $-$0.313 & 0.090 \\ 
  & (0.203) & (0.216) & (0.224) \\ 
  College grad & $-$0.120 & $-$0.446 & $-$0.029 \\ 
  & (0.215) & (0.228) & (0.234) \\ 
  Postgrad & 0.224 & 0.150 & 0.114 \\ 
  & (0.233) & (0.280) & (0.259) \\ 
  Republican & 1.340$^{***}$ & 0.660$^{*}$ & 1.030$^{**}$ \\ 
  & (0.268) & (0.315) & (0.335) \\ 
  Independent & 0.654$^{**}$ & 0.235 & 0.668$^{**}$ \\ 
  & (0.199) & (0.249) & (0.253) \\ 
  Moderate & 0.082 & $-$0.333 & $-$0.528 \\ 
  & (0.375) & (0.422) & (0.428) \\ 
  Conservative & 0.769 & $-$0.075 & 0.302 \\ 
  & (0.464) & (0.497) & (0.553) \\ 
  Midwest & $-$0.161 & 0.378 & 0.076 \\ 
  & (0.228) & (0.240) & (0.263) \\ 
  South & $-$0.004 & 0.556$^{**}$ & $-$0.084 \\ 
  & (0.202) & (0.215) & (0.248) \\ 
  West & $-$0.098 & 0.183 & $-$0.002 \\ 
  & (0.217) & (0.240) & (0.266) \\ 
  Suburb & 0.161 & $-$0.309 & $-$0.054 \\ 
  & (0.418) & (0.487) & (0.452) \\ 
  Town & $-$0.286 & $-$0.764 & $-$0.504 \\ 
  & (0.518) & (0.647) & (0.665) \\ 
  Rural area & 0.318 & $-$0.346 & 0.568 \\ 
  & (0.525) & (0.633) & (0.525) \\ 
  Follows politics some of the time & $-$0.093 & 0.081 & $-$0.006 \\ 
  & (0.176) & (0.197) & (0.199) \\ 
  Follows politics only now and then & 0.025 & $-$0.183 & 0.132 \\ 
  & (0.273) & (0.292) & (0.311) \\ 
  Follows politics hardly at all & $-$0.334 & 0.300 & 0.180 \\ 
  & (0.350) & (0.328) & (0.363) \\ 
  Populism Score & 0.046$^{*}$ & 0.061$^{**}$ & 0.050$^{*}$ \\ 
  & (0.021) & (0.023) & (0.021) \\ 
  Conspiracy Score & 0.138$^{***}$ & 0.179$^{***}$ & 0.188$^{***}$ \\ 
  & (0.012) & (0.015) & (0.016) \\ 
  Suburb: Moderate & 0.189 & 0.148 & 0.195 \\ 
  & (0.494) & (0.578) & (0.567) \\ 
  Town: Moderate & $-$0.365 & 0.225 & $-$0.106 \\ 
  & (0.539) & (0.575) & (0.583) \\ 
  Rural area: Moderate & 0.713 & 0.527 & 0.598 \\ 
  & (0.617) & (0.765) & (0.790) \\ 
  Suburb: Conservative & 0.199 & 0.502 & 0.288 \\ 
  & (0.658) & (0.748) & (0.783) \\ 
  Town: Conservative & 0.074 & 0.054 & $-$0.354 \\ 
  & (0.616) & (0.725) & (0.629) \\ 
  Rural area: Conservative & $-$0.947 & $-$0.715 & $-$1.353$^{*}$ \\ 
  & (0.639) & (0.712) & (0.642) \\ 
  Constant & $-$4.833$^{***}$ & $-$5.817$^{***}$ & $-$6.074$^{***}$ \\ 
  & (0.609) & (0.678) & (0.749) \\ 
 \hline \\[-1.8ex] 
Observations & 1,869 & 1,779 & 1,769 \\ 
\hline 
\hline \\[-1.8ex] 
\textit{Note:}  & \multicolumn{3}{r}{$^{*}$p$<$0.05; $^{**}$p$<$0.01; $^{***}$p$<$0.001} \\ 
\end{tabular} 
\end{table} 

}

\clearpage

\newpage
{ \let\oldcentering\centering \renewcommand\centering{\tiny\oldcentering}

\begin{table}[h] \centering 
  \caption{Logistic Regression Results for Election Crime Questions} 
  \label{tab:law_regs_table3_int2} 
\footnotesize 
\begin{tabular}{@{\extracolsep{5pt}}lccc} 
\\[-1.8ex]\hline 
\hline \\[-1.8ex] 
 & \multicolumn{3}{c}{\textit{Dependent variable:}} \\ 
\cline{2-4} 
\\[-1.8ex] & Paying Voters & Registration Fraud & Dropbox Fraud \\ 
\\[-1.8ex] & (1) & (2) & (3)\\ 
\hline \\[-1.8ex] 
 30-44 & $-$0.044 & $-$0.244 & $-$0.051 \\ 
  & (0.265) & (0.283) & (0.292) \\ 
  45-64 & $-$0.237 & $-$0.109 & $-$0.216 \\ 
  & (0.247) & (0.266) & (0.281) \\ 
  65+ & $-$0.230 & 0.103 & $-$0.430 \\ 
  & (0.260) & (0.271) & (0.286) \\ 
  Female & 0.357$^{*}$ & 0.329$^{*}$ & $-$0.008 \\ 
  & (0.143) & (0.157) & (0.153) \\ 
  Black & 0.662$^{*}$ & 0.662$^{*}$ & 0.408 \\ 
  & (0.262) & (0.272) & (0.266) \\ 
  Hispanic & 0.320 & 0.222 & $-$0.130 \\ 
  & (0.248) & (0.274) & (0.312) \\ 
  Other & 0.238 & 0.268 & 0.294 \\ 
  & (0.288) & (0.301) & (0.309) \\ 
  Some college & $-$0.083 & $-$0.445$^{*}$ & $-$0.050 \\ 
  & (0.195) & (0.200) & (0.210) \\ 
  College grad & $-$0.299 & $-$0.317 & $-$0.100 \\ 
  & (0.203) & (0.215) & (0.217) \\ 
  Postgrad & 0.118 & $-$0.323 & 0.207 \\ 
  & (0.251) & (0.238) & (0.245) \\ 
  Republican & 0.596$^{*}$ & 1.081$^{***}$ & 1.432$^{***}$ \\ 
  & (0.260) & (0.294) & (0.271) \\ 
  Independent & 0.104 & 0.618$^{**}$ & 0.973$^{***}$ \\ 
  & (0.201) & (0.224) & (0.205) \\ 
  Moderate & $-$0.149 & 0.131 & $-$0.319 \\ 
  & (0.352) & (0.382) & (0.357) \\ 
  Conservative & 0.164 & 0.569 & 0.492 \\ 
  & (0.418) & (0.474) & (0.463) \\ 
  Midwest & 0.112 & 0.095 & $-$0.153 \\ 
  & (0.213) & (0.243) & (0.230) \\ 
  South & 0.043 & 0.005 & $-$0.279 \\ 
  & (0.191) & (0.224) & (0.216) \\ 
  West & $-$0.261 & $-$0.007 & $-$0.129 \\ 
  & (0.213) & (0.238) & (0.216) \\ 
  Suburb & $-$0.064 & $-$0.081 & $-$0.404 \\ 
  & (0.383) & (0.404) & (0.431) \\ 
  Town & 0.365 & $-$0.123 & $-$0.235 \\ 
  & (0.407) & (0.505) & (0.444) \\ 
  Rural area & $-$0.593 & $-$0.377 & $-$0.062 \\ 
  & (0.574) & (0.543) & (0.483) \\ 
  Follows politics some of the time & 0.103 & $-$0.074 & $-$0.018 \\ 
  & (0.171) & (0.180) & (0.186) \\ 
  Follows politics only now and then & $-$0.204 & 0.206 & 0.093 \\ 
  & (0.280) & (0.283) & (0.255) \\ 
  Follows politics hardly at all & 0.257 & $-$0.288 & $-$0.174 \\ 
  & (0.336) & (0.326) & (0.349) \\ 
  Populism Score & 0.021 & 0.016 & 0.018 \\ 
  & (0.020) & (0.019) & (0.022) \\ 
  Conspiracy Score & 0.141$^{***}$ & 0.162$^{***}$ & 0.137$^{***}$ \\ 
  & (0.012) & (0.014) & (0.013) \\ 
  Suburb: Moderate & 0.685 & 0.036 & 0.661 \\ 
  & (0.480) & (0.496) & (0.515) \\ 
  Town: Moderate & 0.441 & $-$0.332 & 0.151 \\ 
  & (0.475) & (0.527) & (0.571) \\ 
  Rural area: Moderate & $-$0.396 & 0.148 & 0.205 \\ 
  & (0.550) & (0.630) & (0.570) \\ 
  Suburb: Conservative & $-$0.049 & 0.032 & $-$0.193 \\ 
  & (0.543) & (0.642) & (0.610) \\ 
  Town: Conservative & 1.199 & 0.400 & $-$0.225 \\ 
  & (0.661) & (0.659) & (0.584) \\ 
  Rural area: Conservative & 0.071 & $-$0.563 & $-$1.096 \\ 
  & (0.650) & (0.655) & (0.624) \\ 
  Constant & $-$3.777$^{***}$ & $-$4.429$^{***}$ & $-$3.572$^{***}$ \\ 
  & (0.579) & (0.612) & (0.578) \\ 
 \hline \\[-1.8ex] 
Observations & 1,763 & 1,824 & 1,771 \\ 
\hline 
\hline \\[-1.8ex] 
\textit{Note:}  & \multicolumn{3}{r}{$^{*}$p$<$0.05; $^{**}$p$<$0.01; $^{***}$p$<$0.001} \\ 
\end{tabular} 
\end{table} 

}

\clearpage

{ \let\oldcentering\centering \renewcommand\centering{\tiny\oldcentering}

\begin{table}[h] \centering 
  \caption{Regression Results for Aggregate Answers to Election Crime Questions} 
  \label{tab:total_law_reg_int2} 
\footnotesize 
\begin{tabular}{@{\extracolsep{5pt}}lcc} 
\\[-1.8ex]\hline 
\hline \\[-1.8ex] 
 & \multicolumn{2}{c}{\textit{Dependent variable:}} \\ 
\cline{2-3} 
\\[-1.8ex] & Number of Election Crimes Believed & Proportion of Election Crimes Believed \\ 
\\[-1.8ex] & \textit{survey-weighted} & \textit{survey-weighted} \\ 
 & \textit{normal} & \textit{logistic} \\ 
\\[-1.8ex] & (1) & (2)\\ 
\hline \\[-1.8ex] 
 30-44 & 0.004 & $-$0.014 \\ 
  & (0.276) & (0.189) \\ 
  45-64 & $-$0.213 & $-$0.151 \\ 
  & (0.272) & (0.185) \\ 
  65+ & $-$0.375 & $-$0.250 \\ 
  & (0.270) & (0.187) \\ 
  Female & 0.133 & 0.115 \\ 
  & (0.166) & (0.107) \\ 
  Black & 0.169 & 0.274 \\ 
  & (0.277) & (0.183) \\ 
  Hispanic & 0.057 & 0.087 \\ 
  & (0.295) & (0.202) \\ 
  Other & 0.641 & 0.408 \\ 
  & (0.356) & (0.226) \\ 
  Some college & 0.091 & 0.022 \\ 
  & (0.236) & (0.142) \\ 
  College grad & 0.120 & 0.041 \\ 
  & (0.247) & (0.151) \\ 
  Postgrad & 0.433 & 0.247 \\ 
  & (0.260) & (0.175) \\ 
  Republican & 1.400$^{***}$ & 0.893$^{***}$ \\ 
  & (0.346) & (0.201) \\ 
  Independent & 0.367 & 0.397$^{**}$ \\ 
  & (0.198) & (0.152) \\ 
  Moderate & $-$0.020 & 0.130 \\ 
  & (0.326) & (0.278) \\ 
  Conservative & 0.652 & 0.467 \\ 
  & (0.493) & (0.331) \\ 
  Midwest & 0.116 & 0.038 \\ 
  & (0.242) & (0.160) \\ 
  South & 0.059 & 0.013 \\ 
  & (0.224) & (0.146) \\ 
  West & 0.139 & 0.061 \\ 
  & (0.234) & (0.155) \\ 
  Suburb & 0.090 & $-$0.002 \\ 
  & (0.275) & (0.326) \\ 
  Town & 0.074 & $-$0.032 \\ 
  & (0.301) & (0.376) \\ 
  Rural area & 0.152 & $-$0.038 \\ 
  & (0.359) & (0.418) \\ 
  Follows politics some of the time & $-$0.564$^{**}$ & $-$0.251$^{*}$ \\ 
  & (0.199) & (0.126) \\ 
  Follows politics only now and then & $-$0.777$^{**}$ & $-$0.331 \\ 
  & (0.280) & (0.181) \\ 
  Follows politics hardly at all & $-$0.946$^{*}$ & $-$0.418 \\ 
  & (0.400) & (0.234) \\ 
  Populism Score & 0.043 & 0.023 \\ 
  & (0.023) & (0.015) \\ 
  Conspiracy Score & 0.249$^{***}$ & 0.138$^{***}$ \\ 
  & (0.015) & (0.010) \\ 
  Suburb: Moderate & 0.004 & 0.067 \\ 
  & (0.428) & (0.384) \\ 
  Town: Moderate & $-$0.422 & $-$0.151 \\ 
  & (0.489) & (0.383) \\ 
  Rural area: Moderate & 0.145 & 0.168 \\ 
  & (0.527) & (0.456) \\ 
  Suburb: Conservative & $-$0.204 & $-$0.076 \\ 
  & (0.594) & (0.450) \\ 
  Town: Conservative & $-$0.026 & 0.134 \\ 
  & (0.516) & (0.467) \\ 
  Rural area: Conservative & $-$1.642$^{**}$ & $-$0.684 \\ 
  & (0.566) & (0.470) \\ 
  Constant & $-$2.493$^{***}$ & $-$4.429$^{***}$ \\ 
  & (0.610) & (0.464) \\ 
 \hline \\[-1.8ex] 
Observations & 2,075 & 2,075 \\ 
Log Likelihood & $-$5,351.092 &  \\ 
Akaike Inf. Crit. & 10,766.180 &  \\ 
\hline 
\hline \\[-1.8ex] 
\textit{Note:}  & \multicolumn{2}{r}{$^{*}$p$<$0.05; $^{**}$p$<$0.01; $^{***}$p$<$0.001} \\ 
\end{tabular} 
\end{table} 

}

\newpage
\clearpage

\section{Marginal Effect Plots}

\subsection*{Proportion of All Election Fraud Believed}

\includegraphics[width=0.8\textwidth]{Election_Crimes_Survey_2024/data/law_me_plot.png}

\newpage
\clearpage

\subsection*{Voting More Than Once}
\includegraphics[width=0.8\textwidth]{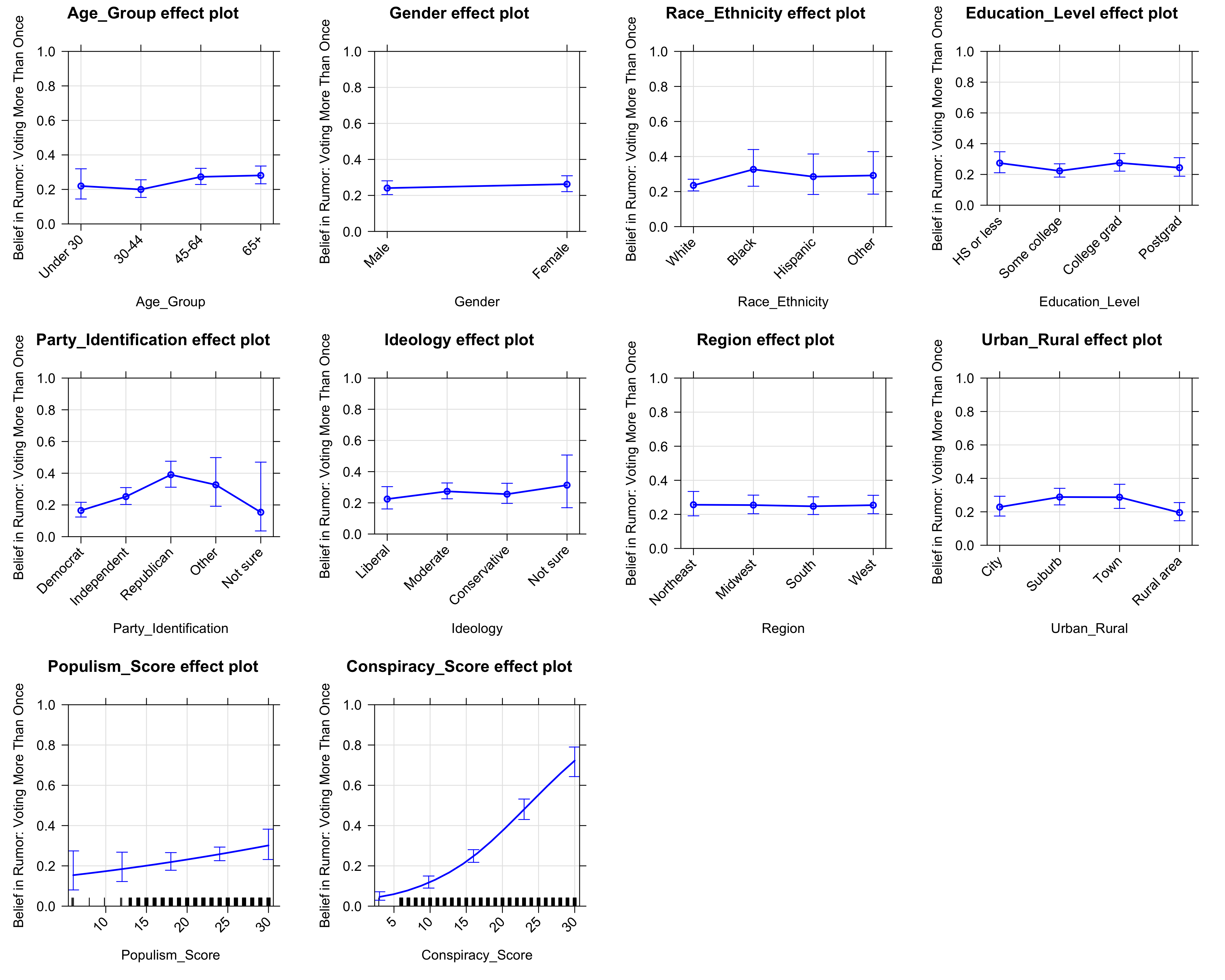}

\newpage
\clearpage

\subsection*{Ballot Tampering}
\includegraphics[width=0.8\textwidth]{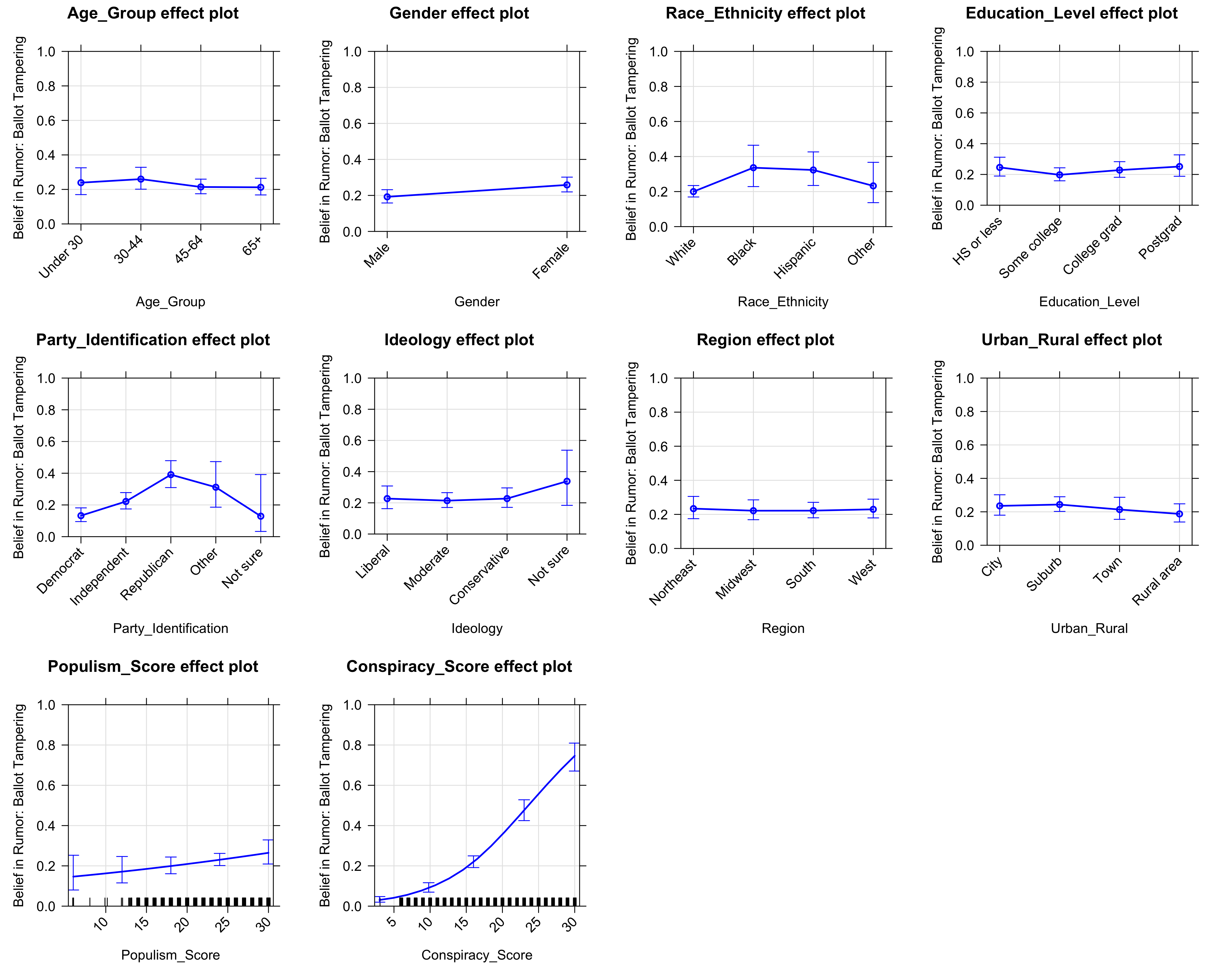}

\newpage
\clearpage

\subsection*{Voter Impersonation}
\includegraphics[width=0.8\textwidth]{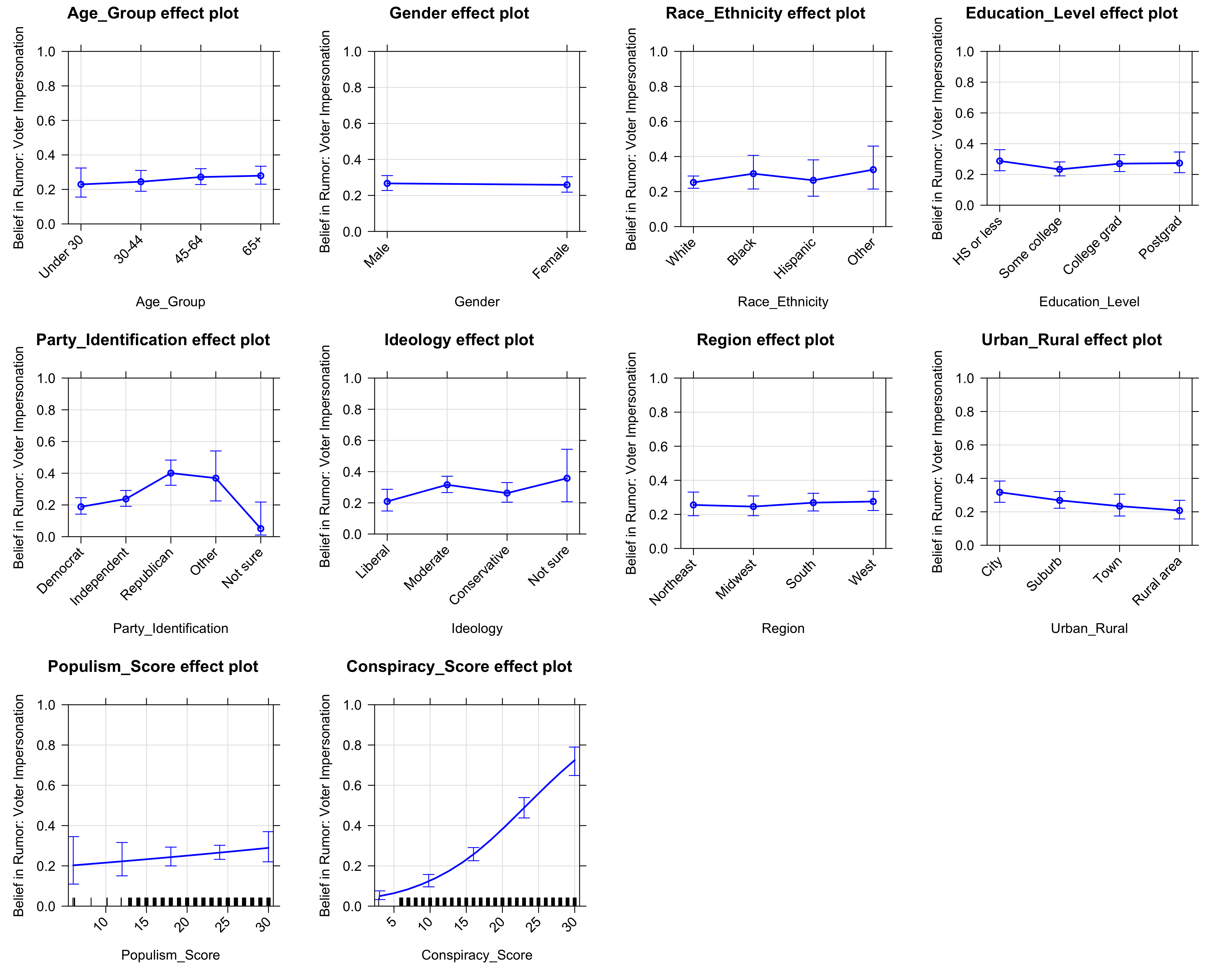}

\newpage
\clearpage

\subsection*{Non-Citizen Voting}
\includegraphics[width=0.8\textwidth]{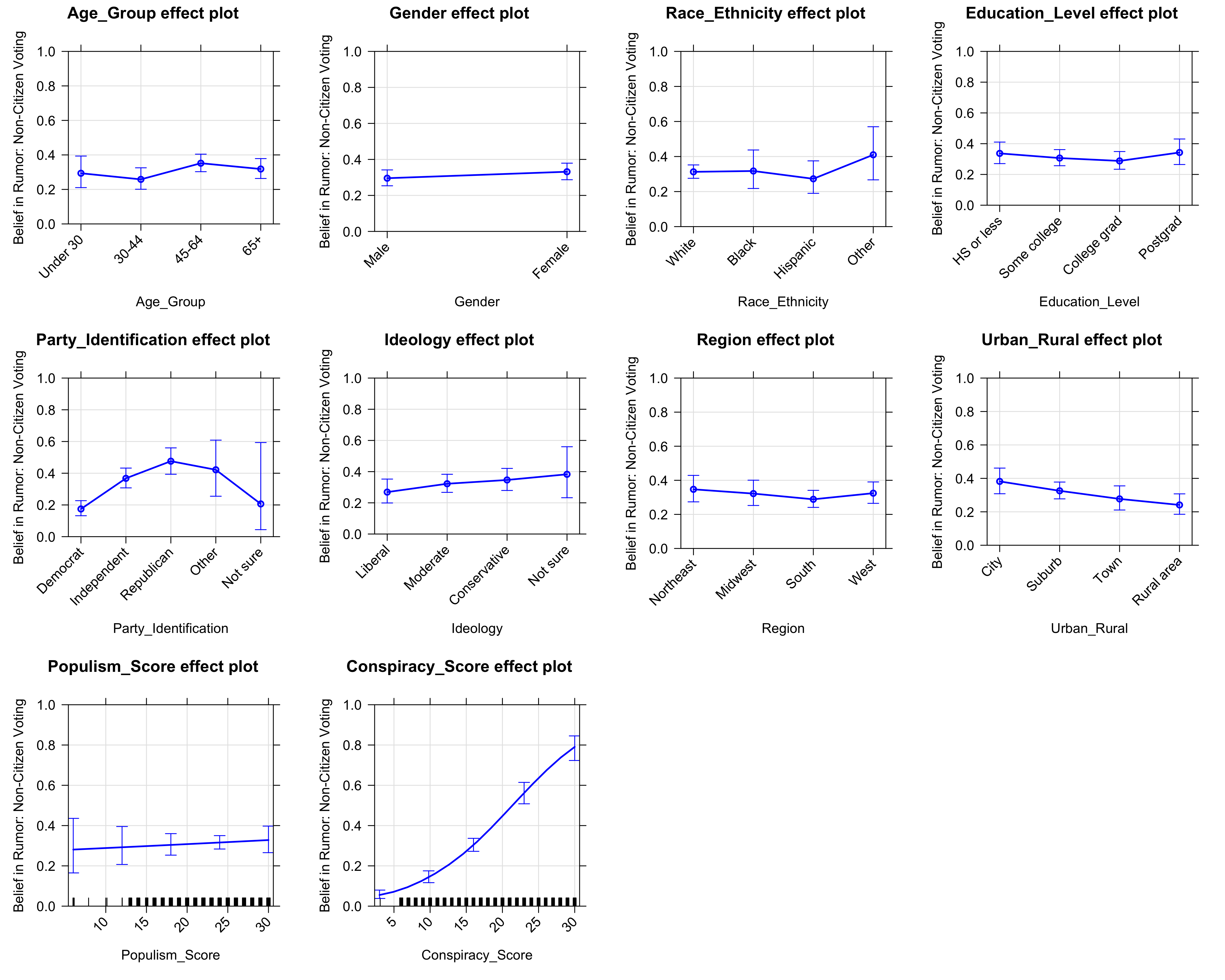}

\newpage
\clearpage

\subsection*{Mail Voting Fraud}
\includegraphics[width=0.8\textwidth]{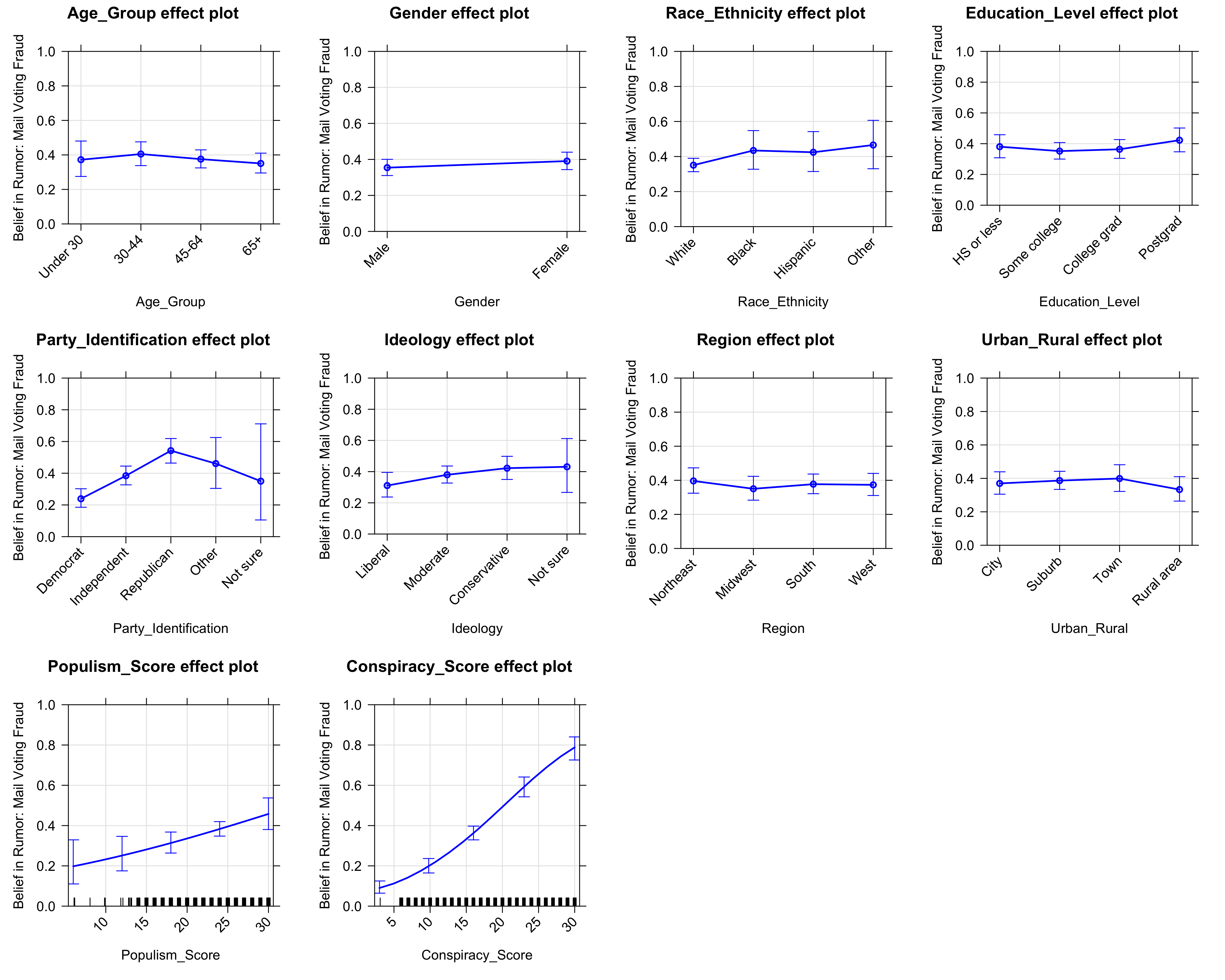}

\newpage
\clearpage

\subsection*{Tampering With Results}
\includegraphics[width=0.8\textwidth]{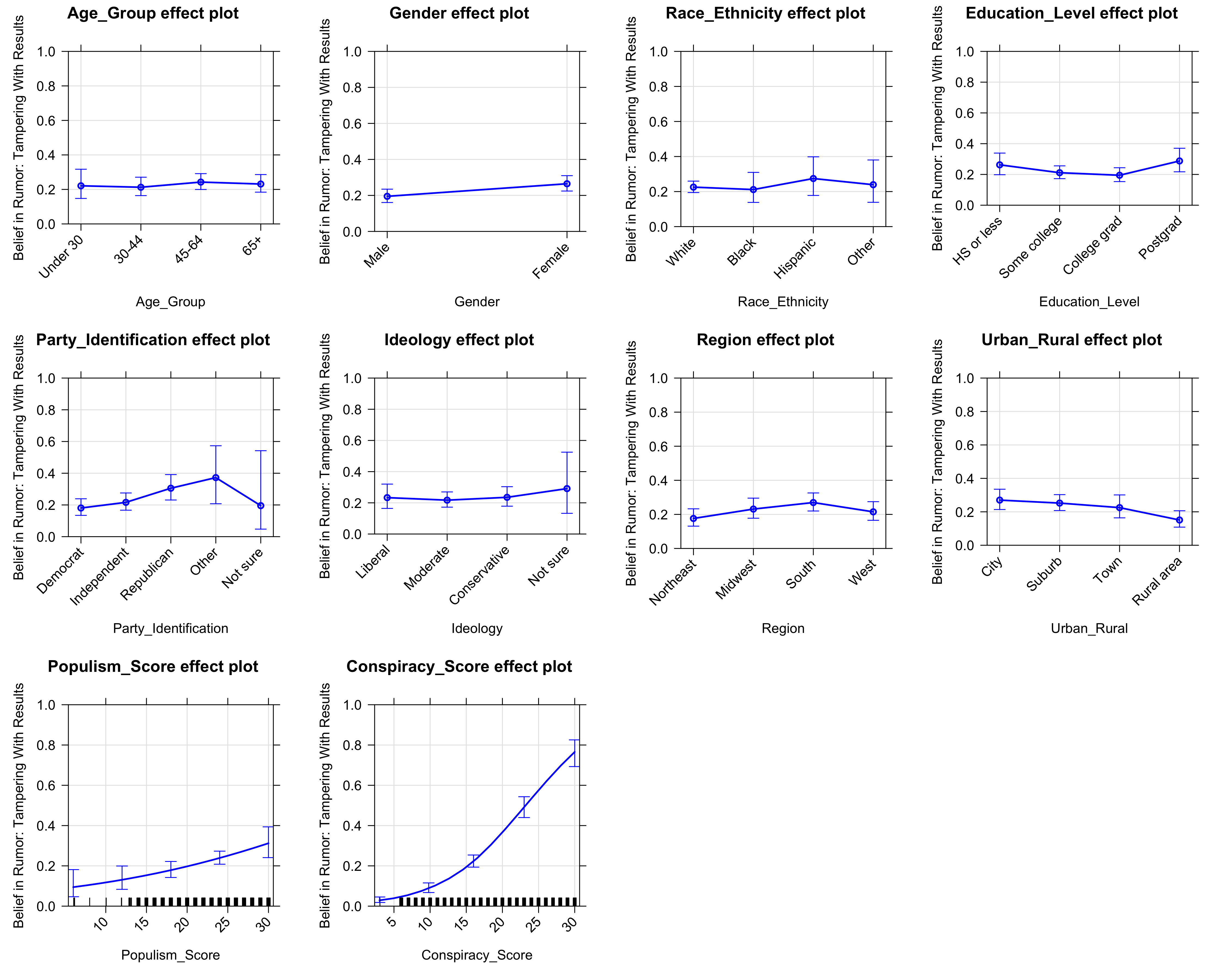}

\newpage
\clearpage

\subsection*{Software Manipulation}
\includegraphics[width=0.8\textwidth]{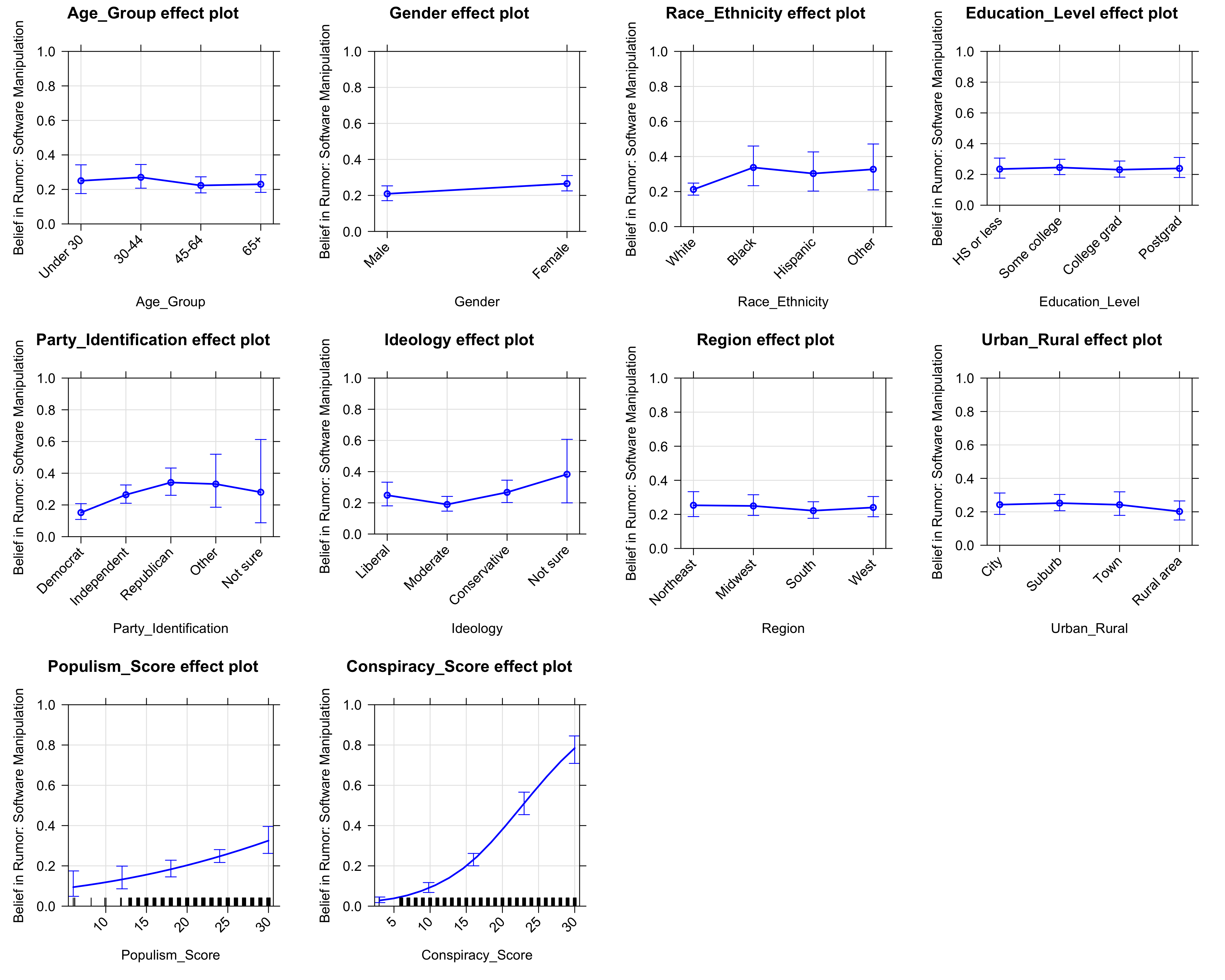}

\newpage
\clearpage

\subsection*{Paying For Votes}
\includegraphics[width=0.8\textwidth]{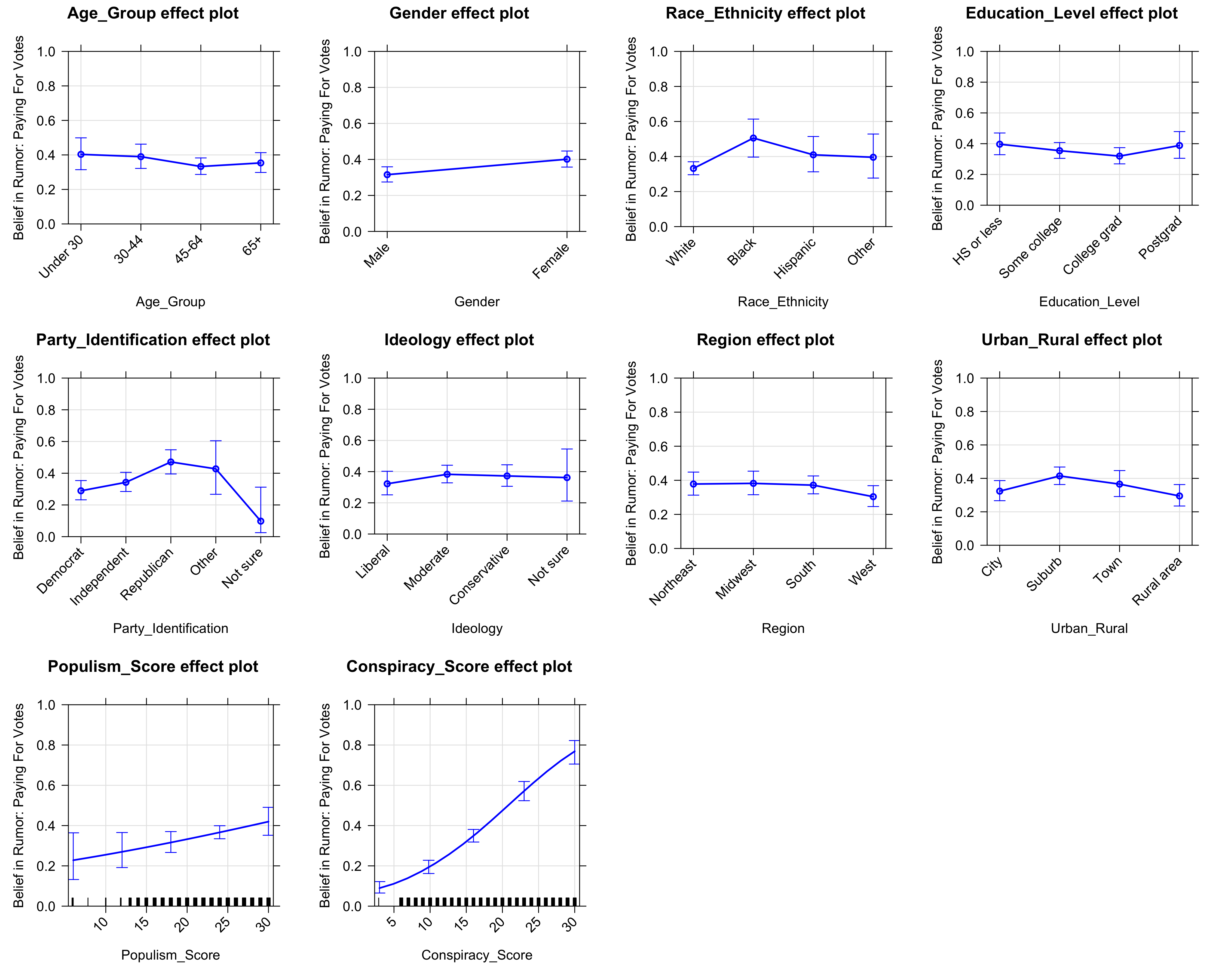}

\newpage
\clearpage

\subsection*{Voter Registration Fraud}
\includegraphics[width=0.8\textwidth]{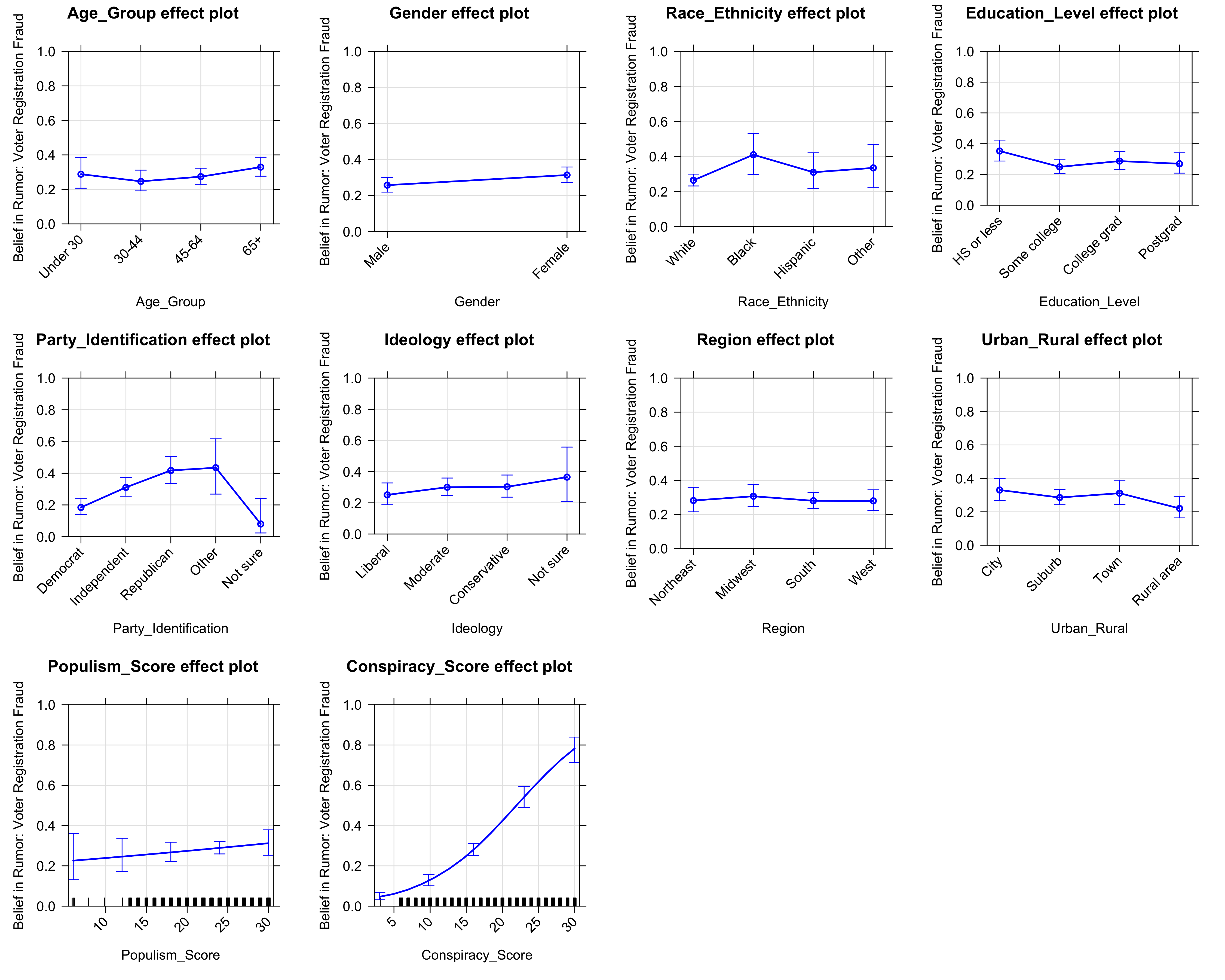}

\newpage
\clearpage

\subsection*{Dropbox Stuffing}
\includegraphics[width=0.8\textwidth]{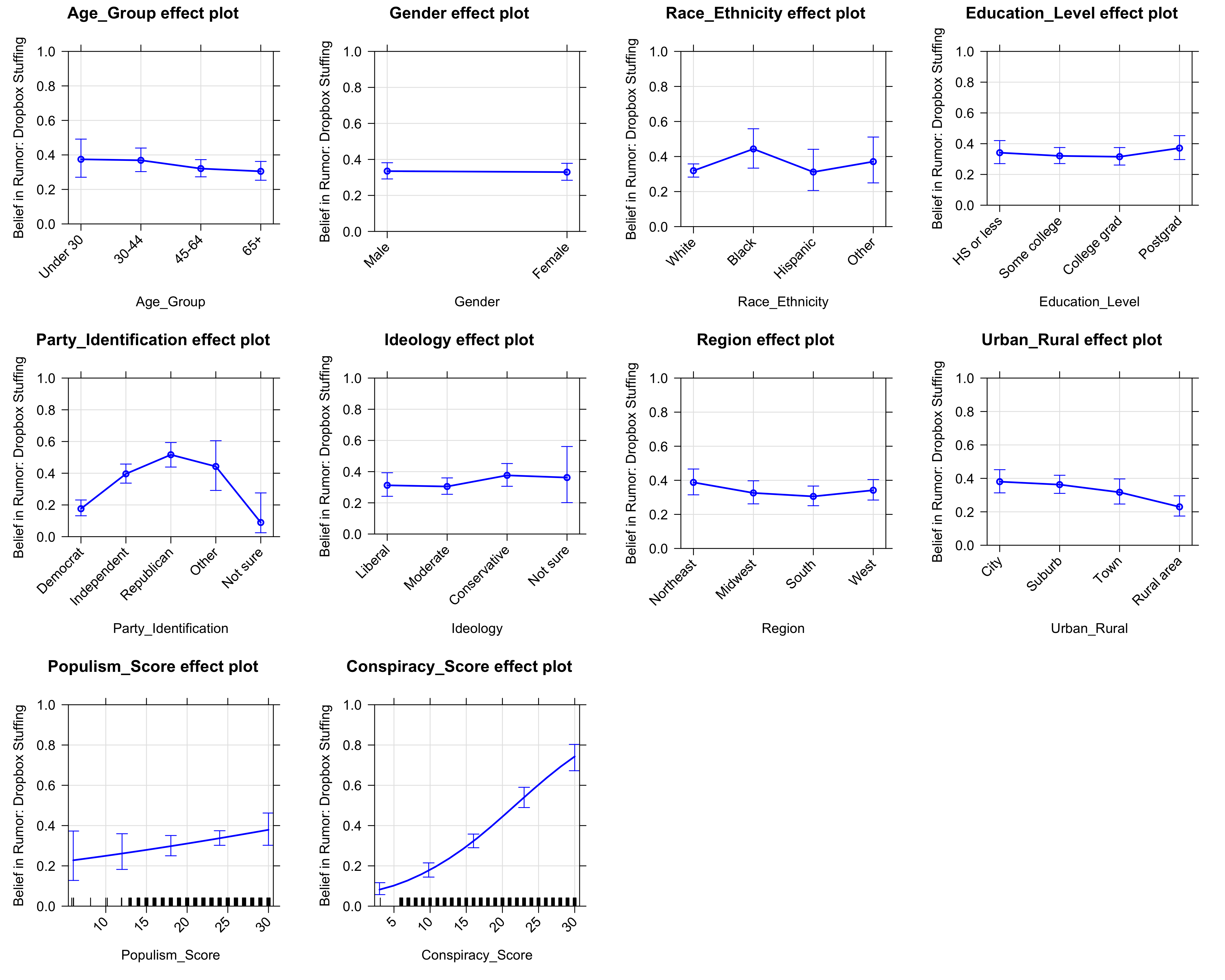}